\documentclass[11pt]{article}

\usepackage{amsfonts}
\usepackage{amsmath}
\usepackage{amssymb}
\usepackage{amsthm}
\usepackage[mathscr]{eucal}
\usepackage{bbold}
\usepackage{bm}
\usepackage{braket}
\usepackage{color}
\usepackage{comment}
\usepackage{dsfont}
\usepackage{framed}
\usepackage{jheppub}
\usepackage{mathtools}
\usepackage{multirow}
\usepackage{physics}
\usepackage[normalem]{ulem}
\usepackage{tensor}
\usepackage{thmtools}
\usepackage{thm-restate}
\usepackage{enumitem}
\usepackage{pifont}
\usepackage{ulem}
\usepackage{tikz}
\usetikzlibrary{math}
\usepackage{graphicx}
\usepackage{caption}
\usepackage{subcaption}
\usepackage{hyperref}

\makeatletter\def\@fpheader{~}\makeatother
\makeatletter
\newcommand*{\defeq}{\mathrel{\rlap{%
                     \raisebox{0.3ex}{$\m@th\cdot$}}%
                     \raisebox{-0.3ex}{$\m@th\cdot$}}%
                     =} 
\makeatother

\newcommand{\rp}{r_+}

\newcommand{\yp}{y_+}

\newcommand{\be}{
\begin{eqnarray}
}

\newcommand{\nl}{
\nonumber \\
}

\newcommand{\ee}{
\end{eqnarray}
}

\allowdisplaybreaks

\numberwithin{equation}{section}

\usepackage{titlesec}

\titleclass{\subsubsubsection}{straight}[\subsection]

\newcounter{subsubsubsection}[subsubsection]
\renewcommand\thesubsubsubsection{\thesubsubsection.\arabic{subsubsubsection}}

\titleformat{\subsubsubsection}
  {\normalfont\normalsize\bfseries}{\thesubsubsubsection}{1em}{}
\titlespacing*{\subsubsubsection}
{0pt}{3.25ex plus 1ex minus .2ex}{1.5ex plus .2ex}

\makeatletter
\renewcommand\paragraph{\@startsection{paragraph}{5}{\z@}%
  {3.25ex \@plus1ex \@minus.2ex}%
  {-1em}%
  {\normalfont\normalsize\bfseries}}
\renewcommand\subparagraph{\@startsection{subparagraph}{6}{\parindent}%
  {3.25ex \@plus1ex \@minus .2ex}%
  {-1em}%
  {\normalfont\normalsize\bfseries}}
\def\toclevel@subsubsubsection{4}
\def\toclevel@paragraph{5}
\def\toclevel@paragraph{6}
\def\l@subsubsubsection{\@dottedtocline{4}{7em}{4em}}
\def\l@paragraph{\@dottedtocline{5}{10em}{5em}}
\def\l@subparagraph{\@dottedtocline{6}{14em}{6em}}

\title{New Well-Posed Boundary Conditions for Semi-Classical Euclidean Gravity}

\author{Xiaoyi Liu,$^a$} \emailAdd{xiaoyiliu@ucsb.edu} 
\author{Jorge E. Santos,$^b$} \emailAdd{jss55@cam.ac.uk}
\author{Toby Wiseman$^c$}
\emailAdd{t.wiseman@imperial.ac.uk }

\affiliation{$^a$Department of Physics, University of California, Santa Barbara, CA 93106, USA}
\affiliation{$^b$Department  of  Applied  Mathematics  and  Theoretical  Physics,  University  of  Cambridge, Wilberforce Road, Cambridge, CB3 0WA, UK}
\affiliation{$^c$Theoretical Physics Group, Blackett Laboratory, Imperial College, London SW7 2AZ, United Kingdom}

\abstract{

We consider four-dimensional Euclidean gravity in a finite cavity.  
Dirichlet conditions do not yield a well-posed elliptic system, and Anderson has suggested boundary conditions that do. 
Here we point out that there exists a one-parameter family of boundary conditions, parameterized by a constant $p$, where a suitably Weyl rescaled boundary metric is fixed, and all give a well-posed elliptic system.
Anderson and Dirichlet boundary conditions can be seen as the limits $p \to 0$ and $\infty$ of these.
Focussing on static Euclidean solutions, we derive a thermodynamic first law. Restricting to a spherical spatial boundary, the infillings are flat space or the Schwarzschild solution, and have similar thermodynamics to the Dirichlet case.
We consider smooth Euclidean fluctuations about the flat space saddle; for $p > 1/6$ the spectrum of the Lichnerowicz operator is stable -- its eigenvalues have positive real part. Thus we may regard large $p$ as a regularization of the ill-posed Dirichlet boundary conditions.
However for $p < 1/6$ there are unstable modes, even in the spherically symmetric and static sector.
We then turn to Lorentzian signature. For $p < 1/6$ we may understand this spherical Euclidean instability as being paired with a Lorentzian instability associated with the dynamics of the boundary itself. However, a mystery emerges when we consider perturbations that break spherical symmetry. Here we find a plethora of dynamically unstable modes even for $p > 1/6$, contrasting starkly with the Euclidean stability we found.
Thus we seemingly obtain a system with stable thermodynamics, but unstable dynamics, calling into question the standard assumption of smoothness that we have implemented when discussing the Euclidean theory.

}

\begin{document}

\maketitle
\large
\section{Introduction}

York originally considered the Einstein equations in a cavity whose boundary is a product of time with a sphere in order to obtain a sensible canonical thermodynamic ensemble for gravity \cite{York:1986it}. Gross, Perry and Yaffe had previously discovered that the Schwarzschild solution in asymptotically flat spacetime suffers from a Euclidean negative mode \cite{Gross:1982cv}, later shown to be directly related to its negative specific heat capacity \cite{Whiting:1988qr,Prestidge:1999uq,Reall:2001ag}, and thus it can never be a dominant saddle point in the Euclidean path integral which computes the canonical partition function. Introducing a finite boundary, York then found that the Schwarzschild solution continuously connected to the asymptotically flat one when the the cavity size is increased, the `small black hole', remains unstable with a negative mode. However a second `large' black hole is now found to exist, always with a size comparable to the sphere of the cavity boundary, but now stable, and for sufficient temperature it dominates the partition function. Shortly afterwards, Hawking and Page discovered precisely the same phenomenon occurs in Anti de Sitter (AdS) spacetime, with the AdS length scale playing an analog role to the size of the cavity in York's construction. For many years this was a curiosity, but with the advent of AdS-CFT it took on a crucial role \cite{Maldacena:1997re,Gubser:1998bc,Witten:1998qj}. A CFT on a sphere should have a sensible partition function, and at high temperature it should have a free energy which scales with the number of local degrees of freedom in the theory. It is precisely the large black hole which  provides the gravitational dual to this behaviour \cite{Witten:1998zw}.

Here we restrict our focus to four-dimensional pure gravity, possibly including a cosmological constant. Thus solutions to the Einstein equations are Einstein metrics, and we are primarily interested in Euclidean geometries with a boundary that takes the form of $S^1 \times \Sigma$, where $S^1$ is interpreted as being Euclidean time, $\tau$, with length $\beta$ so $\tau \sim \tau + \beta$, and $\Sigma$ is a two-dimensional compact Riemannian metric. Then $\beta$ has the thermodynamic interpretation of the inverse temperature. The canonical cavity to consider is a spherical one, so that $\Sigma = S^2$, a round two sphere with a radius $R$, but we wish to consider here the more general setting where the gravitational path integral is thought of as a functional of $\beta$ and $\Sigma$.

An important issue was pointed out by Avramiki and Esposito \cite{Avramidi:1997sh,Avramidi:1997hy} who considered Euclidean quantum gravity on manifolds with boundary, and then discussed by Anderson \cite{Anderson:2006lqb} in the context of Einstein metric infillings in Riemannian geometry. The natural boundary condition one would wish to impose, namely fixing the induced metric on the cavity boundary, so here fixing $\beta$ and the two-geometry $\Sigma$, does not yield a well-posed infilling problem. In Euclidean signature the Einstein equation locally has an elliptic character (after dealing with the issue of coordinate freedom), but this analog of Dirichlet boundary conditions, so fixing the metric of the boundary, does not preserve elliptic regularity of the problem. In \cite{Anderson:2006lqb} it was argued that this may be understood in terms of the Gauss constraint of the boundary. 
Anderson proposed a well-posed set of boundary conditions, which we refer to as `Anderson boundary condition', which entails fixing the conformal class of the boundary (so the induced metric up to a Weyl rescaling) together with the trace of the extrinsic curvature. The lack of well-posedness presents an obstruction to York's original formulation of a stable canonical ensemble for gravity, where the induced metric of the spatial cavity wall, $\beta$ and $\Sigma$, is given as data. In the semi-classical limit we expect  a continuous map from the boundary data at the surface of the cavity to infilling saddle point solutions, and lack of well-posedness for Dirichlet boundary conditions exactly implies the lack of such a map. 
As emphasized in \cite{Avramidi:1997sh,Avramidi:1997hy} this leads to pathologies when computing the Euclidean partition function at  1-loop  as the lack of well-posedness may lead to infinitely many zero modes.
Hence one is naively motivated to consider instead the thermodynamic ensemble given by Anderson's well-posed boundary condition. 

There has been limited study of Anderson boundary condition in the physics context. The implications for gravity were discussed in~\cite{Figueras:2011va} and reviewed in detail in~\cite{Witten:2018lgb}. Recently~\cite{Anninos:2023epi} has shown that in the Lorentzian setting there are linear dynamical instabilities for static spherical cavities, without cosmological constant, that are filled with flat spacetime. They also considered a first law of thermodynamics for static cavities with spherical boundary. In fact stationary rotating black holes in such cavities were actually constructed some time ago in~\cite{Adam:2011dn} -- these are deformed from Kerr due to the requirement the boundary is a round sphere. Perhaps unsurprisingly given the dynamical instabilities found in~\cite{Anninos:2023epi}, in this work we will later see that Anderson boundary condition already yields a pathological partition function for a static spherical cavity filled with flat spacetime due to the existence of Euclidean negative modes.

The main purpose of our paper is to point out a natural generalization of Anderson boundary condition. The gravitational actions relevant for Dirichlet and Anderson boundary conditions differ by the coefficient of the Gibbons-Hawking-York term~\cite{Gibbons:1976ue}. Our generalized boundary conditions derive from considering the coefficient of this term to parameterize a family of choices. 
For these the conformal class is again fixed, with the trace of the extrinsic curvature weighted by a power of the volume element of the induced metric. 
Specifically letting the boundary induced metric be $\gamma$, and the trace of the extrinsic curvature be $K$, these fix the conformal class $[ \gamma ]$ and,
\be
\gamma^p K = \mathrm{fixed}
\ee
where the constant $p$ is related to the coefficient of the Gibbons-Hawking-York term. The case of $p \to 0$ gives the Anderson boundary condition, and $p \to \infty$ corresponds to Dirichlet conditions. 
Alternatively (provided we wish $K$ to be non-vanishing) we may define the Weyl rescaled boundary metric,
\be
\Gamma_{\mu\nu} = K^{\frac{1}{(d-1) p}} \gamma_{\mu\nu}
\ee
and then the boundary condition neatly corresponds to fixing this rescaled metric. 

We will thus call this family of boundary conditions the `generalized conformal boundary conditions'. We show that these are all well-posed in the same sense as Anderson boundary condition is, and further one can regard the (ill-posed) Dirichlet case as a limit of this family of boundary conditions. 

We then explore the physical consequences of these boundary conditions. We begin with the Euclidean theory.
Firstly we show that the first law of thermodynamics can be naturally formulated for static Euclidean cavities, where the rescaled boundary metric $\Gamma$ above is a product of a Euclidean time circle, whose size is the inverse to temperature, and a spatial 2-geometry. 
Secondly we study the Euclidean stability of a static spherical cavity, without cosmological constant, filled with flat spacetime. After an extensive mode analysis we find that for $p < 1/6$ (which includes the case of Anderson) there are Euclidean negative modes, but for $p > 1/6$ there are none. The Euclidean mode that becomes unstable at $p = 1/6$ is in fact spherically symmetric. We also look at the Euclidean stability of black holes, restricted to static spherical fluctuations, and find a similar pattern of instability to the Dirichlet case -- small black holes have an additional unstable mode relative to that of the flat space cavity -- and the thermodynamics reflect this with a York-Hawking-Page first order phase transition between the flat cavity and the large black holes. 
Thus we are lead to conclude that a sensible Euclidean partition function exists for our well-posed boundary conditions provided we take $p > 1/6$. However, there is one mystery. When we consider the specific heat capacity, for $p > 1/6$ and sufficiently large black holes this becomes negative even though they dominate the thermodynamic ensemble and seemingly have no Euclidean negative modes.

We then move on to consider the Lorentzian dynamical stability, as done for Anderson boundary condition in~\cite{Anninos:2023epi}.
We begin with a spherical cavity filled with flat spacetime and consider the spherically symmetric sector. Here
Birkhoff's theorem enforces that the interior geometry is static. However, the boundary may dynamically move in this geometry in a manner similar to the $Z_2$-symmetric branes of the Randall-Sundrum models~\cite{Randall:1999vf}. In fact we may simply solve this dynamics of the boundary non-linearly. In the case $p < 1/6$ we find an instability which acts to shrink the boundary to zero size in finite time, or expand it forever. For $p > 1/6$ the cavity is stable, and at the marginal point $p = 1/6$ this Lorentzian mode is static and precisely corresponds the marginal mode in the Euclidean analysis. For cavities filled with black holes, we find an additional instability for $p > 1/6$ and sufficiently large (depending on the value of $p$) black holes, which precisely correlates with when the specific heat capacity becomes negative. 
Thus we see that such black holes appear well behaved from a Euclidean perspective, but unstable from the Lorentzian one, which is unexpected.
The situation becomes yet more confusing when we consider Lorentzian perturbations that break spherical symmetry. Restricting to a spherical cavity filled with flat spacetime we now find a plethora of unstable (\emph{i.e.} exponentially growing) dynamical modes \emph{even for $p > 1/6$.} 
Thus while these generalized boundary conditions apparently give a sensible Euclidean partition function for $p > 1/6$, they are apparently unstable in the Lorentzian setting. This is a physically unusual situation because one would not normally expect sensible thermodynamics from an unstable dynamical system. It suggests to us that one of the assumptions we have made in considering the Euclidean fluctuations, such as the assumption of smoothness of the geometry, may be invalid.

The plan of this paper is as follows. We begin in section~\ref{sec:newbc} by introducing our new boundary conditions, which include the case of Anderson, and confirm that they are well-posed in the Euclidean setting. 
In section~\ref{sec:Euclidean} we derive a first law of the thermodynamics and then 
compute the Euclidean stability of vacuum filled spherical cavities. We then consider the Lorentzian stability of spherical cavities in section~\ref{sec:Lorentzian}, before concluding with a discussion.

\section{A New One-Parameter Family of Boundary Conditions}
\label{sec:newbc}

In this section, we first introduce our new one-parameter family of boundary conditions in section \ref{sec:BCs}, which is adapted to general coefficients of the Gibbons-Hawking-York term. Then we prove they yield a well-posed elliptic infilling problem in section \ref{sec:Ellipticity}.

\subsection{The Boundary Conditions}
\label{sec:BCs}
Before we detail our novel boundary conditions, let us introduce some terminology. We will be interested in spacetimes $(\mathcal{M}, g)$, where $\mathcal{M}$ is a $d$-dimensional manifold-with-boundary. We are interested in determining which choice of boundary conditions we can impose on the boundary $\partial \mathcal{M}$ so that we have a well-defined variational problem. For the moment, we will work in Euclidean signature, so that $g$ is Riemannian, since it is for this class of spacetimes that we can prove the well-posedness of the new boundary conditions.

Let us first review the action with respect to which variations of the metric satisfying Dirichlet boundary conditions on the boundary $\partial \mathcal{M}$ are stationary. It is well known that for the Dirichlet problem the action reads
\begin{equation}
S=-\frac{1}{16\pi G}\int_{\mathcal{M}}\mathrm{d}^{d}x \sqrt{g}\,(R-2\Lambda)-\frac{1}{8\pi G}\int_{\partial \mathcal{M}}\mathrm{d}^{d-1}x \sqrt{\gamma}\,K\,,
\end{equation}
with $K$ the trace of the extrinsic curvature $K_{\mu\nu}$ computed using an outward pointing unit normal $n$, $\gamma$ the induced metric on $\partial \mathcal{M}$, $R$ the Ricci scalar associated with $g$, $G$ the Newton's constant and $\Lambda$ a cosmological constant. The first term is the standard Einstein-Hilbert action, whereas the second is the Gibbons-Hawking-York term \cite{Gibbons:1976ue}. Varying the action with respect to $g$ yields
\begin{equation}
\delta S=\frac{1}{16\pi G}\int_{\mathcal{M}}\mathrm{d}^d x\sqrt{g} E^{ab}\delta  g_{ab}+\frac{1}{16\pi G}\int_{\partial \mathcal{M}}\mathrm{d}^{d-1}x\,\sqrt{\gamma}\,T^{\mu\nu}\delta \gamma_{\mu\nu}\,\label{eq:firstvariation}
\end{equation}
where lower case Latin indices are bulk indices, lower case Greek indices run over the boundary $\partial \mathcal{M}$ and
\begin{equation}
T_{\mu\nu} \equiv  K_{\mu\nu}-\gamma_{\mu \nu}K\quad\text{and}\quad E_{ab}\equiv R_{ab}-\frac{1}{2}g_{ab}R+\Lambda g_{ab}\,,
\end{equation}
with $R_{ab}$ the components of the Ricci tensor associated with $g$. One can see from (\ref{eq:firstvariation}) that imposing Dirichlet conditions on the boundary $\partial \mathcal{M}$ and the bulk equations of motion $E_{ab}=0$, do give rise to a well-defined variational problem.

We will now ask what boundary conditions we should choose to render the first variation of the action
\begin{equation}
\label{eq:newaction}
S_\Theta=-\frac{1}{16\pi G}\int_{\mathcal{M}}\mathrm{d}^{d}x \sqrt{g}\,(R-2\Lambda)-\frac{\Theta}{8\pi G}\int_{\partial \mathcal{M}}\mathrm{d}^{d-1}x \sqrt{\gamma}\,K\,,
\end{equation}
stationary with respect to metrics satisfying $E_{ab}=0$ for fixed values of $\Theta$. To our knowledge such general class of boundary term has never been considered in the literature and is the main objective of this section of our paper.

We now vary the action (\ref{eq:newaction}) with respect to $g$ and keep track of all boundary terms. This yields
\begin{subequations}
\begin{align}
\delta S_\Theta =& \frac{1}{16\pi G}\int_{\mathcal{M}}\mathrm{d}^d x \sqrt{g}\,E^{ab}\delta g_{ab}+\frac{1}{16\pi G}\int_{\partial \mathcal{M}}\mathrm{d}^{d-1}x\sqrt{\gamma}\,T^{\mu\nu}\delta \gamma_{\mu\nu} \nonumber\\
&+\frac{1}{16\pi G}\int_{\partial \mathcal{M}}\mathrm{d}^{d-1} x\,\delta(2\sqrt{\gamma}  K )-\frac{\Theta}{8\pi G}\int_{\partial \mathcal{M}}\mathrm{d}^{d-1}x\,\delta(\sqrt{\gamma}\,K)\,.
\label{eq:var1}
\end{align}
with
\begin{equation}
E^{ab}\equiv R^{ab}-\frac{1}{2}g^{ab}R+\Lambda g^{ab}\quad\text{and}\quad T^{\mu\nu}\equiv K^{\mu\nu}-K\,\gamma^{\mu\nu}\,,
\end{equation}
\end{subequations}
with the latter being the so called Brown-York tensor \cite{Brown:1992br}. Since we are interested in spacetimes that satisfy the bulk Einstein equation $E_{ab}=0$, the first term in (\ref{eq:var1}) vanishes.

To proceed, we decompose the variation of the boundary metric $\delta \gamma_{\mu\nu}$ as
\begin{equation}
\delta \gamma_{\mu\nu}=\delta \tilde{\gamma}_{\mu\nu}+\frac{1}{d-1}\frac{\delta \gamma}{\gamma}\gamma_{\mu\nu}\,,
\label{eq:conformal}
\end{equation}
with $\delta \tilde{\gamma}_{\mu\nu}$ being traceless with respect to $\gamma^{\mu\nu}$, \emph{i.e.}
\begin{equation}
\gamma^{\mu\nu}\delta \tilde{\gamma}_{\mu\nu}=0\,.
\label{eq:traceless}
\end{equation}
One might wonder why we chose to introduce a factor of $1/\gamma$ in the last term of the decomposition (\ref{eq:conformal}). This was done to align variations of the determinant of $\gamma_{\mu\nu}$ with $\delta \gamma$, thereby justifying the nomenclature.

Using our decomposition (\ref{eq:conformal}) we can write the first order variation as
\begin{subequations}
\begin{align}
\delta S_\Theta 
= & \frac{1}{16\pi G}\int_{\partial \mathcal{M}}\mathrm{d}^{d-1}x\sqrt{\gamma}\, \tilde{K}^{\mu\nu}\delta \tilde{\gamma}_{\mu\nu}+\frac{1-\Theta}{8\pi G}\int_{\partial \mathcal{M}}\mathrm{d}^{d-1}x \sqrt{\gamma}\,\gamma^{-p}\delta\left(\gamma^p K\right)\,.
\end{align}
where we identify 
\begin{equation}
p=\frac{1}{2}\frac{1}{1-\Theta}\left(1-\Theta-\frac{d-2}{d-1}\right)\quad \Leftrightarrow \quad \Theta = \frac{2 p (d-1)-1}{(d-1) (2 p-1)}
\label{eq:finalp}
\end{equation}
\end{subequations}
and where we have defined $\tilde{K}^{\mu\nu}$ as the traceless part of the extrinsic curvature, which equals the traceless part of the Brown-York stress tensor, 
\begin{equation}
\tilde{K}^{\mu\nu}=K^{\mu\nu}-\frac{\gamma_{\alpha\beta}K^{\alpha\beta}}{d-1}\gamma^{\mu\nu}\,.
\end{equation}
Note that $\Theta<0$ when $\frac{1}{2(d-1)}<p<\frac{1}{2}$.

Now we see that we have a good variational principle if we require boundary conditions on $\partial \mathcal{M}$ such that,
\begin{equation}
\delta\left({\gamma}^p K \right)=0
\end{equation}
together with either,
\begin{equation}
\delta \tilde{\gamma}_{\mu\nu}=0 
\end{equation}
or,
\begin{equation}
\tilde{K}^{\mu\nu} = 0 \; .
\end{equation}
In this paper we will study the first class of boundary conditions.
The second class of boundary conditions is also of interest, 
providing a generalization of the Neumann condition $K_{\mu\nu} = 0$, but will be studied elsewhere. 

The condition $\delta \tilde{\gamma}_{\mu\nu}=0 $  implies that we are making a statement about the conformal structure of the boundary metric, since $\delta \tilde{\gamma}_{\mu\nu}$ has a well defined conformal weight under Weyl transformations. We can formulate our  boundary condition using this language. 
\begin{enumerate}[leftmargin=*,labelindent=2em]
\item 
Fix the conformal class of the boundary metric $\gamma_{\mu\nu}$ -- that is to say we fix the boundary metric $\gamma_{\mu\nu}$ to be some given metric up to a Weyl transformation. This precisely yields $\delta \tilde{\gamma}_{\mu\nu}=0 $.
\item
Fix,
\begin{equation}
\gamma^p K = \mathfrak{f}
\end{equation}
where $\mathfrak{f}$ is a scalar density  on $\partial \mathcal{M}$ with the appropriate weight (given by $p$) such that this expression transforms correctly under coordinate changes.
\end{enumerate}

We see that for $p = 0$, so $\Theta = 1/(d-1)$, then $\mathfrak{f}$ is simply a function and these reduce to pure Anderson boundary conditions as described in \cite{Anderson:2006lqb}. Furthermore, it is evident from the expression for $\Theta$ (see Eq. (\ref{eq:finalp})) that Dirichlet boundary conditions can be attained as a somewhat singular limits when $p\to \pm\infty$ when $\Theta \to 1$. In this scenario, we may more conveniently write the second condition as fixing $\gamma K^{\frac{1}{p}}$, and then taking $p \to \infty$ we see that this is essentially equivalent to fixing $\gamma$, the determinant of the metric. Then in combination with fixing the conformal class, this yields standard Dirichlet boundary conditions. 
As we will soon show, these boundary conditions are well-posed for any finite $p$, and hence we may regard large positive $p$ as a regulator for Dirichlet boundary conditions that makes them well-posed. For large $|p|$ and fixed length scales, the boundary conditions tend to the behaviour of Dirichlet. For example, the spectrum of the Lichnerowicz operator will agree for modes of eigenvalue with magnitude below some fixed large value as $p \to \pm \infty$. However, a subtle point is that there is no guarantee that there are no instabilities that are introduced at short distances, or high eigenvalues. Indeed as $p \to -\infty$ we will see exactly this, that in the Euclidean setting, there are negative modes of the Lichnerowicz operator that have very large magnitude as $p \to - \infty$. In the positive $p \to + \infty$ the high-lying spectrum appears to be stable, and thus we may regard $p \to + \infty$ as a good regulator of Dirichlet boundary conditions, whereas $p \to - \infty$ is not.

We noted above that two special values of $p$ are $p = \frac{1}{2(d-1)}$ and $p = \frac{1}{2}$, where $\Theta$ vanishes and diverges, respectively.
For $p = \frac{1}{2(d-1)}$, which corresponds to $p = \frac{1}{6}$ in four spacetime dimensions, we observe that $\Theta$ vanishes. Therefore, the action in~\eqref{eq:newaction} consists solely of the bulk Einstein-Hilbert and cosmological terms. In the case where there is no cosmological constant, this implies that the solutions exhibit scale invariance, given that the Einstein-Hilbert term possesses this property. Taking any solution, $\dd s^2 = g_{ab}(x) \dd x^a \dd x^b$, then for any constant $\lambda > 0$, $\dd s^2 = \lambda^2 g_{ab}(x) \dd x^a \dd x^b$ is also a solution to the bulk Einstein equation and boundary conditions. This can be explicitly observed in the boundary conditions mentioned above. Fixing the conformal class is a scale-invariant condition. Given the scale transformation $\gamma \to \lambda^{d-1} \gamma$ and $K \to \lambda^2 K$, fixing $\gamma^p K$ is scale-invariant precisely for $p = \frac{1}{2(d-1)}$.

Now, consider the special case $p \to \frac{1}{2}$ where instead $\Theta \to \infty$. Fixing $\gamma^p K$ becomes fixing $\sqrt{\gamma} K$, which is precisely the integrand of the boundary term. As a consequence, according to equation~\eqref{eq:var1}, it is evident that $\Theta$ must diverge to counterbalance the boundary term arising from the integration by parts of the variation of the Einstein-Hilbert term in the action.

In the case that we impose the trace of the extrinsic curvature is non-vanishing, so the scalar density $\mathfrak{f}$ is chosen not to vanish, then we may reformulate the boundary condition by defining,
\begin{equation}
\label{eq:GammaDefn}
\Gamma_{\mu\nu}= K^{\frac{1}{(d-1)p}}\gamma_{\mu\nu} 
\end{equation}
and noting that since $K \ne 0$ then $\Gamma_{\mu\nu}$ is also a metric on the boundary, in the same conformal class as $\gamma_{\mu\nu}$. We will refer to this as the \emph{rescaled boundary metric}. 
\footnote{
Strictly we see that $\Gamma_{\mu\nu}$ as defined in~\eqref{eq:GammaDefn} doesn't have the correct units for a metric. We can correct this simply by instead defining,
\begin{equation}
\Gamma_{\mu\nu}= \left( \frac{K}{K_0} \right)^{\frac{1}{(d-1)p}}\gamma_{\mu\nu} \nonumber
\end{equation}
for an appropriate constant $K_0$. However since this constant plays no role except to rescale $\Gamma_{\mu\nu}$, for convenience we will always set it to one in our units.}
One can then show that the variation of the action can neatly be written as,
\begin{subequations}
\begin{align}
\delta S_\Theta = \frac{1}{2}\int_{\partial \mathcal{M}}\mathrm{d}^{d-1}x\sqrt{\Gamma}\,\mathcal{T}^{\mu\nu}\delta \Gamma_{\mu\nu}\,.
\label{eq:finalaction}
\end{align}
with,
\begin{equation}
\mathcal{T}^{\mu\nu}\equiv \frac{1}{8\pi G}K^{-\frac{d+1}{2(d-1)p}}(K^{\mu\nu}-\Theta \gamma^{\mu\nu}K)\,.
\end{equation}
\end{subequations}
being a generalization of the Brown-York stress tensor. Now we may simply state the boundary condition as being a Dirichlet condition on this rescaled boundary metric $\Gamma_{\mu\nu}$. We will find this perspective very useful when considering thermodynamics in a cavity with these boundary conditions later.

It is important to emphasize that this formulation of the boundary condition, while elegant, cannot capture the case where we wish to impose that $K$ vanishes or changes sign. However, we note that particularly for spherical cavities, it is natural to wish to impose a strictly positive extrinsic curvature.

\subsection{Ellipticity of the Conformal Boundary Conditions}\label{sec:Ellipticity}
The main advantage of conformal boundary conditions is that these give rise to well-posed elliptic infilling problem. In what follows we will show that this is the case for any finite value of $p$.

To demonstrate the well-posedness of our novel boundary conditions, we will work at the linear level by setting $g_{ab}=\hat{g}_{ab}+h_{ab}$, where the hatted quantities denote a background metric, and $h_{ab}$ is assumed to be infinitesimally small compared to $\hat{g}_{ab}$. Next, we will choose a suitable gauge for our purposes, which, as it turns out, is the de Donder gauge:
\begin{equation}\label{eq:dedonderg}
C_b\equiv \hat{\nabla}^a h_{ab}-\frac{1}{2}\hat{\nabla}_b h = 0,
\end{equation}
where $h=\hat{g}^{ab}h_{ab}$ represents the trace of the metric perturbation. 

We will follow the approach presented in \cite{Witten:2018lgb} to review why Dirichlet boundary conditions are not well-posed in this gauge. Since the issue of well-posedness pertains to high-frequency behavior, we will consider our manifold with boundary to be the half-space $\mathbb{R}_+^d$ within a flat Euclidean space $\mathbb{R}^d$, with $x_{\perp}\geq 0$. The boundary is simply defined by $x_{\perp}=0$, and has vanishing extrinsic curvature.

A general plane wave solution takes the form:
\begin{equation}
h_{ab} = \alpha_{ab}e^{i \mathbf{y}\cdot \mathbf{K}},
\end{equation}
where $\mathbf{y}\in\mathbb{R}^d$. We aim to explore solutions that may have non-trivial behavior along the boundary while decaying away from it. For this reason, we consider $\mathbf{y}=\{\vec{x},x_{\perp}\}$ and $\mathbf{K}=\{\vec{k},i k_{\perp}\}$, where $\vec{x}$ and $\vec{k}$ belong to $\mathbb{R}^{d-1}$ and $k_{\perp}>0$. This choice yields
\begin{equation}
h_{ab} = \alpha_{ab}\,e^{i \vec{x}\cdot \vec{k}-|\vec{k}|x_{\perp}},
\label{eq:simplesol}
\end{equation}
with the perturbed Einstein equation locking $k_{\perp}=|\vec{k|}$. Our goal is to investigate whether nontrivial solutions of the above form can persist when we impose Dirichlet boundary conditions and the de Donder gauge for arbitrary wavevectors $\vec{k}$. If this turns out to be the case, it indicates that the boundary conditions are not well-posed in the chosen gauge. Physically we may view this as saying that we would then have a class of modes that have arbitrarily short wavelength along the boundary direction, while decaying immediately away from the boundary into the bulk. Hence the bulk solution would not correspond to a unique boundary geometry.

We can now analyze this component by component. When $x_{\perp}=0$, the Dirichlet boundary conditions imply $\alpha_{ij}=0$, where $i$ and $j$ denote the directions parallel to the boundary. Let $\vec{\alpha}_i=\alpha_{i\,\perp}$ and $\beta=\alpha_{\perp\perp}$, with $i=1,2,\ldots,d-1$. It then follows that
\begin{subequations}
\begin{equation}
C_{\perp}=0\Rightarrow i\,\vec{k}\cdot \vec{\alpha}-\frac{1}{2}\beta |\vec{k}|=0\,
\end{equation}
and
\begin{equation}
C_{i}=0\Rightarrow -\vec{\alpha} |\vec{k}|-\frac{1}{2}\beta\,i\,\vec{k}=0\,.
\end{equation}
\end{subequations}
Both of the above can be solved if we take
\begin{equation}
\vec{\alpha}=-\frac{i}{2}\beta\frac{\vec{k}}{|\vec{k}|}\,,
\end{equation}
thus showing that Dirichlet conditions are not well-posed in the de Donder gauge.

We would like to repeat the above for our general class of boundary conditions and show that for any non-trivial $\vec{k}$ we necessarily have $\alpha_{\mu\nu}=0$. While the conformal class is fixed, our boundary conditions do allow the perturbation in the induced metric, $h_{ij}$, to be non-zero at the boundary $x_{\perp}=0$, so long as $h_{ij}=\frac{\delta \gamma}{d-2}\hat{g}_{ij}$ at $x_{\perp}=0$, with $\delta \gamma$ now being infinitesimally small. This is a fluctuation resulting from an infinitesimal Weyl rescaling.

The linearisation of the condition $\delta(\gamma^p K)=0$,
in coordinates where $\hat{g}_{\perp\perp}=1$ and $\hat{g}_{\perp i}=0$, about our simple wall yields
\begin{equation}
\partial_{x_{\perp}}h^{i}_{\phantom{i}i}-2 \partial^i h_{\perp i}=0\,.
\end{equation}
Next we consider solutions of the form (\ref{eq:simplesol}), which brings the above to
\begin{subequations}\label{eq:EllipticConformal}
\begin{equation}
2i\vec{\alpha}\cdot \vec{k}+|\vec{k}|\delta \gamma=0\,,
\end{equation}
while from the de Donder gauge condition we get
\begin{equation}
i\,\vec{k}\cdot \vec{\alpha}-\frac{1}{2}\beta |\vec{k}|+\frac{1}{2}\delta \gamma |\vec{k}|=0
\end{equation}
and
\begin{equation}
-\vec{\alpha} |\vec{k}|-\frac{1}{2}\beta\,i\,\vec{k}+\frac{1}{2}\delta \gamma\,i\,\vec{k}=0\,.
\end{equation}
\end{subequations}
It is a simple exercise to check that all the above conditions can only be satisfied if $\vec{\alpha}=0$, $\beta=0$, and $\delta \gamma=0$, thus enforcing $\alpha_{ab}=0$ and showing that our new boundary conditions give rise to a well-posed elliptic problem. The reader might note that the exponent $p$ does not make an appearance in the above calculations.
This is due to the fact that when discussing ellipticity we focus on short wavelengths, so that the boundary effectively has zero extrinsic curvature, so then $\delta(\gamma^p K) = \gamma^p \delta( K )$, which simply implies $\delta K = 0$ -- this is the same condition as for the Anderson case ($p=0$), and hence our generalized conformal boundary conditions are well-posed for the same reason as for Anderson boundary condition, with no dependence on the (finite) value of $p$.

We note that the ill-posed Dirichlet boundary conditions correspond to the limit $p \to \infty$, and then one must take care considering the short wavelength limit as we do here -- in fact the order of limits is important, and taking $p \to \infty$ first leads to the last term in each of the equations~\eqref{eq:EllipticConformal} vanishing, and yields the usual ill-posedness of Dirichlet conditions.

\section{Euclidean Gravitational Path Integral}
\label{sec:Euclidean}
Next we delve into the realm of Euclidean gravitational path integrals, drawing inspiration from the perspective introduced by Gibbons and Hawking \cite{Gibbons:1976ue}. Again we consider a $d$-dimensional manifold $\mathcal{M}$ with boundary $\partial \mathcal{M}$ equipped with a metric $g_{ab}$ and boundary metric $\gamma_{\mu\nu}$. On $\partial \mathcal{M}$ we impose the boundary conditions discussed in section \ref{sec:newbc} where (provided $K$ is non-vanishing) we fix the geometry $( \Gamma, \partial \mathcal{M})$ defined by the rescaled boundary metric, $\Gamma_{\mu\nu}$. In the event $K$ does vanish somewhere on the boundary, we should use the alternate formulation where we fix the conformal class together with $\gamma^p K$. The discussion below can be simply rewritten using that formulation, but is less elegant, and thus we prefer to phrase it in terms fixing $(\Gamma,\partial \mathcal{M})$.

The Euclidean gravitational path integral
\begin{equation}
    \mathcal{Z}[\Gamma_{\mu\nu}]=\int\mathcal{D}g\, e^{-S[g]},
    \label{eq:pathi}
\end{equation}
is then computed by integrating over all metrics $g$, satisfying the prescribed boundary conditions, where $S[g]$ is the action given in equation~\eqref{eq:newaction}.

Regrettably, it is well-known that the Euclidean gravitational action is unbounded from below, owing to the so-called conformal factor problem \cite{Gibbons:1976ue}. Consequently, the integral cannot be taken over Riemannian metrics $g$. Instead, one is compelled to choose a contour of integration that ensures the convergence of the path integral. Alternatively, one could initiate the analysis with Lorentz signature (where the path integral is always well-defined, at least in a distributional sense), leading to implications for Euclidean path integrals determined through a functional Cauchy-type theorem and a careful study of the analytic properties of the gravitational path integral \cite{Marolf:2022ybi}. Unfortunately, we do not understand the Lorentzian path integral well enough to derive consequences for the Euclidean path integral.

Nevertheless, progress can be made by focusing on saddle points. In \emph{semiclassical} quantum gravity, the Euclidean gravitational path integral is calculated through saddle point approximation
\begin{equation}
\mathcal{Z}[\Gamma_{\mu\nu}]\sim e^{-S[\hat{g}]}
\label{eq:saddle}
\end{equation}
with $\hat{g}$ satisfying the classical equations of motion derived from $S[g]$, subject to the appropriate boundary conditions. The Euclidean stability of such saddles is then investigated by expanding the metric $g_{ab}=\hat{g}_{ab}+h_{ab}$ and computing the action to quadratic order in $h_{ab}$. For the case of pure gravity, possibly in the presence of a cosmological constant, the resulting quadratic action can be written as an expectation value\cite{Headrick:2006ti,Marolf:2022ntb}
\begin{equation}
S^{(2)}[h]=\frac{1}{32\pi G}\int_{\mathcal{M}} {\rm d}^d x\sqrt{\hat{g}}\,h_{ab}\, \hat{\mathcal{G}}^{ab\,cd}\,(L h)_{cd}\,
\end{equation}
where $\hat{\mathcal{G}}^{ab\,cd}$ is a particular DeWitt ultralocal metric
\begin{equation}
\label{eq:DW-1}
 \hat{\mathcal{G}}^{ab\;cd}=\frac{1}{2}\left(\hat{g}^{ac}\hat{g}^{bd}+\hat{g}^{ad}\hat{g}^{bc}-\hat{g}^{ab}\hat{g}^{cd}\right)\, ,
\end{equation}
and
\begin{equation}
(L h)_{ab} = (\hat{\Delta}_L h)_{ab}+2 \hat{\nabla}_{(a}\hat{\nabla}^p \bar{h}_{b)p}
\label{eq:actionungaugedL}
\end{equation}
where
\begin{equation}
\bar{h}_{ab}=h_{ab}-\frac{\hat{g}_{ab}}{2}h\, ,
\end{equation}
and
\begin{equation}
(\hat{\Delta}_L h)_{ab}=-\hat{\nabla}_p \hat{\nabla}^p h_{ab}-2\,\hat{R}_{acbd}\,h^{cd}\,
\end{equation}
is the Lichnerowicz operator. In particular, $L$ reduces to the usual Lichnerowicz operator $\hat{\Delta}_L$ on perturbations that satisfy the de Donder gauge~\eqref{eq:dedonderg}, \emph{i.e.,} $\hat{\nabla}^a \bar{h}_{ab}=0$.

Unlike the leading saddle point approximation, the quadratic contribution to the Euclidean path integral does exhibit a linearized version of the conformal factor problem \cite{Gibbons:1978ac}. Fortunately, this issue has been studied in detail in \cite{Marolf:2022ntb,Marolf:2022jra,Liu:2023jvm}, where a rule of thumb to investigate the stability of a given saddle has been presented and thoroughly examined. For the type of boundary conditions under consideration, the rule of thumb in \cite{Marolf:2022ntb,Marolf:2022jra,Liu:2023jvm} is equivalent to studying the following eigenvalue problem
\begin{equation}
(\Delta_L h)_{ab}=\lambda h_{ab}\,,
\label{eq:lichne}
\end{equation}
where $h_{ab}$ is assumed to be in the de Donder gauge. A given mode is then stable when it satisfies ${\rm Re}\,\lambda>0$, with marginal stability for ${\rm Re}\,\lambda=0$ and strict instability for ${\rm Re}\,\lambda<0$.
We note that stability of saddle points can also be determined by their stability under Ricci flow since they are fixed point of the flow, as discussed in~\cite{Headrick:2006ti}, and the same condition for stability is found.

\subsection{First law of black hole mechanics}\label{sec:FirstLaw}
The canonical ensemble for gravity is defined by the semiclassical Euclidean path integral via an analysis of its saddle points.
Before delving into the on-shell action of well-known saddles in a spherical cavity, the black hole saddles and the hot flat space infilling,
 let us take a moment to discuss a version of the first law of black hole mechanics that will prove useful in the upcoming discussion. We will observe that our boundary conditions give rise to a natural definition of an ensemble, accompanied by a corresponding thermodynamic potential, temperature, and energy. Interestingly, these quantities are naturally defined with respect to the rescaled boundary metric, $\Gamma_{\mu\nu}$ rather than the usual induced metric, $\gamma_{\mu\nu}$.

We first note that general \emph{infinitesimal} diffeomorphisms $x^a\to x^a+v^a$ induce boundary diffeomorphisms $x^\mu\to x^\mu+v^\mu$. Naturally, these should leave the action invariant. Since $\Gamma_{\mu\nu}$ is a symmetric two-tensor, we also know that
\begin{equation}
\delta \Gamma_{\mu\nu} = (\pounds_v \Gamma)_{\mu\nu}=2\nabla^{\Gamma}_{(\mu}v_{\nu)}\,,
\end{equation}
where $\nabla^\Gamma$ is a connection that preserves the metric $\Gamma_{\mu\nu}$. This implies that variations of the action generated by infinitesimal coordinate transformations $v^a$ should satisfy
\begin{equation}
0=\delta S_\Theta=\int_{\partial \mathcal{M}}{\rm d}^{d-1}\sqrt{\Gamma} \,\mathcal{T}^{\mu\nu}\nabla^\Gamma_{(\mu} v_{\nu)}=-\int_{\partial \mathcal{M}}{\rm d}^{d-1}\sqrt{\Gamma}\, v_{\nu}\nabla_{\mu}^\Gamma\mathcal{T}^{\mu\nu}\,
\end{equation}
where in the last step we assumed that the boundary is compact so that boundary terms can be discarded. Since the above holds for arbitrary choices of $v^{\mu}$ it must be that $\mathcal{T}^{\mu\nu}$ is covariantly conserved with respect to $\nabla^{\Gamma}$, \emph{i.e.,}
\begin{equation}
\nabla_{\mu}^\Gamma\mathcal{T}^{\mu\nu}=0\,.
\end{equation}
Thus we interpret $\mathcal{T}^{\mu\nu}$ as a boundary stress tensor with respect to the rescaled boundary geometry $\Gamma_{\mu\nu}$, and define all thermodynamic quantities with respect to the pair $(\Gamma_{\mu\nu},\mathcal{T}^{\mu\nu})$ on $\partial\mathcal{M}$.

To derive the first law, we therefore assume that $\Gamma_{\mu\nu}$ can be written as a thermal ultrastatic metric, \emph{i.e.}
\begin{equation}
\Gamma_{\mu\nu}\mathrm{d}x^{\mu}\mathrm{d}x^{\nu}=\frac{\mathfrak{b}^2}{(2\pi)^2}\mathrm{d}\tau^2+\mathfrak{n}_{ij}\mathrm{d}x^i\mathrm{d}x^j 
\end{equation}
where $\mathfrak{b}$ is a constant, and $\mathfrak{n}_{ij}$ is a $(d-2)$-dimensional Riemannian metric on the space ${\widetilde{\partial \mathcal{M}}}$ spanned by the non-thermal boundary spatial coordinates $x^i$, and $\tau$ varies over the range $\tau\sim \tau+2\pi$, denoting an angle along the thermal circle. 
The thermal circle in this rescaled geometry $(\Gamma, \partial \mathcal{M})$ then has length $\mathfrak{b}$, and we define its inverse, $\mathfrak{t} = 1/\mathfrak{b}$, as the natural measure of temperature in this ensemble.
We will assume that the infilling metric has a Killing field $\partial/\partial \tau$ that remains hypersurface orthogonal, so that in particular $\mathcal{T}^{\tau i}=0$.
It is worth noting that while the time-time component of the rescaled metric, $\Gamma_{\tau\tau}$, is constant, this will not generally be true for the induced metric, where $\gamma_{\tau\tau} \sim K^{\frac{1}{(d-1)p}}$ might spatially vary over the boundary. 

Let us define an energy density $\rho = \Gamma_{\tau\tau}\mathcal{T}^{\tau\tau}$, so that one can expand the variations of the action as
\begin{align}
\delta S_\Theta &= \frac{1}{2} \int_{\partial \mathcal{M}} {\rm d}^{d-1}x \sqrt{\Gamma}\,  \mathcal{T}^{\mu\nu}\,\delta\Gamma_{\mu\nu} \nonumber \\
& = \delta \mathfrak{b} \int_{\widetilde{\partial \mathcal{M}}} {\rm d}^{d-2}x\, \sqrt{\mathfrak{n}}\,\rho   +  \frac{\mathfrak{b}}{2} \int_{\widetilde{\partial \mathcal{M}}} {\rm d}^{d-2}x\,\sqrt{\mathfrak{n}}\,\mathcal{T}^{ij} \delta \mathfrak{n}_{ij}\,.
\end{align}
At this stage we introduce a potential $\mathfrak{F}$, and an energy $\mathfrak{E}$,
\begin{equation}
S_\Theta = \mathfrak{b}\,\mathfrak{F} \quad\text{and} \quad \mathfrak{E} = \int_{\widetilde{\partial \mathcal{M}}} {\rm d}^{d-2}x\,\sqrt{\mathfrak{n}}\,\rho
\end{equation}
We note that the energy we define here is defined with respect to the Killing vector $\partial/\partial \tau$ in the rescaled geometry $(\Gamma, \partial\mathcal{M})$, and hence differs from the energy defined in~\cite{Odak:2021axr} that applies to general dynamical cavity geometries and hence is not defined with respect to a bulk or boundary Killing vector.
Then using $\mathfrak{b} = 1/\mathfrak{t}$ yields
\begin{equation}
\label{eq:nearlyfirstlaw}
 \delta \mathfrak{F} =  - \frac{\mathfrak{E} - \mathfrak{F}}{\mathfrak{t}} \delta \mathfrak{t} +  \frac{1}{2} \int_{\widetilde{\partial \mathcal{M}}} {\rm d}^{d-2}x\,\sqrt{\mathfrak{n}}\,\mathcal{T}^{ij}\,\delta \mathfrak{n}_{ij}  \; .
\end{equation}
This formula for the variation $\delta \mathfrak{F}$ has a striking resemblance with the first law of black hole thermodynamics if one could identify $\mathfrak{b}(\mathfrak{E} - \mathfrak{F})$ as the Bekenstein-Hawking entropy \cite{Bekenstein:1973ur,Hawking:1975vcx}. We shall shortly see that this will be the case.

To show this we start by looking at the on-shell action which determines our potential $\mathfrak{F}$ as,
\begin{align}
\mathfrak{b}\,\mathfrak{F} = S_\Theta &= - \frac{\Theta }{8 \pi G} \int_{\partial \mathcal{M}} {\rm d}^{d-1}x \sqrt{\gamma} K-\frac{1}{16\pi G}\int_{\mathcal{M}}\mathrm{d}^{d}x \sqrt{g}\,(R-2\Lambda) \nonumber \\
& =  - \frac{\Theta }{8 \pi G} \mathfrak{b}\,\int_{\widetilde{\partial \mathcal{M}}} {\rm d}^{d-2}x \sqrt{\mathfrak{n}} K^{1-\frac{1}{2p}}-\frac{1}{16\pi G}\int_{\mathcal{M}}\mathrm{d}^{d}x \sqrt{g}\,(R-2\Lambda)\, .
\end{align}
On the other hand, for the energy we find
\begin{align}
\mathfrak{b}\,\mathfrak{E} &=  \int_{\widetilde{\partial \mathcal{M}}} {\rm d}^{d-2}x\,\sqrt{\mathfrak{n}}\,\rho \nonumber \\
& = \mathfrak{b} \frac{1}{8 \pi G}  \int_{\widetilde{\partial \mathcal{M}}} {\rm d}^{d-2}x\,\sqrt{\mathfrak{n}}\,K^{-\frac{1}{2p}}  K^{\tau\tau}   \gamma_{\tau\tau} -  \mathfrak{b} \frac{\Theta}{8 \pi G}  \int_{\widetilde{\partial \mathcal{M}}} {\rm d}^{d-2}x \,\sqrt{\mathfrak{n}}\,K^{1-\frac{1}{2p}}\,.  
\end{align}
Combining these expressions gives,
\begin{align}
\frac{ \mathfrak{E} - \mathfrak{F} }{\mathfrak{t}}
& =  \frac{1}{8 \pi G}  \int_{\partial \mathcal{M}} {\rm d}^{d-1}x \sqrt{\gamma}  K^{\tau\tau}   \gamma_{\tau\tau}+\frac{1}{8\pi G}\frac{2\Lambda}{d-2}\int_{\mathcal{M}}\mathrm{d}^{d}x \sqrt{g}\,,
\label{eq:long}
\end{align}
where we have used that on-shell $R=2 d \Lambda/(d-2)$. Due to our symmetry assumptions, the infilling metric can be written as
\begin{equation}
\mathrm{d}s^2 = N(x)^2\mathrm{d}\tau^2+H_{IJ}(x)\mathrm{d}x^I\,\mathrm{d}x^J\,,
\end{equation}
where $H_{IJ}$ is the metric on the $(d-1)$-dimensional space $H$ spanned by all the bulk non-thermal directions, and we recall that $\tau\sim \tau +2\pi$. Using the above form of the metric, we can write the $\tau\tau$ component of the Einstein equation as
\begin{equation}
R_{\tau\tau}-\frac{2 \Lambda}{d-2}g_{\tau\tau}=0\Rightarrow - N \left( \nabla_{(H)}^2 N + \frac{2}{d-2} \Lambda N \right)=0\Rightarrow \nabla_{(H)}^I J_I+\frac{2 \Lambda}{d-2}\Lambda N =0\,,
\end{equation}
where we defined $J^I\equiv H^{IJ}{\nabla_{(H)}}_IN$ and $\nabla_{(H)}$ is the metric preserving connection on $H$. Let us look at the last term appearing in Eq.~(\ref{eq:long}) and write it as a function of $J^I$
\begin{equation}
\frac{2\Lambda}{d-2}\int_{\mathcal{M}}\mathrm{d}^{d}x \sqrt{g}=2\pi \frac{2\Lambda}{d-2}\int_{H}\mathrm{d}^{d-1}x \sqrt{H}N=-2\pi \int_{H}\mathrm{d}^{d-1}x \sqrt{H}\nabla_{(H)}^I J_I\,.
\end{equation}
We see that this is a total divergence, and so can be readily evaluated on $\partial H$. Let us now assume that the cavity  is filled with a black hole with a single horizon component.
The boundary of $\partial H$ then has two parts: the cavity boundary, $\widetilde{\partial \mathcal{M}}$, and the bifurcating Killing horizon $\mathcal{H}$ of the black hole. Then each component yields a contribution to the boundary terms as,
\begin{equation}
\frac{2\Lambda}{d-2}\int_{\mathcal{M}}\mathrm{d}^{d}x \sqrt{g}=-2\pi \int_{\partial H}\mathrm{d}^{d-2}x\sqrt{\bar{\gamma}}\,n_i\,J^i=-2\pi \int_{H}\mathrm{d}^{d-2}x\sqrt{\bar{\gamma}}\,n_i\,J^i-2\pi \int_{\widetilde{\partial \mathcal{M}}}\mathrm{d}^{d-2}x\sqrt{\bar{\gamma}}\,n_i\,J^i\,,
\end{equation}
where $n_i$ is the unit outer normal to a given boundary and $\bar{\gamma}_{ab}$ is the spatial $(d-2)$-dimensional metric on that boundary. In order to evaluate the contributions at the boundaries it is convenient to use Gaussian normal coordinates, that is to say in the neighbourhood of a boundary we choose coordinates,
\begin{equation}
\mathrm{d}s^2 = \mathrm{d}r^2+\bar{\gamma}_{ij}\mathrm{d}x^i \mathrm{d}x^j\,.
\end{equation}
so that the boundary is at $r=0$ and the interior of the geometry is $r>0$.  

It then follows that
\begin{equation}
\sqrt{\bar{\gamma}}n^i J_i=\sqrt{\bar{\gamma}}n^r J_r = -\sqrt{\bar{\gamma}}\partial_r N\,.
\end{equation}
For a bifurcating Killing horizon, $N=r+o(r)$, since we have fixed the period of $\tau$ to be $2\pi$ and so,
\begin{equation}
-2\pi \int_{\mathcal{H}}\mathrm{d}^{d-2}x\sqrt{\bar{\gamma}}\,n_i\,J^i = 2\pi \int_{\mathcal{H}}\mathrm{d}^{d-2}x\sqrt{\bar{\gamma}}=2\pi A\,,
\end{equation}
where $A$ is the horizon area. For the cavity boundary, in these normal coordinates,
\begin{equation}
K_{\tau\tau}=\left.-N \partial_r N\right|_{r=0}
\end{equation}
and thus,
\begin{equation}
-2\pi \int_{\widetilde{\partial \mathcal{M}}}\mathrm{d}^{d-2}x\sqrt{\bar{\gamma}}\,n_i\,J^i=-2\pi \int_{\widetilde{\partial \mathcal{M}}}\mathrm{d}^{d-2}x\sqrt{\gamma}K_{\tau\tau}\gamma^{\tau\tau}=-\int_{\partial \mathcal{M}}\mathrm{d}^{d-1}x\sqrt{\gamma}K_{\tau\tau}\gamma^{\tau\tau}\,,
\end{equation}
where we used that $\gamma^{\tau\tau}=\left.1/N^2\right|_{r=0}$. We have thus found that
\begin{equation}
\frac{2\Lambda}{d-2}\int_{\mathcal{M}}\mathrm{d}^{d}x \sqrt{g}=2\pi A-\int_{\partial \mathcal{M}}\mathrm{d}^{d-1}x\sqrt{\gamma}K_{\tau\tau}\gamma^{\tau\tau}\,,
\end{equation}
which we can substitute into Eq.~(\ref{eq:long}) to confirm our claim above, concluding that for a black hole in a cavity,  
\begin{equation}
\label{eq:EminusF}
\frac{ \mathfrak{E} - \mathfrak{F} }{\mathfrak{t}} =  \mathcal{S}_{BH}
\end{equation}
where $\mathcal{S}_{\rm BH}=A/(4G)$ is the Bekenstein-Hawking entropy.
Then from~\eqref{eq:nearlyfirstlaw} we obtain the first law,
\begin{equation}
 \delta \mathfrak{F} = - \mathcal{S}_{\rm BH}\,\delta \mathfrak{t} +  \frac{1}{2} \int {\rm d}^{d-2}x \,\sqrt{\mathfrak{n}}\,\mathcal{T}^{ij}\,\delta \mathfrak{n}_{ij}  \; .
\end{equation}
We emphasize that the temperature $\mathfrak{t}$ and metric $\mathfrak{n}_{ij}$ are those associated to the rescaled metric $\Gamma_{\mu\nu}$, and not those associated with induced boundary metric $\gamma_{\mu\nu}$. The second term above is a work term associated to deforming the boundary spatial geometry. It is a generalization of the familiar $\sim P \dd V$ term, and reduces to that if we restrict to a round sphere cavity so that the change in volume solely comes from a change of radius.
We also note that the first law above also applies to empty cavities which contain no black hole horizons, where one would simply drop the horizon entropy term. Likewise if interior solutions with multiple horizon components were to exist -- for example with a positive cosmological term~\cite{Dias:2023rde} -- then the horizon entropy term would be the sum of that from each horizon component.
A reduced first law was considered for Anderson boundary condition in~\cite{Anninos:2023epi}. This restricted to cavities with a round sphere spatial geometry, and only considered changes in dimensionless quantities, so in that context a dimensionless free energy as a function of the conformal class. We believe our first law is consistent with their results in that special situation.

\subsection{Saddle points for spherical cavities: hot space and black holes}\label{sec:3saddles}
To evaluate the gravitational path integral (\ref{eq:pathi}), we will use the saddle point approximation (\ref{eq:saddle}). We want to consider spherical cavities, and we will assume that spherical symmetry is preserved in the interior of the cavity. The metric $\gamma_{\mu\nu}$ on the boundary $\partial \mathcal{M}$ will be $S_{\beta}^1 \times S_{R_0}^2$, with $S_{\beta}^1$ having length $\beta$ and parametrizing a periodic Euclidean time coordinate $\tau\sim\tau+2\pi$, while $S^2_{R_0}$ represents the metric on a round two-sphere with a radius of $R_0$. As in the Dirichlet case, the infilling saddle points are either black holes, or the vacuum spacetime with compact Euclidean time circle which we term `hot space'.

We will compare actions for the various saddles while keeping
\begin{equation}
\mathcal{B}\equiv \frac{\beta}{R_0}\quad \text{and} \quad \mathcal{K}\equiv \sqrt{\gamma(R_0)}K(R_0)^{\frac{1}{2p}}
\end{equation}
constant, noting that $\mathcal{B}$ determines the conformal class of the static spherical boundary. It is important to note that the different saddles we may consider might have distinct values for $R_0$ and $\beta$, but equal values for $\mathcal{B}$ and $\mathcal{K}$. Note that from our discussion in section \ref{sec:newbc} this is equivalent to fixing $\Gamma_{\mu\nu}$ under spherical symmetry. An important point is that while $\beta$ gives the size of the Euclidean time circle as measured in the metric $\gamma_{\mu\nu}$, we will find that it is the size of the circle in the rescaled metric $\Gamma_{\mu\nu}$, which we denote as $\mathfrak{b}$, that will determine the notion of inverse temperature in the thermodynamics with our conformal boundary conditions. We may compute this rescaled boundary metric to find,
\begin{equation}
\Gamma_{\mu\nu}\mathrm{d}x^\mu\mathrm{d}x^\nu=\frac{\mathfrak{b}^2}{(2\pi)^2}\mathrm{d}\tau^2+\mathfrak{r}^2\mathrm{d}\Omega_2^2\quad \text{with}\quad \mathfrak{b} = (2\pi \mathcal{K})^{1/3}\mathcal{B}^{2/3}\quad\text{and}\quad \mathfrak{r}=\left(\frac{2\pi \mathcal{K}}{\mathcal{B}}\right)^{1/3}\,.
\end{equation}

We will ignore the cosmological constant here, although we do give a discussion of the case with cosmological constant in the Appendix~\ref{sec:ccthermo}.
Thus our infilling solutions must be Ricci flat and, in particular, have a vanishing Ricci scalar $R$.
As a result, the bulk action $S_{\rm EH}$ vanishes and the action is given by the boundary term only:
\begin{equation}
S[\mathcal{B},\mathcal{K}]=-\frac{\Theta}{8\pi G}\int_{\partial \mathcal{M}}\mathrm{d}^3x\,\sqrt{\gamma}\,K\,.
\end{equation}

By virtue of Birkhoff's theorem \cite{1923rmp..book.....B}, there are only two different line elements to investigate, corresponding to Euclidean space and Euclidean Schwarzschild black hole, which we will analyse in turn.

Let us first consider Euclidean space with a cavity of radius $r=r_E$
\begin{equation}
\mathrm{d}s_E^2=\frac{\beta_E^2}{(2\pi)^2}\,\mathrm{d}\tau^2+\mathrm{d}r^2+r^2\,\mathrm{d}\Omega_2^2\,,
\label{eq:eufla}
\end{equation}
for which the Euclidean action reads
\begin{equation}
S_E[\mathcal{B},\mathcal{K}]=-\frac{r_E\,\beta_E\,\Theta}{G}\,.
\end{equation}
Furthermore, it a simple exercise to compute
\begin{equation}
\mathcal{B}_E=\frac{\beta_E}{r_E}\quad\text{and}\quad \mathcal{K}_E=2^{\frac{1}{2 p}} r_E^{2-\frac{1}{2 p}} \frac{\beta_E}{2\pi}\,,
\end{equation}
which we could use to rewrite $S_E[\mathcal{B},\mathcal{K}]$ as a function of $\mathcal{B}$ and $\mathcal{K}$ explicitly.

Let us turn our attention now to the line element of Euclidean Schwarzschild inside a cavity of radius $r=r_B$
\begin{subequations}
\begin{equation}\label{eq:BHsol}
\mathrm{d}s_B^2=\frac{\beta_B^2\,F(r)^2}{(2\pi)^2F(r_B)^2}\,\mathrm{d}\tau^2+\frac{\mathrm{d}r^2}{F(r)^2}+r^2\,\mathrm{d}\Omega_2^2
\end{equation}
with
\begin{equation}\label{eq:BHf_flat}
F(r)^2=1-\frac{r_+}{r}\,.
\end{equation}
\end{subequations}
The horizon, located at $r=r_+$, is an $S^2$ bolt \cite{Gibbons:1979xm} of area $4\pi r_+^2$ and regularity there requires the standard choice \cite{Zumino,Gibbons:1976pt,Hawking:1976jb}  for the length of the thermal circle
\begin{equation}
\beta_B = \frac{2\pi}{F^\prime(r_+)}\,.
\end{equation}
The on-shell action now reads
\begin{equation}
S_B[\mathcal{B},\mathcal{K}]=-\frac{r_B \beta _B \Theta}{4 G}\frac{4-3x}{\sqrt{1-x}}\,
\end{equation}
and furthermore
\begin{equation}
\mathcal{B}_B=\frac{\beta_B}{r_B}=4\pi x\sqrt{1-x}\quad \text{and}\quad \mathcal{K}_B=2^{-\frac{1}{2 p}} r_B^{2-\frac{1}{2 p}} (4-3 x)^{\frac{1}{2 p}} (1-x)^{-\frac{1}{4 p}} \frac{\beta _B}{2\pi}\,,
\end{equation}
where we defined $x\equiv r_+/r_B\in(0,1)$. We see from above that, at a fixed and sufficiently small $\mathcal{B}_B$ there are exactly \emph{two} black hole saddles with distinct values of $x$, one with $x > 2/3$ and the other with $x < 2/3$, as first observed by York~\cite{York:1986it}. The saddle with larger (smaller) values of $x$ is denoted by large (small) Euclidean Schwarzschild black hole. We note that since $x$ is determined by the conformal class of the boundary it is not sensitive to the value of $p$, and thus the boundary between small and large black holes, $x = 2/3$, is the same for all the boundary conditions we consider.

We now impose our boundary conditions $\Gamma^E_{\mu\nu}=\Gamma^B_{\mu\nu}$, which in turn demands
\begin{equation}
\frac{\beta_B}{r_B}=\frac{\beta_E}{r_E}=4\pi x\sqrt{1-x}\quad \text{and}\quad r_B=4^{-\frac{1}{1-6 p}} (4-3 x)^{\frac{1}{1-6 p}} (1-x)^{-\frac{1}{2 (1-6 p)}}\,r_E\,,
\end{equation}
and automatically ensures $\mathcal{B}_E=\mathcal{B}_H\equiv \mathcal{B}$ and $\mathcal{K}_E=\mathcal{K}_H\equiv \mathcal{K}$. 
The difference between the on-shell actions can then be readily computed and we find
\begin{subequations}
\begin{equation}
\Delta S[\mathcal{B},\mathcal{K}] \equiv S_B[\mathcal{B},\mathcal{K}]-S_E[\mathcal{B},\mathcal{K}]= -2^{\frac{2}{1-6 p}} \mathcal{B}^{1+\frac{4 p}{1-6 p}} (2\pi\mathcal{K})^{-\frac{4 p}{1-6 p}}\,\frac{\Theta}{G}\,H(p,x)
\end{equation}
with
\begin{equation}
H(p,x)\equiv\left(1-\frac{3 x}{4}\right)^{1+\frac{2}{1-6 p}} (1-x)^{-\frac{1}{2}-\frac{1}{1-6 p}}-1\,.
\end{equation}
\end{subequations}
For any $p$ we see that $H(p,8/9)=0$ indicating a first order transition between the large black hole and hot flat space. For small $x$ we have,
\begin{equation}
\Delta S[\mathcal{B},\mathcal{K}]=2^{\frac{2}{1-6 p}} \mathcal{B}^{1+\frac{4 p}{1-6 p}} \mathcal{K}^{-\frac{4 p}{1-6 p}}\,\frac{1}{4G}\,\left[x+\mathcal{O}(x^2)\right]\,
\end{equation}
showing that the small Euclidean Schwarzschild saddle is never dominant. Furthermore, since
\begin{equation}
\Theta H\sim-\frac{x}{4}
\end{equation}
near $x=0$ and has a single vanishing derivative at $x=2/3$ in the interval $x\in(0,1)$, we conclude that $\Theta H$ has a minimum at $x=2/3$. This shows that the large Euclidean Schwarzschild saddle becomes dominant for $\mathcal{B}<8\pi/(3\sqrt{3})$, whereas hot Euclidean space dominates in the complementary regime. Indeed, this result is independent of $p$! This is the same result that York found for Dirichlet boundary conditions~\cite{York:1986it} and analogous to the Hawking-Page transition in AdS spacetimes \cite{Hawking:1982dh}. 
One might wonder why the point of transition is independent of $p$ -- this can be understood as being due to the on-shell action reducing simply to a boundary term, and changing $p$ only changes the coefficient of this term. Thus the action of the competing saddles are all scaled together as $p$ is varied, and so the value of $\mathcal{B}$ where the large black hole and hot space have equal action is not changed.

Now let us have a look at the first law of black hole thermodynamics derived in section~\ref{sec:FirstLaw} again. In the black hole background~\eqref{eq:BHsol}, we have
\begin{equation}
    \mathfrak{b}=2^{2-\frac{1}{6p}}\pi x(4-3x)\left(\frac{4-3x}{\sqrt{1-x}}\right)^{\frac{1}{6p}-1}r_B^{1-\frac{1}{6p}}\,,\quad \mathfrak{r}=2^{-\frac{1}{6p}}\left(\frac{4-3x}{\sqrt{1-x}}\right)^{\frac{1}{6p}}r_B^{1-\frac{1}{6p}}\,,
\end{equation}
while for the potential $\mathfrak{F}$ and the energy $\mathfrak{E}$ we find
\begin{subequations}
\begin{align}
    \mathfrak{F}=\frac{S_B}{\mathfrak{b}}&=-\frac{2^\frac{1}{6p}\Theta}{4G}\cdot r_B^{1+\frac{1}{6p}}\cdot\left(\frac{4-3x}{\sqrt{1-x}}\right)^{1-\frac{1}{6p}}\,,
    \\
    \mathfrak{E}=\int_{\widetilde{\partial \mathcal{M}}} {\rm d}^{d-2}x\,\sqrt{n}\,\rho&=\frac{2^\frac{1}{6p}\Theta}{4G}\cdot r_B^{1+\frac{1}{6p}}\cdot(x-4\Theta+3x\Theta)\cdot\frac{(1-x)^{\frac{1}{12}(-6+\frac{1}{p})}}{(4-3x)^{\frac{1}{6p}}}\,.
\end{align}
\end{subequations}
Combining these together gives an explicit check of our earlier equation~\eqref{eq:EminusF},
\begin{equation}
    \mathfrak{b}(\mathfrak{E}-\mathfrak{F})=\frac{\pi r_B^2 x^2}{G}=\frac{\pi\rp^2}{G}=\frac{A}{4G}\,,
\end{equation}
where $A=4\pi\rp^2$ is the area of the black hole horizon, recovering the standard Bekenstein-Hawking entropy of a black hole.

Given the inverse temperature $\mathfrak{b}$ and entropy $\mathcal{S}_{\mathrm{BH}}$, we can calculate the black hole specific heat via
\begin{equation}
    C=-\mathfrak{b}\frac{\partial \mathcal{S}_{\mathrm{BH}}}{\partial\mathfrak{b}}.
\end{equation}
This is simple to compute, but we are interested only in its sign, which is governed by the conformal class as,
\begin{equation}
\label{eq:SpecificHeat}
    C=\mathrm{(some\ positive\ expression)}\cdot \frac{16x-8-9x^2+12p(1-x)(4-3x)}{(6p-1)(3x-2)}\,.
\end{equation}
We note that for $p=0$ this agrees with the specific heat computed in~\cite{Anninos:2023epi}. As for Dirichlet boundary conditions,
for the small black hole, $x<\frac{2}{3}$, the specific heat is always negative, indicating thermodynamic instability. For the large black hole, $x>\frac{2}{3}$, if $p<\frac{1}{6}$ then the specific heat is positive as in the Dirichlet case. However for $p > \frac{1}{6}$ then the specific changes from positive to negative when the black hole is large enough, as can be seen because $C \sim \mathrm{(positive\ expression)} \frac{1}{1-6p} \left(1 + (1 - 12p) (1-x) \right) + O(1-x)^2$ near $x = 1$.
This transition happens at $x\to 1$ if $p\to +\infty$ so we still recover the Dirichlet result that the large black hole is thermodynamically stable.

Thus in the case $p > 1/6$ we obtain the novel behaviour that the specific heat capacity of sufficiently large large black holes is negative, naively indicating thermodynamic instability. We will later explore whether this potential instability shows up in either the fluctuations about this Euclidean saddle point, or its dynamical stability.

\subsection{Fluctuations about the empty spherical cavity}\label{sec:Fieldtheoretic}
The aim of this subsection is to demonstrate that an empty spherical cavity filled with hot Euclidean space is devoid of fluctuation negative modes if $p$ is sufficiently large, specifically if $p > 1/6$. Our findings can be extended to $\Lambda\neq0$, and we defer  discuss this explicitly in appendix \ref{app:ads}. While we will primarily concentrate on four spacetime dimensions for the sake of presentation, we have verified that our results generalize to $d\geq4$ \emph{mutatis mutandis}. Note however that in $d\geq5$ one should also consider tensor modes, but these are the same as in the Dirichlet case.  The eigenvalue problem to study is simply that given in Eq.~(\ref{eq:lichne})
\begin{equation}
(\Delta_L h)_{ab}=\lambda\,h_{ab}\quad\text{with}\quad \nabla_a h^{a}_{\phantom{a}b}=\frac{1}{2}\nabla_b h
\label{eq:lichneagain}
\end{equation}
around hot Euclidean space with a cavity of radius $r=r_E$ (see Eq.~(\ref{eq:eufla})), so,
\begin{equation}
\mathrm{d}s_E^2=\frac{\beta_E^2}{(2\pi)^2}\,\mathrm{d}\tau^2+\mathrm{d}r^2+r^2\,\mathrm{d}\Omega_2^2\,,
\end{equation}
where recall that $\tau\sim\tau+2\pi$ so that the length of the thermal circle is $\beta_E$.

As emphasized in~\cite{Marolf:2022ntb}, when gravity is placed in a cavity, due to the mixing of the transverse traceless and conformal modes the Lichnerowicz operator is self-adjoint with respect to the non-positive DeWitt metric, and thus its eigenvalues need not be real. It was argued that stability is determined by the real part of the eigenvalues being positive. This precisely correlates with stability of a saddle point under Ricci flow -- a saddle is a fixed point of Ricci flow, and a negative real eigenvalue corresponds to an unstable mode under the flow that takes one away from this fixed point in flow time. Thus for stability we require,
\be
\mathrm{Re}\,\lambda \ge 0 
\ee
for all eigenvalues $\lambda$.

Since $\partial/\partial \tau$ is a Killing isometry, we can decompose all the perturbations into Fourier modes $e^{i n \tau}$ with $n\in \mathbb{Z}$. Furthermore, we can exploit the background $SO(3)$ symmetry and decompose all modes based on their transformations under diffeomorphisms on the $S^2$. In four spacetime dimensions, these fall into two classes: scalar-derived gravitational perturbations and vector-derived gravitational perturbations. Scalar-derived modes are built from spherical harmonics $Y^{\ell_S\,m_S}$ on $S^2$, while vector-derived modes are built from vector harmonics on $S^2$ $Y^{\ell_V\,m_V}_I$, where upper case Latin indices will run over the sphere directions. In dimensions $d\geq5$, there are also tensor harmonics to consider. There is one further simplification that occurs for four-spacetime dimensions. Let $G_{IJ}$ be the metric on a round two sphere of unit radius. Let also $\mathcal{D}_I$ be the metric-preserving covariant derivative on the unit radius $S^2$. It is easy to show that $Y^{\ell_V\,m_V}_I$ are simply two-dimensional Hodge duals of gradients of $Y_{\ell_S\,m_S}$. For spherical harmonics, we have $\ell_S\geq0$, while for vector harmonics $\ell_V\geq1$.
For later convenience, we define
\begin{equation}
\tilde{\tau}=\frac{\beta_E}{2\pi}\,\tau\quad \text{and}\quad e^{i\,n\,\tau}=e^{i\,\tilde{n}\,\tilde{\tau}}\,,
\end{equation}
where we note that $n\in\mathbb{Z}$, but $\tilde{n}$ is generally \emph{not}  an integer.

The analysis of the scalar fluctuations is rather technical, with various special cases to consider. The full details are reserved for the Appendix~\ref{app:EucPT}. The upshot is that the vector modes of the Lichnerowicz operator for our new conformal boundary condition have precisely the same spectrum as in the Dirichlet case -- there is no dependence on the boundary condition, and further they are all real and stable. Thus we will concentrate on the scalar modes which need not be real, and depend in detail on the boundary conditions.

Non-spherical scalar-derived gravitational modes are constructed from scalar harmonics $Y^{\ell_S\,m_S}$, which satisfy the usual eigenvalue equation
\begin{equation}
\label{eq:SphHarm}
\mathcal{D}_I \mathcal{D}^I Y^{\ell_S\,m_S}+\ell_S(\ell_S+1)Y^{\ell_S\,m_S}=0\,,
\end{equation}
where we recall that $I$ is an index on the two sphere and $\mathcal{D}$ is the metric preserving covariant derivative on the unit round two-sphere.  Let lower case hatted Latin indices run over $\tilde{\tau}$ and $r$. Scalar derived perturbations are then written as
\begin{subequations}
\begin{equation}
\delta{\rm d}s^2\equiv \sum_{\ell_S=0}^{+\infty}\sum_{m_S=-\ell_S}^{\ell_S}h^{\ell_S\,m_S}_{ab}(\tilde{\tau},r,\theta,\phi){\rm d}x^a{\rm d}x^b
\end{equation}
with $\theta$ and $\phi$ the usual polar and azimuthal angles on the two sphere and
\begin{align}
&h^{\ell_S\,m_S}_{\hat{a}\hat{b}}=\hat{f}_{\hat{a}\hat{b}}^{\ell_S\,m_S}(\tilde{\tau},r)\,Y^{\ell_S\,m_S}
\\
& h^{\ell_S\,m_S}_{\hat{a}I}=\hat{f}_{\hat{a}}^{\ell_S\,m_S}(\tilde{\tau},r)\,\mathcal{D}_I Y^{\ell_S\,m_S}
\\
& h^{\ell_S\,m_S}_{IJ}=\hat{h}^{\ell_S\,m_S}_L(\tilde{\tau},r)\,G_{IJ}\,Y^{\ell_S\,m_S}+\hat{h}_T^{\ell_S\,m_S}(\tilde{\tau},r)\,S^{\ell_S\,m_S}_{IJ}
\end{align}
where $\mathcal{S}^{\ell_S\,m_S}_{IJ}$ is a traceless symmetric two tensor defined as
\begin{equation}
\mathcal{S}^{\ell_S\,m_S}_{IJ}=\mathcal{D}_I\mathcal{D}_J Y^{\ell_S\,m_S}+\frac{\ell_S(\ell_S+1)}{2}G_{IJ}Y^{\ell_S\,m_S}
\label{eq:SIJ}
\end{equation}
and
\begin{equation}
    G_{IJ}\dd x^I\dd x^J=\dd \Omega_2^2\,.
\end{equation}
\end{subequations}%
Since all the angular dependence is fixed, we are left with determining $\hat{f}_{\hat{a}\hat{b}}$, $\hat{f}_{\hat{a}}$, $\hat{h}_L$ and $\hat{h}_T$. At this point we use the fact that $\partial/\partial \tilde{\tau}$ is a Killing vector field of the background and further expand $\hat{f}_{\hat{a}\hat{b}}$, $\hat{f}_{\hat{a}}$, $\hat{h}_L$ and $\hat{h}_T$ as
\begin{multline}
\label{eq:EucCmpts}
\hat{f}^{\ell_S\,m_S}_{\hat{a}\hat{b}}(\tilde{\tau},r)=e^{i\tilde{n}\tilde{\tau}}\,f^{\ell_S\,m_S\,\tilde{n}}_{\hat{a}\hat{b}}(r)\,,\quad \hat{f}^{\ell_S\,m_S}_{\hat{a}}(\tilde{\tau},r)=e^{i\tilde{n}\tilde{\tau}}\,f^{\ell_S\,m_S\,\tilde{n}}_{\hat{a}}(r)\,,
\\
\hat{h}^{\ell_S\,m_S}_L(\tilde{\tau},r)=e^{i\tilde{n}\tilde{\tau}}\,h^{\ell_S\,m_S\,\tilde{n}}_L(r)\quad\text{and}\quad \hat{h}^{\ell_S\,m_S}_T(\tilde{\tau},r)=e^{i\tilde{n}\tilde{\tau}}\,h^{\ell_S\,m_S\,\tilde{n}}_T(r)
\end{multline}
Modes with different values of $\ell_S$, $m_S$  and/or $\tilde{n}$ decouple from each other at the quadratic level in the action. For this reason, we shall drop the subscript $\ell_S\,m_S\,\tilde{n}$ in what follows. 

In the analysis of these fluctuations, given in detail in the Appendix~\ref{app:EucPT}, the spherically symmetric modes with $\ell_S =0$, and those with $\ell_S=1$ are special, since for those $S_{IJ}^{1\,m_S}=0$, and must be treated as separate cases. Furthermore the modes that are static with $n = 0$ must be treated separately. In that Appendix we give analytic expressions for the various fluctuation component functions in equation~\eqref{eq:EucCmpts}.
In all cases, included those that must be treated separately, the result of the scalar perturbation mode analysis may be summarized in the following condition on the eigenvalue:
\be
\label{eq:pexpression}
p(\ell_S, n, \lambda) & = & \frac{A_{\ell_S}(\tilde{\Lambda}) + \varpi^2 C_{\ell_S}(\tilde{\Lambda})}{B_{\ell_S}(\tilde{\Lambda}) + \varpi^2 D_{\ell_S}(\tilde{\Lambda})}
\ee
where the lefthand side encodes the choice of boundary condition.  Furthermore, we defined
\be
\tilde{\Lambda}\equiv  \tilde{\lambda} - \varpi^2 \; , \qquad \tilde{\lambda} \equiv \lambda\,r_E^2 \; , \qquad \varpi\equiv\tilde{n}\,r_E
\ee
and we have somewhat complicated expressions for $A_{\ell_S}$, $B_{\ell_S}$, $C_{\ell_S}$ and $D_{\ell_S}$.
These take the form,
\be
A_{\ell_S}(\tilde{\Lambda}) & = & 
\sum_{q=0}^4 a_q(\sqrt{\tilde{\Lambda}}) \mathrm{J}_{\frac{1}{2} + \ell_{S}}(\sqrt{\tilde{\Lambda}})^q \mathrm{J}_{\frac{3}{2} + \ell_{S}}(\sqrt{\tilde{\Lambda}})^{4-q} 
\ee
where the five coefficients $a_q(\sqrt{\tilde{\Lambda}},\ell_S)$ are polynomials in the arguments $\sqrt{\tilde{\Lambda}}$ and $\ell_S$, up to  quadratic order in $\ell_S$, but have no explicit $p$ dependence. These can be explicitly read off from the expressions in equations~\eqref{eq:generalLambda}. The expressions $B_{\ell_S}$, $C_{\ell_S}$ and $D_{\ell_S}$ take the same form. Ideally we would solve the above expression~\eqref{eq:pexpression} to obtain the eigenvalue $\lambda$ as a function of $\ell_S$, $n$ and the cavity radius $r_E$. Unfortunately this appears not to be analytically tractable. However the above form is convenient to study the Euclidean stability of this background, hot Euclidean space, and we now discuss this.

We note that for Dirichlet boundary conditions, a similar analysis yields the condition on the modes,
\be
B_{\ell_S}(\tilde{\Lambda}) + \varpi^2 D_{\ell_S}(\tilde{\Lambda}) = 0
\ee
which corresponds to the vanishing of the denominator in~\eqref{eq:pexpression}, and hence to the infinite $| p |$ limit. This demonstrates that in the infinite $| p |$ limit, modes with finite eigenvalue $\tilde{\lambda}$ will limit to those of the Dirichlet problem. Conversely there will be modes whose magnitude of their eigenvalue diverges in the $p \to \pm \infty$ limit, and are not shared by the Dirichlet boundary conditions.

\subsubsection{Static spherical modes}\label{sec:EucStaticSpherical}

\begin{figure}[ht]
    \centering
    \includegraphics[width=0.43\textwidth]{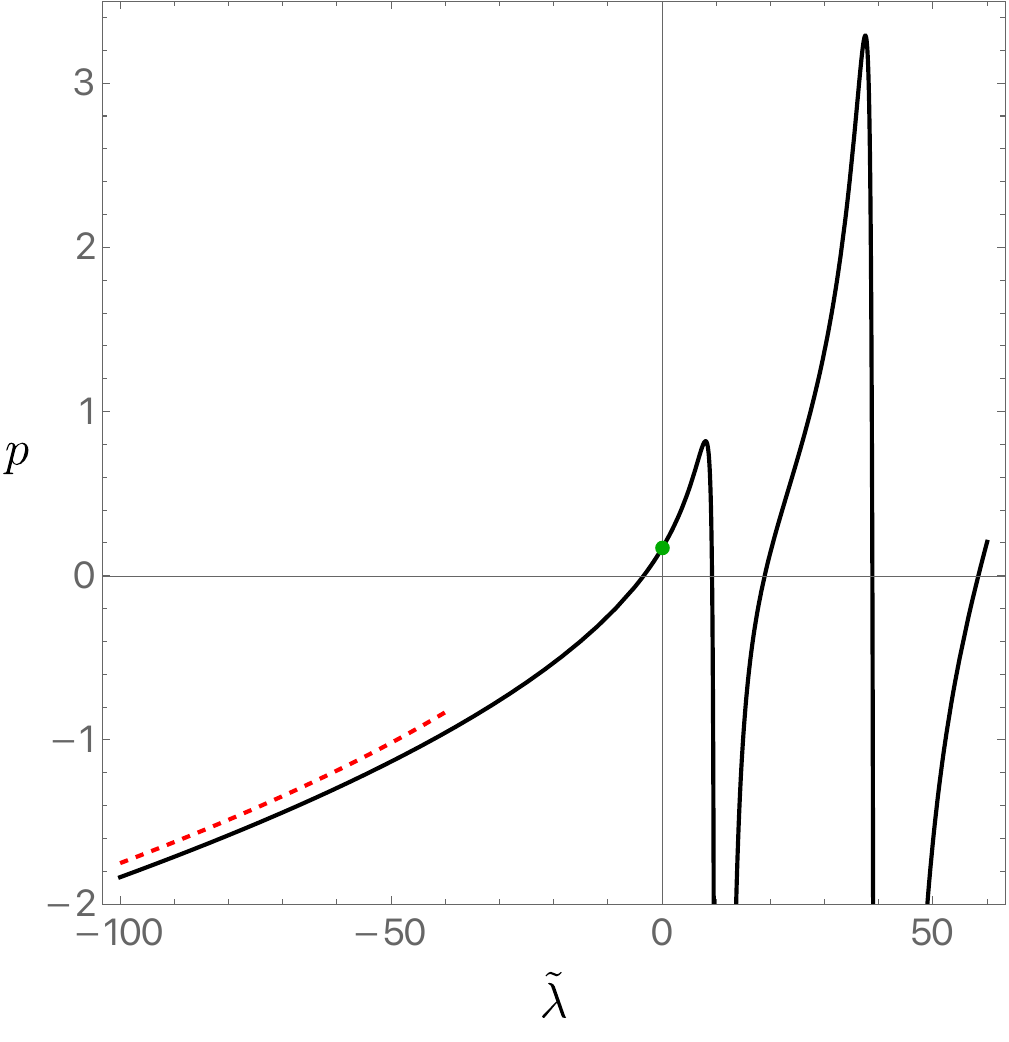}
    \caption{$p$ as a function of $\tilde\lambda$ for the static spherical modes $n=\ell_S=0$. The green dot marks $p=\frac{1}{6}$, where $\tilde\lambda$ is precisely $0$. The red dashed curve gives the asymptotic behavior of $p(\tilde\lambda)$ as $\tilde\lambda\to-\infty$.}
    \label{fig:p_Euc_static_sph}
\end{figure}

In the case that $p < 1/6$, we indeed find eigenmodes with $\mathrm{Re}\,\lambda < 0$. The simplest sector to see these is the static spherically symmetric scalar sector, $n = \ell_S = 0$, where the condition above yields,
\be
p\left(\tilde{\lambda}\right) & = & \frac{1}{12} \left[ \frac{
1+5 \tilde{\lambda}-3 \tilde{\lambda}^{3/2} \cot( \sqrt{\tilde{\lambda}})
-\tilde{\lambda} \cot( \sqrt{\tilde{\lambda}})^2
}
{
1+\tilde{\lambda
   } - \sqrt{\tilde{\lambda}} \cot( \sqrt{\tilde{\lambda}}) 
} \right] \; .
\ee
Now expanding about $\tilde{\lambda} = 0$ we find,
\be
\label{eq:linearp}
p = \frac{1}{6} + \frac{1}{18} \tilde{\lambda} + O\left(  \tilde{\lambda}^2\right) \; .
\ee
Thus for $p = 1/6$ the spectrum has a marginal mode, with $\lambda = 0$, and near to this value, for $p < 1/6$ we see that $\tilde{\lambda}$ becomes real and negative, and hence we see that an unstable mode with $\mathrm{Re}\,\lambda < 0$ exists. For real $\tilde{\lambda}$ then $p$ is real and is given in figure~\ref{fig:p_Euc_static_sph}. We may then follow the unstable mode above. Asymptotically one finds,
\be
\label{eq:asympstaticsph}
p = \mp \frac{\sqrt{-\tilde{\lambda}}}{4} + \frac{3}{4} + O\left(\tilde{\lambda}^{-1/2}\right)
\ee
with the upper minus sign for ${\rm Im}\,\tilde{\lambda} \ge 0$ and the lower positive sign for ${\rm Im}\,\tilde{\lambda}<  0$. This is shown in red on the figure, and so $p \to - \infty$ as $\tilde{\lambda} \to - \infty$. From the figure we see the relation is monotonic for $\tilde{\lambda} < 0$, and hence the negative mode exists for all $p < 1/6$. This asymptotics precisely illustrates the behaviour mentioned above, namely that there are modes in the eigenspectrum with diverging eigenvalue as $p \to \pm \infty$. Here we see that for $p \to - \infty$ this is a negative mode, whereas for $p \to + \infty$ it is a stable mode.

For $p > 1/6$ there are no unstable modes in this static spherical sector. We see from the figure that modes with positive real $\tilde{\lambda}$ exist for $p > 1/6$. Indeed for all values of $p$ we may also find modes with complex $\tilde{\lambda}$, but as we will soon argue, these are stable having ${\rm Re}\,\tilde{\lambda} > 0$. In figure~\ref{fig:n0ell0} in Appendix~\ref{sec:n0ell0}, a variety of modes are shown for different $p$.

\subsubsection{Instabilites for $p < 1/6$ for the static non-spherical modes}\label{sec:EucInstability}

For $p < 1/6$ we have seen that there is an unstable spherically symmetric static mode. However for sufficiently small $p$ there are other unstable modes, both with real but also complex $\tilde{\lambda}$. This can already be seen in the static sector with $\varpi = 0$ (and so $n = 0$). 

\begin{figure}[ht]
    \centering
    \includegraphics[width=\textwidth]{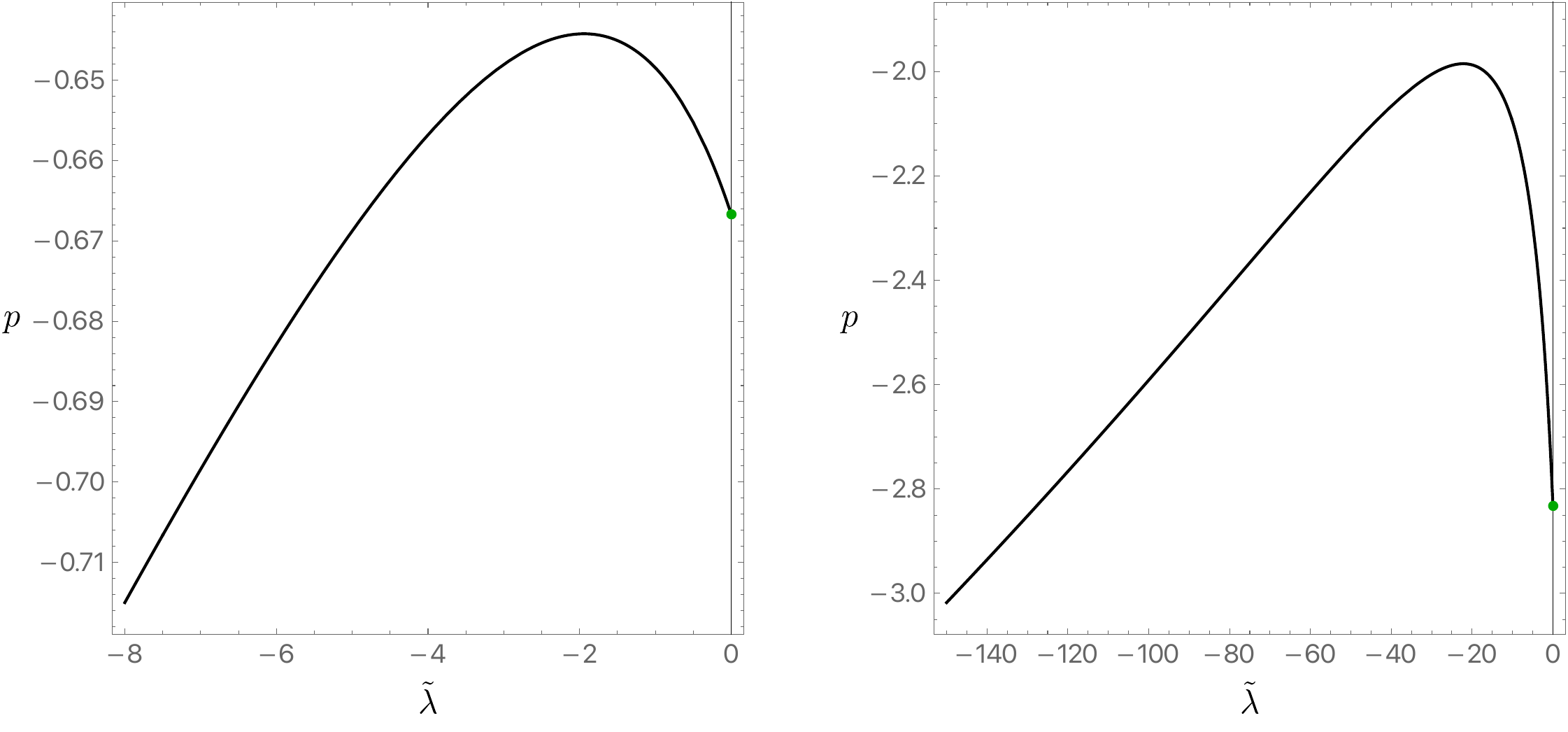}
    \caption{$p$ as a function of $\tilde\lambda$ for the static non-spherical modes with $\ell_S=2$ (left panel) and $\ell_S=4$ (right panel). The green dots mark $\left(0,\frac{1}{12}(2-\ell_S-2\ell_S^2)\right)$.}
    \label{fig:unstable_l2and4_real}
\end{figure}

In figure~\ref{fig:unstable_l2and4_real} we plot $p$ against negative $\tilde{\lambda}$ for both $\ell_S = 2$ and $4$. 
For $\ell_S = 2$ then $p = -2/3$ at $\tilde{\lambda} = 0$, and we see that for the range $- \frac{2}{3} < p < -0.64422$ there are a pair of unstable modes with real $\tilde{\lambda}$. For $p < - \frac{2}{3}$ there persists a single unstable mode. Likewise for $\ell_S = 4$ then $p = -\frac{17}{6}$ at $\tilde{\lambda} = 0$ 
and for the range $-\frac{17}{6} < p < -1.98491$ there is a pair of unstable modes with real $\tilde{\lambda}$, and for $p < - \frac{17}{6}$ there is a single real unstable mode.
This phenomena holds for all $\ell_S$, each giving rise to a pair of real unstable modes for sufficiently small $p$, and a single mode for smaller $p \le \frac{1}{12}(2-\ell_S-2\ell_S^2)$, where the righthand side of this inequality derives from the value of $p$ at $\tilde{\lambda} = 0$.

\begin{figure}[ht]
    \centering
    \includegraphics[width=\textwidth]{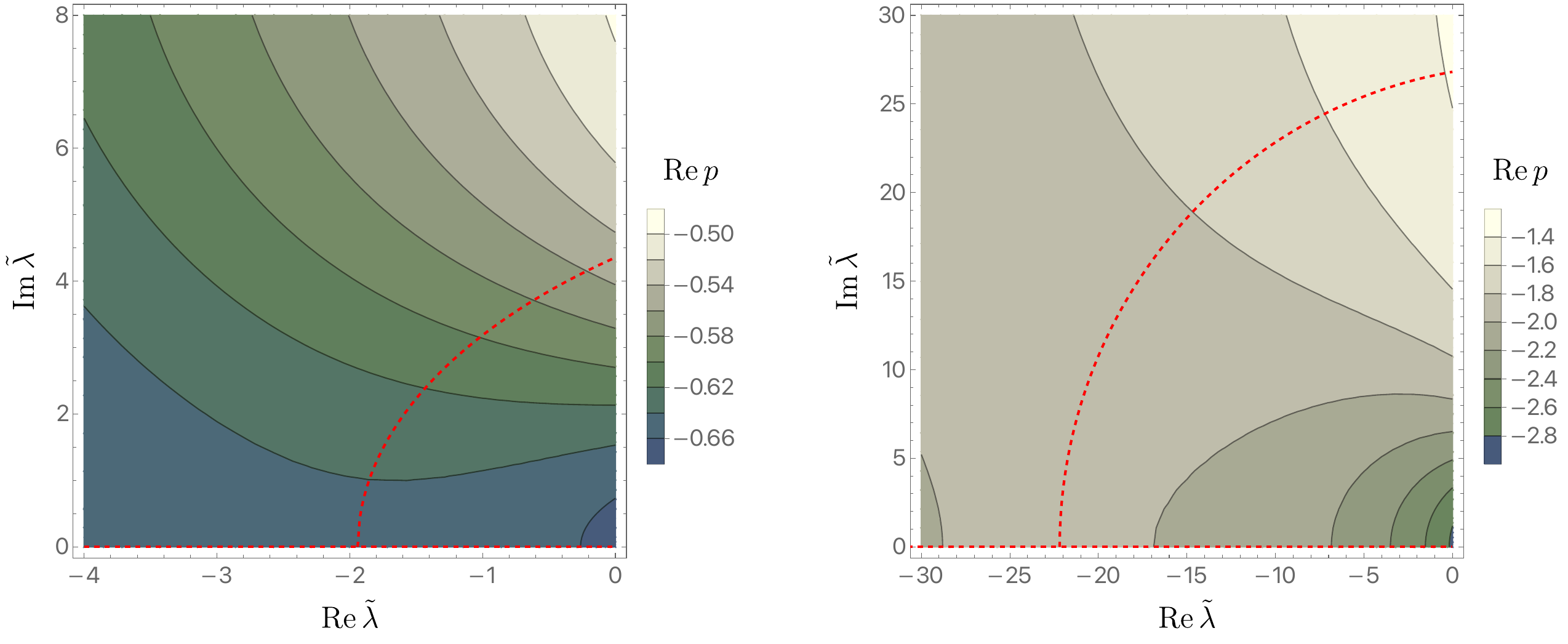}
    \caption{Contour plot of $\mathrm{Re}\,p$ in the complex $\tilde\lambda$ plane for $\ell_S=2$ (left panel) and $\ell_S=4$ (right panel). The red dashed curves are where $\mathrm{Im}\,p=0$ that corresponds to a physical unstable mode. }
    \label{fig:unstable_l2and4_complex}
\end{figure}

However the situation is more complicated as for certain ranges of $p$ there may also exist complex unstable modes, so modes with complex $\tilde{\lambda}$ where ${\rm Re}\,\tilde{\lambda} < 0$. We may see these in the example of $\ell_S = 2$ and $4$ depicted in figure~\ref{fig:unstable_l2and4_complex}. Here a contour plot of the real part of $p$, ${\rm Re}\,p$, is shown in the complex $\tilde{\lambda}$ plane in the potentially unstable region, ${\rm Re}\,\tilde{\lambda} < 0$. Now $p$ must be a real value, and so only if $\mathrm{Im}(p) = 0$ can this value of $p$ correspond to a physical unstable mode. On the figures we have plotted a red curve showing where $\mathrm{Im}(p)$ vanishes. While it vanishes along the negative real axis, for $\ell_S = 2,4$ it vanishes also on a curve that then connects to the imaginary $\tilde{\lambda}$ axis. This leaves the negative real axis precisely at the point which maximizes $p$, so at $p = -0.644$ and $-1.98$ for $\ell_S = 2$ and $4$ respectively. We find that $p$ increases along the curve towards the imaginary axis where $p = -0.549$ and $-1.39$ for $\ell_S = 2$ and $4$ respectively.
Therefore we conclude that for $\ell_S = 2$ and $p \in (-0.644,-0.549)$ we have unstable modes with complex eigenvalue, and likewise for $\ell_S = 4$ we have $p \in (-1.98, -1.39)$.
In fact, we observe this same feature for all $\ell_S \ge 2$. The lowest few modes for $\ell_S=2$ are given in figure~\ref{fig:ell2} in appendix~\ref{sec:staticModesS}. The green dashed curve below the line $\mathrm{Re}\,\tilde\lambda=0$ is consistent with our analysis here.

\subsubsection{Stability for $p > 1/6$ for all modes}\label{sec:EucStability}
Having seen that for $p < 1/6$ there is at least one unstable mode, and that these modes may have complex eigenvalue, we now argue that for $p > 1/6$ there are no unstable modes for any $\ell_S$ and $\varpi$. By inspecting the function for $p(\tilde{\Lambda})$ for different $\ell_S$ and $\varpi$ in the complex $\tilde{\Lambda}$ plane one can verify this. 
However, the presence of unstable modes off the real $\tilde{\lambda}$ axis means that it is challenging to rule out unstable modes for all $\ell_S$ and $\varpi$. Thus rather than relying on inspection of the function, we give a neater argument based on analyticity properties of the function $p$.
The details of the argument are presented in Appendix~\ref{app:analyticity_p}. The main claims are the following:
\\

\noindent
{\bf Claim 1.}: ${\rm Re}\,p$ is maximized in the complex half plane  ${\rm Re}\,\tilde{\lambda} \le 0$ by its value on the imaginary axis ${\rm Re}\,\tilde{\lambda} = 0$. \\

We are not able to give a rigorous proof of this, but are able to give a convincing demonstrate of this by a combination of analytic and numerical work presented in that Appendix. It follows from showing that the function $p$ is analytic in the complex half plane ${\rm Re}\,\tilde{\Lambda} < 0$ and further observing that ${\rm Re} \,p$ goes to $- \infty$ asymptotically in all directions in that half plane. Recalling that $\tilde{\Lambda} = \tilde{\lambda} - \varpi^2$ with $\varpi > 0$ and real, then this implies the same results hold in the complex half plane ${\rm Re}\,\tilde{\lambda} < 0$ of the eigenvalue $\tilde{\lambda}$, with this half plane being the one of interest for stability.
Being analytic then writing $\tilde{\lambda} = x + i y$, it obeys the Laplace equation in the variables $x, y$, and by the maximum principle must take its greatest value on the boundary of the region. 

A corollary of this result is that if ${\rm Re}\,p$ is bounded from above on the imaginary $\tilde{\lambda}$ axis, then that bounds ${\rm Re}\,p$ for all $\tilde{\lambda}$ in the complex half plane where ${\rm Re}\,\tilde{\lambda} < 0$ and thus putative instabilities might lie.
\\

\noindent
{\bf Claim 2.}: ${\rm Re}\,p < 1/6$ for all $\ell_S$ and $\varpi$ on the  imaginary axis $\tilde{\lambda}$ axis. \\

The combination of these two claims then imply that ${\rm Re}\,p < 1/6$ for all $\tilde{\lambda}$ with ${\rm Re}\,\tilde{\lambda} < 0$. While of course $p$ must be real for a physical mode to exist, this bound on the real part of $p$ implies that there can be no instabilities for $p > 1/6$.

We relegate showing analyticity to the Appendix, but here simply note that both the numerator and denominator of the expression for $p$ in~\eqref{eq:pexpression}, given by $A_{\ell_S}(\tilde{\Lambda}) + \omega^2 C_{\ell_S}(\tilde{\Lambda})$ and $B_{\ell_S}(\tilde{\Lambda}) + \omega^2 D_{\ell_S}(\tilde{\Lambda})$ respectively, are clearly analytic for ${\rm Re}\,\tilde{\Lambda} < 0$ except for branch cuts along the negative real axis from the origin, from the properties of Bessel functions. Since $\tilde{\Lambda} = \tilde{\lambda} - \varpi^2$, then the same holds for ${\rm Re}\,\tilde{\lambda} < 0$. In fact $p$ is analytic at the origin, the non-analytic behaviour of the numerator and denominator  cancelling there. Hence showing that $p$ is analytic for ${\rm Re}\,\tilde{\lambda} < 0$ comes down to showing that the denominator has no zeros in that region, and this is the work detailed in Appendix \ref{app:analyticity_p}.

Once analyticity is shown then ${\rm Re}\,p$ is maximized on a boundary or asymptotically. Due to the reality of the components of the expression for $p$, we have,
\be
{\rm Re}\,p(x,y) = {\rm Re}\,p(x,-y) \; , \quad \mathrm{Im}\,p(x,y) = - \mathrm{Im}\,p(x,-y) 
\ee
where again $\tilde{\lambda} = x + i y$, so that ${\rm Re}\,p$ is an even function of $y = {\rm Im}\,\tilde{\lambda}$. Further the leading asymptotics  for the static spherical fluctuation case in equation~\eqref{eq:asympstaticsph} are similar to those when $\ell_S, \varpi \ne 0$. We recall that $\ell_S$ is a positive number, and $\varpi$ is a positive real. Using the asymptotic properties of Bessel functions we find in the upper unstable (ie. ${\rm Re}\,\tilde{\lambda} < 0$) plane, so $x \in (-\infty,0]$, $y \in [0,\infty)$, that $p$ goes as,
\be
\label{eq:asymp}
p = - \frac{\sqrt{-\tilde{\Lambda}}}{4} + \frac{3}{4}  + O\left(\tilde{\Lambda}^{-1/2}\right)
\ee
with $\ell_S$ and $\varpi$ only affecting the subleading terms, and since $\tilde{\Lambda} = \tilde{\lambda} - \varpi^2$ we recover the same asymptotics in $\tilde{\lambda}$ as above in~\eqref{eq:asympstaticsph}.
Then for any finite $\varpi$, writing $\tilde{\lambda} = r e^{i \theta}$ then the leading asymptotic behaviour as $r \to \infty$ is,
${\rm Re}\,p \to  - \frac{\sqrt{r}}{4} \sin(\frac{\theta}{2})$
for the unstable region of the upper half complex plane $\frac{\pi}{2} \le \theta \le \pi$. Since $\sin(\frac{\theta}{2}) > \frac{1}{\sqrt{2}}$ then as $r \to \infty$ then ${\rm Re}\,p \to - \infty$.
As a result of this, the only asymptotic region or boundary remaining for $x \le 0$ where ${\rm Re}\,p$ could be maximized is the imaginary $\tilde{\lambda}$ axis $x = 0$ where ${\rm Re}\,\tilde{\lambda} = 0$, or equivalently, ${\rm Re}\,\tilde{\Lambda} = - \varpi^2$.

We now consider the behaviour of ${\rm Re}\,p$ on the imaginary $\tilde{\lambda}$ axis. To address the second claim we will argue that at fixed $\tilde{\lambda}$ on the imaginary axis, ${\rm Re}\,p$ for all positive numbers $\ell_S$ and real $\varpi \ge 0$ takes its maximum value for the static spherical case $\ell_S = \varpi = 0$. This follows from the fact that we find the behaviour of ${\rm Re}\,p$ at fixed $\tilde{\lambda}$ decreases with increasing $\ell_S$ and with increasing $\varpi$.

\begin{figure}[ht]
    \centering
    \includegraphics[width=0.43\textwidth]{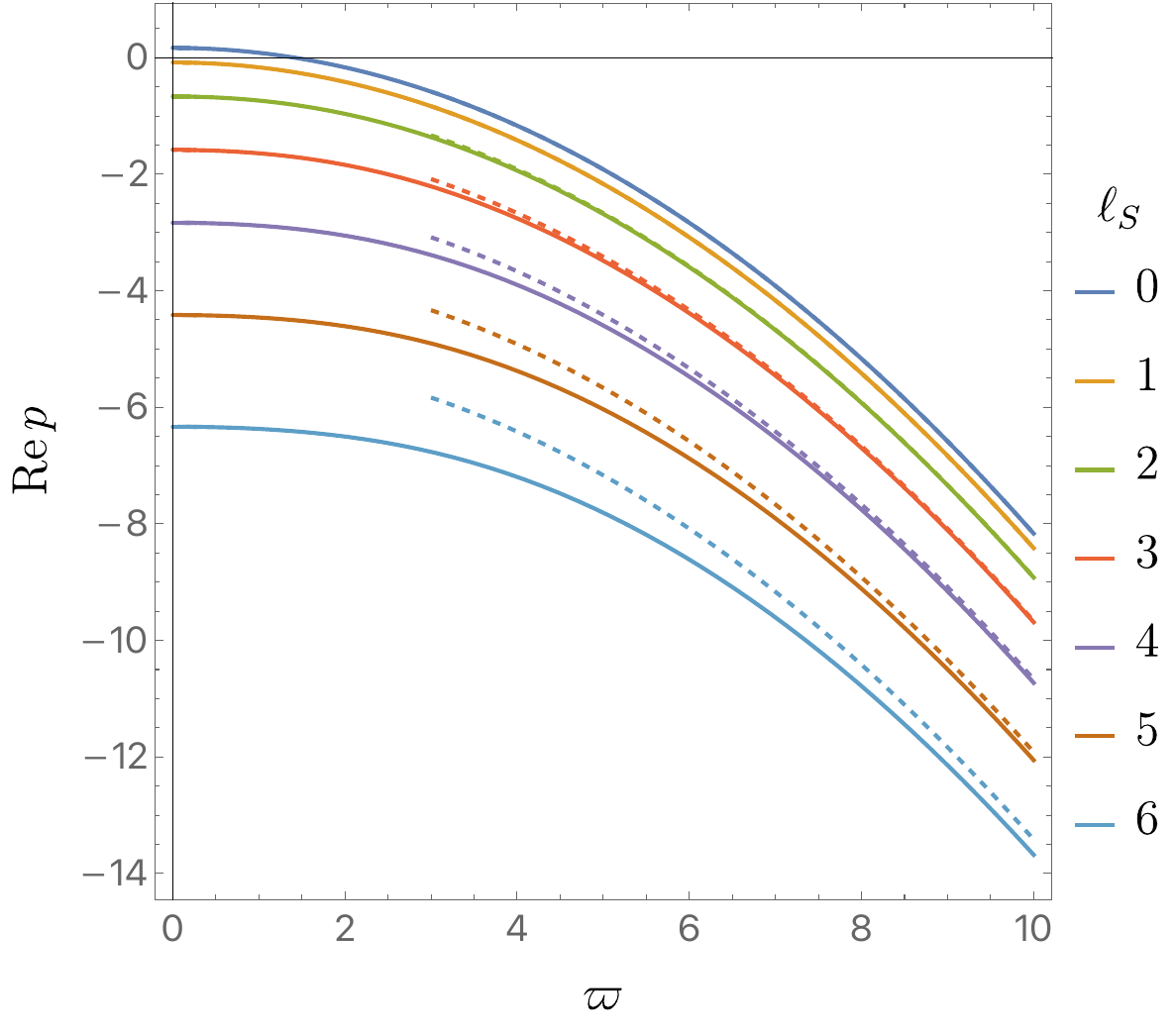}     
    \caption{$\mathrm{Re}\,p$ as a function of $\varpi$ at $\tilde\lambda=0$ for different $\ell_S$'s. The dashed lines give the asymptotic behaviors. The black solid line is $\mathrm{Re}\,p=0$.}
    \label{fig:Euc_ReP_Origin}
\end{figure}

Firstly in figure~\ref{fig:Euc_ReP_Origin} we show the value of ${\rm Re}\,p$ at the origin, $\tilde{\lambda} = 0$, so at ${\rm Re}\,\tilde{\Lambda} = - \varpi^2$, as a function of $\varpi$ for various $\ell_S = 0, \ldots, 6$. We see that indeed ${\rm Re}\,p$ decreases for increasing $\ell_S$ and increasing $\varpi$. The maximum value of $1/6$ is attained for the spherical static case. The asymptotic behaviour of these curves is given by,
\be
{\rm Re}\,p & = & - \frac{\varpi^2}{12} + \frac{4 - 3 \ell_S - 3 \ell_S^2}{24} + O\left(\varpi^{-2}\right)
\ee
and is shown in figure~\ref{fig:Euc_ReP_Origin} by the dashed curves. The value of ${\rm Re}\,p$ at the origin in the figure, $\varpi = 0$, is,
\be
\label{eq:p_at_omega_lambda_origin}
p = \frac{2 - \ell_S - 2 \ell_S^2}{12} \; .
\ee
We see both the value at the origin and the asymptotic behaviour decrease for increasing $\ell_S$.

\begin{figure}[ht]
    \centering
    \includegraphics[width=0.43\textwidth]{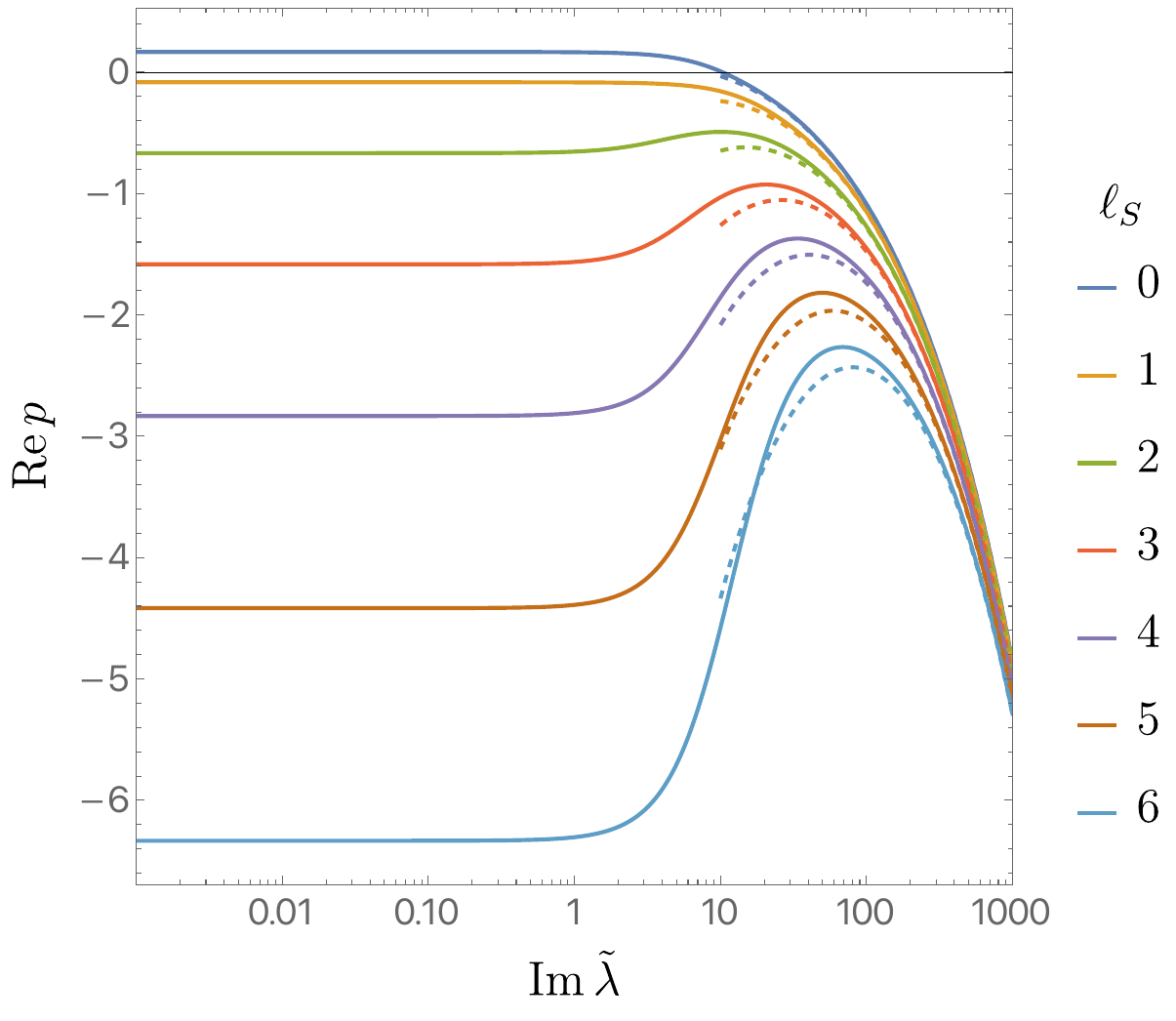}     
    \caption{$\mathrm{Re}\,p$ as a function of $\mathrm{Im}\,\tilde\lambda$ at $\varpi=0$ for different $\ell_S$'s. The dashed lines give the asymptotic behaviors. The black solid line is $\mathrm{Re}\,p=0$.}
    \label{fig:Euc_ReP_ImagAxis}
\end{figure}

Secondly in figure~\ref{fig:Euc_ReP_ImagAxis} we show ${\rm Re}\,p$ with $\varpi = 0$ on the imaginary $\tilde{\lambda}$ axis again for $\ell_S = 0, \ldots , 6$. We show the asymptotic behaviour, at fixed $\ell_S$, which goes as,
\be
{\rm Re}\,p = - \frac{\sqrt{y}}{4 \sqrt{2}} + \frac{3}{4} - \frac{24 + 11 \ell_S + 11 \ell_S^2}{24 \sqrt{2 y}} + O\left(y^{-3/2}\right)
\ee
as dashed lines, which we see decreases in value as $\ell_S$ increases. The plot confirms that increasing $\ell_S$ indeed decreases the maximum value of ${\rm Re}\,p$ on the imaginary axis. For $\ell_S = 0$ the maximum is the value at the origin, where $p = 1/6$. For $\ell_S > 0$, where the value at the origin is given above in~\eqref{eq:p_at_omega_lambda_origin}, in fact the maximum of ${\rm Re}\, p$ occurs away from the origin on the imaginary $\tilde{\lambda}$ axis.

\begin{figure}[ht]
    \centering
    \includegraphics[width=0.71\textwidth]{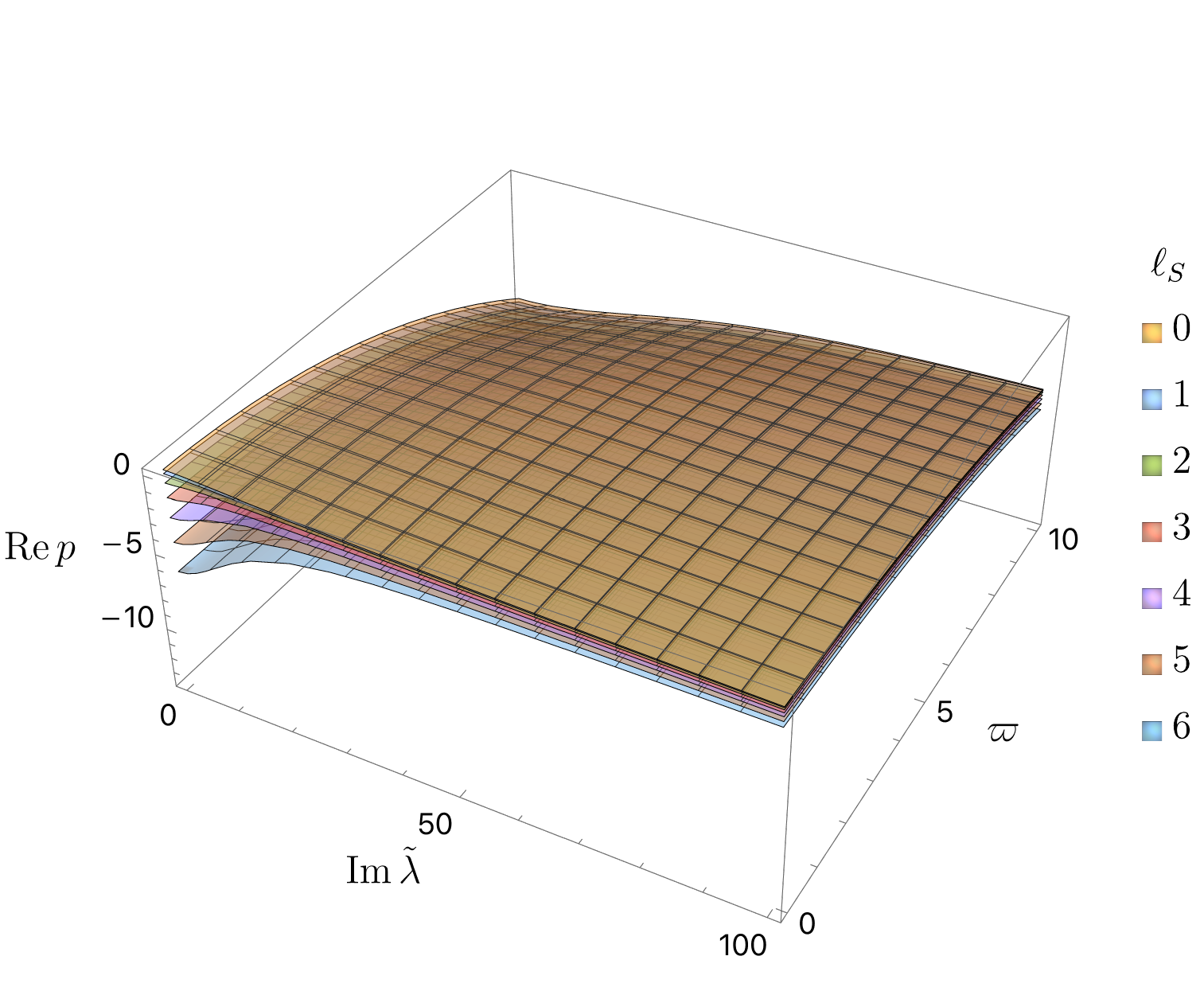}    
    \caption{Plot of ${\rm Re}\,p$ against $\tilde{\lambda}$ on the imaginary axis and also against $\varpi$. This is plotted for various values of $\ell_s = 0, \ldots, 6$. The orange surface is that with $\ell_S = 0$ and bounds all the others from above. At fixed ${\rm Im}\,\tilde{\lambda}$ and $\varpi$ increasing $\ell_S$ decreases ${\rm Re}\,p$. At fixed ${\rm Im}\,\tilde{\lambda}$, increasing $\varpi$ decreases ${\rm Re}\,p$ for all $\ell_S$.}
    \label{fig:Euc_ReP}
\end{figure}

Thus from the previous plots we have seen that ${\rm Re}\,p$ is maximized for $\varpi = 0$ on the imaginary axis by the spherically symmetric modes $\ell_S = 0$. At fixed location on the imaginary axis, increasing $\ell_S$ decreases ${\rm Re}\,p$. Further we have seen that for $\tilde{\lambda} =0$ increasing $\varpi$ also decreases ${\rm Re}\,p$.
Generally we find that increasing $\varpi$ always decreases ${\rm Re}\,p$ on the imaginary axis. In figure~\ref{fig:Euc_ReP} we show surfaces of ${\rm Re}\,p$ plotted on the imaginary axis and also against $\varpi$, with various $\ell_S = 0, \ldots, 6$. For fixed ${\rm Im}\,\tilde{\lambda}$ and $\varphi$ we again see a decrease with increasing $\ell_S$. For fixed ${\rm Im}\,\tilde{\lambda}$ and $\ell_S$ we see a monotonic decrease with increasing $\varphi$.
Hence the $\ell_S = 0$ curve in figure~\ref{fig:Euc_ReP_ImagAxis} bounds all other values of $\ell_S = 0$ and $\varpi = 0$, and its maximum value is at the origin $\tilde{\lambda} = 0$, where $p = 1/6$.
Thus we have that ${\rm Re}\,p < 1/6$ for all $\ell_S$ and $\varpi$ on the imaginary axis, and hence the same is true in the unstable part of the complex $\tilde{\lambda}$ plane (i.e. where ${\rm Re}\,\tilde{\lambda}<0$). Finally then, we arrive at the conclusion that there can be no Euclidean instability for $p > 1/6$.

\subsection{Black hole fluctuations: spherical static negative modes}\label{sec:NegativeModes}

The existence of negative modes for black holes is a sign of local thermodynamic instability \cite{Gross:1982cv}. In \cite{Marolf:2022ntb}, the authors showed that if we compute the spectrum of the Lichnerowicz operator acting on static spherically symmetric perturbations, then the path integral stability matches with thermodynamic stability for Euclidean Schwarzschild (AdS) black holes under Dirichlet boundary conditions. We will now consider the same problem, namely that of computing the eigenvalues of the Lichnerowicz operator for our generalized conformal boundary condition on a black hole background, and comparing the Euclidean stability determined by the existence of unstable modes to the thermodynamic stability.
Unlike the previous section, where we considered the hot flat space saddle point, here we are unable to solve the fluctuations analytically for a black hole background. We therefore proceed numerically, and restrict to considering only the static spherically symmetric sector. 
For simplicity here we present only the case without cosmological constant. However for completeness in the Appendix~\ref{sec:ccthermo} we give results where the cosmological constant is included.
To proceed we take the background Schwarzschild solution in equation~\eqref{eq:BHsol}, and consider the fluctuation,
\begin{equation}
    \delta{\rm d}s^2\equiv h_{ab}\mathrm{d}x^a\mathrm{d}x^b=F(r)^2\,a(r)\,{\rm d}\tau^2+b(r)\,\frac{{\rm d}r^2}{F(r)^2}+r^2\,c(r)\,{\rm d}\Omega_2^2,
\end{equation}
where $a$, $b$ and $c$ are unknown functions of $r$. We impose our boundary conditions given in~\eqref{eq:bc} and the gauge condition~\eqref{eq:dedonder}, and then examine the existence of negative modes. The numerical method we use here is the same as the one proposed in \cite{Marolf:2022ntb} and reviewed in \cite{Liu:2023jvm}. One can also use the traditional operator approach, \emph{e.g.}, in \cite{Headrick:2006ti}.

\begin{figure}
    \centering
    \includegraphics[width=\textwidth]{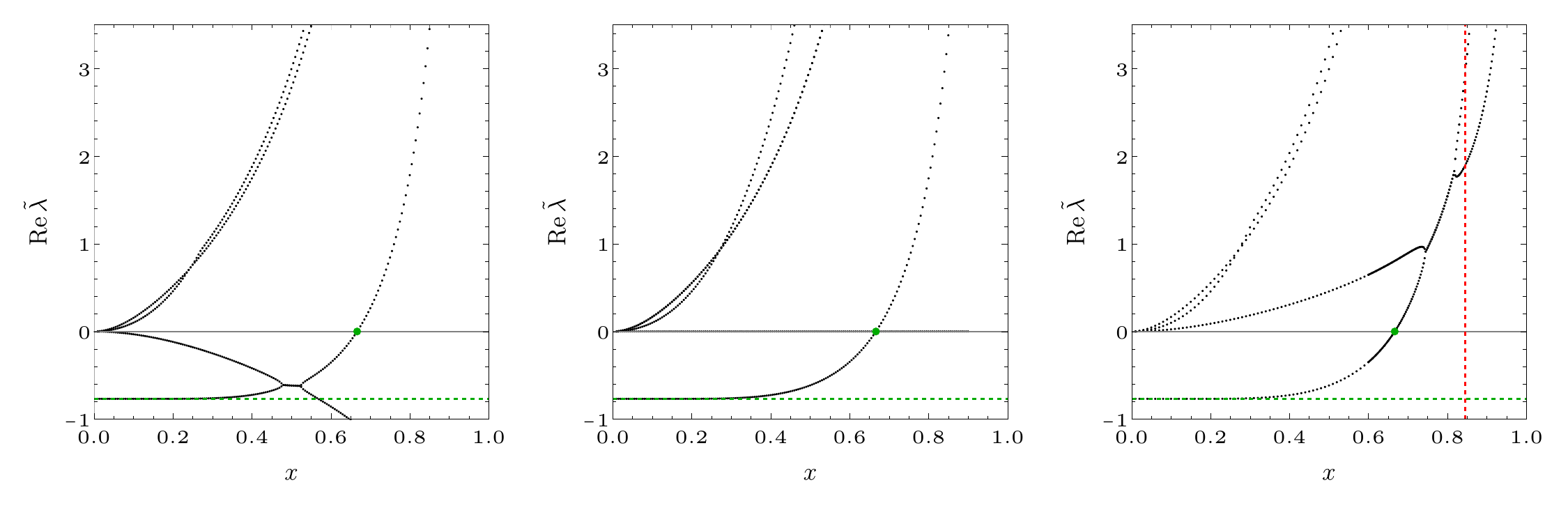}
    \caption{The real part of the four lowest lying eigenvalues of $\hat{\Delta}_L$ as a function of $x$ for the case with $\Lambda$=0. The left, middle, and right panels correspond to $p=0$, $p=\frac{1}{6}$, and $p=\frac{1}{3}$, respectively. The green dots marks $x=\frac{2}{3}$. The gray solid line and the green dashed line are $\mathrm{Re} \,\tilde\lambda=0$ and $\mathrm{Re} \,\tilde\lambda=\tilde\lambda_{\rm GPY}=-0.7677$, respectively. The dashed red line in the red panel denotes $x=\frac{2}{3} \left(3-\sqrt{3}\right)\approx 0.845$, where the specific heat changes sign, according to equation~\eqref{eq:SpecificHeat}. Since only the sign of the real part of the eigenvalues determines the path integral stability, we do not distinguish between real modes and complex modes.}
    \label{fig:flat_BH_modes}
\end{figure}

Figure \ref{fig:flat_BH_modes} gives the lowest four modes of the Lichnerowicz operator in the background of a Schwarzschild black hole inside a cavity with our boundary conditions for $p$ taking three values $p = 0, 1/6$ and $1/3$.
It is worth recapping what happens for Dirichlet boundary conditions \cite{Marolf:2022ntb}. There a small black hole, $x< 2/3$, has a negative mode and large black holes where $x > 2/3$ have none. This negative mode is the generalization of the Gross-Perry-Yaffe negative mode of asymptotically flat Schwarzschild~\cite{Gross:1982cv}, and indeed becomes precisely that mode in the very small black hole limit $x \to 0$. As one moves from the small black hole branch to $x = 2/3$ the negative mode becomes a marginal zero mode, and is simply given by the tangent to the space of Schwarzschild solutions, and so is generated by perturbing the Schwarzschild radius $r_+$. 
This occurs as at this special point the conformal class, and hence the temperature measured in units of the 2-sphere radius, is stationary as $r_+$ is deformed.
This Euclidean stability exactly coincides with thermodynamic stability, the small black holes having negative specific heat and the large ones having positive specific heat.
Now with our generalized boundary conditions we recover what we might have naively expected. The number of static spherical negative modes for a black hole in a cavity can be understood by taking the number of static spherical negative modes in an empty cavity, and adding a negative mode in the case the black hole is small. 
Firstly consider the case $p < 1/6$, illustrated in the figure in the left panel for $p=0$. Here for $p < 1/6$ we find two static spherical negative modes for small black holes, so $x< 2/3$, one associated to the cavity boundary, and one to the `small' horizon. For large black holes with $p < 1/6$ we find only one negative mode, which we take to be associated to the cavity. At $x = 2/3$ there is an exactly marginal zero mode that again is simply tangent to the Schwarzschild family and is generated by perturbing $r_+$, and exists since as we have seen the dimensionless temperature $\mathcal{B} = \beta/R_0$ has a minimum and so is stationary as $r_+$ varies precisely at $x = 2/3$.
Looking back to our thermodynamic calculation of specific heat, the change in the number of negative modes at $x=2/3$ agrees with the change in sign of the specific heat there. We note that while the $p<1/6$ large black holes have a positive specific heat, they suffer from a Euclidean negative mode associated to the cavity, and thus presumably are not good saddle points of our ensemble.

Now we consider the special case $p = 1/6$, illustrated in the middle panel of the figure. Now we have a negative mode associated to the horizon for small black holes and none for large, and for all black holes we have an exact zero mode. The cavity instability for $p < 1/6$ becomes precisely marginal for $p = 1/6$. In fact one can quickly see that this marginal mode simply corresponds to a global scaling of the cavity, where both the black hole radius $r_+$ and the cavity radius $R_0$ are perturbed together, and arises as for the special value $p=1/6$ the boundary condition, $\gamma^p K = $constant is scale invariant, while the conformal class is always scale invariant.

The most interesting case is $p>1/6$. Here we have seen an empty cavity has no unstable modes and thus naively we expect that a small black hole will have one negative mode, a generalization of that of Gross-Perry-Yaffe, and a large black hole will have no unstable modes. This is precisely what we observe in the numerics, and can be seen in the right panel of the figure where the low-lying spectrum for the representative value $p=1/3$ is shown.
But this gives rise to a strange situation -- we have seen that the thermodynamics of black holes for $p>1/6$ has the novelty that very large black holes have negative specific heat capacity, and thus we would expect them to be thermodynamically unstable. However we see no hint of this in the Euclidean fluctuation analysis where the saddle point appears to be stable, at least to static spherically symmetry modes. 
In particular in that figure where $p=1/3$ we denote the range $x \ge 0.845$ for which the large black hole negative specific heat is negative, and we see no hint of negative modes.
We have carefully checked for other values of $p$ for the existence of static spherically symmetric negative modes in the range of $x$ where large black holes are thermodynamically unstable, finding none. \footnote{As a further check for $p > 1/6$ we have applied Ricci flow to static spherical perturbations of large thermodynamically \emph{unstable} Schwarzschild black holes with our boundary conditions, following the methods in~\cite{Headrick:2006ti}. We will not give the full details here, but report that generic static spherical perturbations quickly decay back to the large black hole fixed point, again confirming our conclusion that there are no Euclidean negative modes in the static spherical sector.}
While it is reasonable that the specific heat might be too crude a measure to capture a Euclidean instability -- for example the large black holes with $p < 1/6$ have an unstable mode due to the cavity, but positive specific heat -- it seems far less reasonable that on can have unstable thermodynamics but no sign of this when considering Euclidean fluctuations. It is worth emphasizing that we are only considering static and spherical fluctuations, but naively we might have expected these to capture the negative specific heat of large black holes for $p>1/6$. We will discuss this further when we consider the Lorentzian stability of these solutions.

\section{Dynamical Lorentzian Stability of Generalized Conformal Boundary Conditions}
\label{sec:Lorentzian}

Having discussed a new well-posed set of Euclidean boundary conditions, which apparently give a sensible thermodynamics in the case of $p > 1/6$ for Euclidean spherical cavities, we now consider the Lorentzian stability of these systems.
We will lead to two related puzzles. 

Firstly in section~\ref{sec:spherical},  we consider the Lorentzian dynamics in the spherical subsector of the theory. Due to the Birkhoff theorem the bulk geometry is fixed, and we may analyse the behaviour by considering purely the dynamics of the cavity boundary, which is analogous to the motion of a `brane', and can be solved non-linearly. We find that the Lorentzian stability of flat spacetime filled cavities is the same as that discussed above in the Euclidean theory, namely instability for $p < 1/6$ and stability for $p > 1/6$. The Lorentzian instability we see is identified to the dynamics of the boundary, and gives a physical interpretation for the Euclidean instability observed above. The unstable cavity walls wants to shrink to a point, or to expand forever.

Then in section~\ref{sec:puzzle1}, we consider the spherical dynamics of black holes, which produces a mystery. For $p < 1/6$ the picture is as expected -- there is a dynamical instability associated to the boundary which is paired to a Euclidean instability. However for sufficiently large black holes, for $p > 1/6$ we see a new dynamical instability, again associated to the boundary dynamics. This time this is not seen in the preceding Euclidean analysis, although interestingly it is hinted at in the earlier calculation of the specific heat.
Thus we have our first puzzle; the strange situation where a cavity with a black hole is stable in the Euclidean setting, but is unstable in Lorentzian signature. We should note that our Euclidean analysis of the Euclidean black hole was restricted to static and spherically symmetric modes, and this might be a potential resolution, although we suspect that this is not the case.

Our second puzzle then arises, in section~\ref{sec:puzzle2}, when we go beyond the spherical sector. We may analyse this only using linear perturbation theory. For flat spacetime filled cavities we find Lorentzian instabilities \emph{even} for $p > 1/6$, when they are stable in Euclidean signature. Thus going beyond the spherical sector we find a similar situation to that for the black holes, namely a stable Euclidean behaviour but unstable Lorentzian dynamics, even for an empty cavity.


\subsection{Spherical stability}\label{sec:spherical}

We begin with a spherically symmetric infilling geometry of a spherical cavity, 
\begin{equation}
{\rm d}s^2 = - V(t) F(r)^2 {\rm d}t^2 + \frac{{\rm d}r^2}{F(r)^2} + r^2 {\rm d}\Omega^2 \; .
\end{equation}
Of course in this situation for gravity with a cosmological constant Birkhoff's theorem dictates the infilling solution must be static (as in our ansatz above), and must in fact be Schwarzschild or its generalizations to (A)dS-Schwarzschild. Explicitly then it dictates the general infilling solution is:
\begin{equation}
F(r)^2 = - \frac{r^2 \Lambda}{3} + 1 - \frac{r_0}{r} \; .
\end{equation}
Here we have also included a reparameterization of time for convenience through the introduction of the (strictly positive) function $V(t)$. 
One might then naively imagine that with no gravitational dynamics for such spherical symmetry, there is no possibility to have an instability
\footnote{
We note that in fact one may take any 2-dimensional homogeneous space, so that,
\begin{equation}
{\rm d}s^2 = - V(t) F(r)^2 {\rm d}t^2 + \frac{{\rm d}r^2}{F(r)^2} + r^2 {\rm d}\Omega_k^2
\end{equation}
where ${\rm d}\Omega_k^2$ is the metric on the 2-dimensional homogeneous space, which explicitly we can take as ${\rm d}\Omega_k^2 = \frac{1}{1 - k x^2} {\rm d}x^2 + x^2 {\rm d}y^2$, so that its 2d Ricci scalar curvature is equal to $2 k$. The calculation leads to precisely the same equation~\eqref{eq:boundarydynamics} below, and is independent of $k$ (except through the function $F(r)$).}.

The main point we wish to make here is that while the interior geometry is fixed, the boundary conditions for the cavity wall can induce an instability entirely localized  there. In fact this instability is quite unrelated to the interior Einstein equation, and hence to emphasize this we will leave the function $F(r)$ entirely free, assuming only the static, spherical interior whose line element is above. Let us consider the boundary which shares the spherical symmetry. Hence in our coordinate chart this will follow a trajectory in the $t$-$r$ plane -- we may think of this as the dynamics of a `brane' -- which we parameterize as $r = R(t)$, and we assume the interior of the spacetime near the cavity is $r < R(t)$, so that the two dimensional homogeneous space shrinks inside the cavity.

We will impose our generalized conformal boundary conditions, so fixing the conformal class and $\gamma^p K$, or equivalently for non-vanishing $K$ fixing the rescaled boundary metric $\Gamma_{\mu\nu}$, and ask whether this is stable. The outward directed unit normal form, $n_\mu$, to the boundary is then, 
\begin{equation}
n_\mu = \frac{1}{\sqrt{F(r)^2 - \frac{R'(t)^2}{F(r)^2 V(t)}}} \left( - R'(t), 1, 0, 0 \right) 
\end{equation}
writing the coordinates as $x^\mu = (t, r, x, y)$.
Using this we compute the projection tensor $\perp_{\mu\nu} = g_{\mu\nu} - n_\mu n_\nu$, which can be used to pull the metric back to the cavity hypersurface to get the induced metric $\gamma_{\mu\nu}$, and the extrinsic curvature $K_{\mu\nu} = \perp_{\mu}^{~\alpha} \nabla_{\alpha} n_{\nu}$, from which we can extract its trace, $K$.
We assume that for the bulk infilling there is a static boundary location at radius $r = R_0$. Taking $V(t) = 1/F(R_0)^2$ and $R(t) = R_0$, we term the induced metric, $\gamma_0$, and trace of the extrinsic curvature of this static boundary as $K_0$, where these are given as,
\begin{equation}
{\rm d}s^2_{\gamma_0} =  - {\rm d}t^2 + R_0^2 {\rm d}\Omega^2 \; , \quad K_0 = F'(R_0) + \frac{2}{R_0} F(R_0) \; .
\end{equation}
We wish to consider the dynamics of the cavity, where the conformal class of the boundary $\gamma$ is fixed to that of the induced metric above, $\gamma_0$, and  $\gamma^p K$ is fixed to its value for this static case, so equal to  $\gamma_0^p K_0$.

For a dynamical cavity wall, so $R = R(t)$ we make the choice that,
\begin{equation}
V = \frac{\dot{R}^2}{F(R)^4} + \left( \frac{ R }{F(R) R_0 }\right)^2,
\end{equation}
and then we see that indeed the induced boundary metric is conformal to the static one above,
\begin{equation}
\dd s^2_{\gamma} = \frac{R^2}{R_0^2} \dd s^2_{\gamma_0}.
\end{equation}
 Now the condition on $\gamma^p K$ can be written explicitly as the `brane' equation of motion,
\begin{equation}
\label{eq:boundarydynamics}
\frac{\ddot{R}}{R} + \frac{\dot{R}^2}{R^2} -  \left(\frac{R}{R_0}\right)^{1-6 p}  \left( F'(R_0) + \frac{2}{R_0} F(R_0)  \right) \sqrt{ \frac{\dot{R}^2}{R^2} + \frac{ F(R)^2}{R_0^2}}  = - \frac{R}{R_0^2} F(R) \left( F'(R) + \frac{2}{R} F \right)
\end{equation}
and we may then regard the position of the cavity wall as a `brane' undergoing this dynamics.

If we perturb about the static point, so $R = R_0 + \delta R$, then linearizing we find,
\begin{equation}\label{eq:linear_expo}
\delta \ddot{R} = -  \alpha \delta R \; , \quad \alpha =  F(R_0) \left( F''(R_0) + \frac{2(1 + 3 p)}{R_0} F'(R_0) - \frac{2 ( 1 - 6 p)}{R_0^2} F(R_0)  \right) \; .
\end{equation}
Thus the stability is simply determined by the sign of the coefficient $\alpha$, positive implying stability and negative implying instability. Again we emphasize this depends on the Einstein equation only through the geometry of the static spacetime, and specifically the function $F$. In the case that the fixed point $R = R_0$ is unstable, we may use the dynamics above to solve the full non-linear evolution to see where the instability leads.

We might wonder what happens if we instead take the spacetime interior to be $r > R(t)$ so that the spherical spatial sections increase in size away from the cavity wall. This flips the sign of the extrinsic curvature, but the last two equations above governing the dynamics of cavity boundary, so $R(t)$, are unchanged.

\subsubsection{Linear instabilities for spherical cavities and black holes}

Now let us specialize to the case of a spherical cavity with no cosmological constant filled by flat spacetime, so $F(r) = 1$. Then,
\begin{equation}\label{eq:flat_alpha}
\alpha =  - \frac{2 (1 - 6p)}{R_0^2}  \,,
\end{equation}
so that for cavity boundary conditions with $p < 1/6$ (including Anderson boundary condition) the cavity is unstable, but for $p > 1/6$ we have stability.
We note that for this spherically symmetric sector the stability precisely agrees with that of the Euclidean fluctuation stability. We might expect this since for the marginal case of dynamical stability, $p = 1/6$, the Lorentzian static perturbation continues to a Euclidean zero mode, and hence marks the boundary of Euclidean stability as $p$ varies. We will show this is indeed the case  shortly.
We note that just as in the Euclidean case, we expect the frequency spectrum of perturbations to agree with the Dirichlet case in the $| p | \to \infty$ limit. However there may be modes with diverging frequency that do not  coincide with those of the Dirichlet problem, and we may view this static spherical dynamics exactly as such a perturbation -- we see the frequency, going as $\sqrt{\alpha}$ diverges in magnitude with $| p | \to \infty$. For $p \to - \infty$ this leads to an increasingly rapid instability, whereas for $p \to + \infty$ it is a stable mode. 

We may consider the case with cosmological constant too. Then writing $\tilde{x} = \Lambda R_0^2$, we have that $\tilde{x} < 0$ is the Anti de Sitter case, and $0 < \tilde{x} < 3$ is the de Sitter case where the cavity lies between the origin at the de Sitter horizon. Finally $\tilde{x} = 0$ corresponds to the flat case with zero cosmological constant. Then we find,
\begin{equation}
\alpha R_0^2 =   -  1 + 6 p (2  - \tilde{x}) - \frac{3}{3 - \tilde{x}}  \; .
\label{eq:growthads}
\end{equation}
As $\tilde{x} \to 3$ then $\alpha$ is negative, and hence a cavity sufficiently close to the de Sitter horizon size is always unstable for any $p$. Conversely for a large AdS cavity, so $x \to - \infty$, then for any $p > 0$ the cavity is stable, while for $p > 1/6$ any AdS cavity is stable.
\\

Now let us consider the linear stability of a Schwarzschild black hole background with no cosmological constant, so $F(r)^2=1-\frac{r_+}{r}$. 
Then from equation~\eqref{eq:linear_expo}, we obtain,
\begin{equation}\label{eq:bh_alpha}
    \alpha=\frac{F(R_0)^2}{4R_0^2(1-x)^2}\cdot\left[12 p (x-1) (3 x-4)+x (16-9 x)-8\right]
\end{equation}
where $x\equiv\rp/R_0\in(0,1)$ and so a marginal static perturbation implies
\begin{equation}
\label{eq:pstar}
    p_*=  \frac{8-16x+9x^2}{12(1-x)(4 - 3x)}\,.
\end{equation}
Then $p>p_*$ is the criteria for the black hole to be stable under our one-parameter family of boundary conditions. In the small black hole limit $x \to 0$ this reproduces the flat space criteria. When $0<x<2/3$ then $p_*$ is in the narrow interval $p_* \in (p_{min},1/6)$ with $p_{min}= (\sqrt{6}-2)/3 \simeq 0.150$.
Instead when $2/3<x<1$, then we have $p_*>1/6$, and further we see from the asymptotic behaviour, $p_* \simeq \frac{1}{12(1-x)}$ as $x \to 1$
that $p_*$ diverges to positive infinity as $x$ approaches to $1$, indicating that large enough black holes are never stable unless we have strict Dirichlet boundary conditions.
It appears that the close proximity of the cavity to the horizon leads to instability independent of $p$ similar to the case of the 
de Sitter cavity discussed above, where the cavity boundary becomes unstable if it gets to close to the de Sitter horizon.

Now if we compare the above equation with the black hole specific heat given in equation~\eqref{eq:SpecificHeat}, we see that the minimal value of $x$ for stability for $p>1/6$ is precisely the same as the $x$ where the specific heat changes sign. As a result, there is a dynamical instability associated with the thermodynamically unstable large black holes.

Thus our first mystery is revealed. We have seen a correspondence between the Lorentzian and Euclidean stability of flat space cavities. However, for black hole interiors and taking $p > 1/6$ the thermodynamics, specifically the specific heat, suggests an instability for sufficiently large black holes. We have now seen this is reflected in a Lorentzian dynamical instability. However, at least in the spherically symmetric static Euclidean sector, the corresponding Euclidean saddle point appears to be stable. Therefore this looks to be a system that is dynamically unstable, but nonetheless has a good (\emph{i.e.,} stable) thermodynamic description via the partition function. 
It is worth emphasizing that these large black holes still appear to be the dominant saddle point, as compared to hot flat space or the small black hole.
We might wonder what happened to the na\"ive intuition that the Euclidean and Lorentzian stability should agree if the stability for one changes at a marginal perturbation, since this marginal perturbation, being static, should be shared between both the Euclidean and Lorentzian pictures. It is instructive to see why this is not the case here, so we briefly revisit this Lorentzian linear stability calculation by considering a  spherical perturbation of the interior of the cavity, rather than the `brane' style calculation we have done above, in order that we can make contact with our previous Euclidean analysis.

\subsubsection{Revisiting the spherical stability calculation}\label{sec:puzzle1}

An equivalent way to study the spherically symmetric dynamics is by directly perturbing the interior of the cavity. Consider the background metric
\begin{equation}
\dd s^2 = \bar{g}_{ab} \dd x^a \dd x^b = - F(r)^2 \dd t^2 + \frac{1}{F(r)^2} \dd r^2 + r^2 \dd \Omega^2
\end{equation}
which we take either to be Schwarzschild, with $F(r)^2 =  1- \frac{r_+}{r}$, or flat spacetime, $F(r) = 1$,
and then consider a perturbation induced by the diffeomorphism,
\begin{equation}
r \to r ( 1 + \epsilon f(r) ) \; , \quad t \to t ( 1 + \epsilon q ) \,,
\end{equation}
which induces a new metric, related to the original one by the perturbation,
\begin{equation}
\dd s^2 = g_{ab} \dd x^a \dd x^b = \left( \bar{g}_{ab} + \epsilon h_{ab} \right) \dd x^a \dd x^b\,,
\end{equation}
such that,
\begin{equation}
h_{ab} \dd x^a \dd x^b = - F(r)^2 [ 1+ \epsilon \,\delta T(r) ] \dd t^2 + \frac{1}{F(r)^2}  [ 1+ \epsilon\, \delta A(r) ]  \dd r^2 + r^2   [ 1+ \epsilon\, \delta S(r) ]  \dd \Omega^2\,,
\end{equation}
with
\begin{eqnarray}
\label{eq:perturbation}
\delta T(r) & = & - 2 q - 2 r f(r)  \frac{F'(r) }{F(r) }\,, \nl
\delta A(r)  & = & - 2 r f'(r)  + f(r)  \left( 2 r \frac{F'(r) }{F(r) } - 2 \right)\,, \nl
\delta S(r)  & = & - 2 f(r)\,. 
\end{eqnarray}
Since this new metric is related to the original metric by a diffeomorphism, the new metric also solves the bulk equations of motion. Furthermore, this metric is static for any choice of $f(r)$.

However, now we do \emph{not} perturb the boundary location to $r = R_0 ( 1 + \epsilon f(R_0))$, because otherwise we have not done anything! Instead, we keep the boundary fixed at $r = R_0$, and focus on $f(r)$ such that $f(R_0) \ne 0$.
Then if we obey the boundary conditions at $r = R_0$ this is a genuine physical static perturbation of the system.

Without loss of generality, we choose units such that $R_0 = 1$. The boundary induced metric is given by
\begin{equation}
\dd s^2_\gamma = \gamma_{\mu\nu} \dd x^\mu \dd x^\nu = - F(R_0)^2 [ 1 + \epsilon \delta T(R_0)  ] \dd t^2 + R_0^2 [ 1 + \epsilon \delta S(R_0) ] \dd \Omega^2\,.
\end{equation}
Now we wish to fix,
\begin{equation}
\Gamma_{\mu\nu} = \left. K^{\frac{1}{3p}} \gamma_{\mu\nu} \right|_{r = 1}\,,
\end{equation}
where at the boundary $r = 1$,
\begin{equation}
K_{r=1} = K_0 + \epsilon \delta K \, , 
\end{equation}
with
\begin{equation}
    K_0 = 2 F(1) + F'(1) \, , \quad \delta K = F(1) \left( \delta S'(1) + \frac{1}{2} \delta T'(1) - \delta A(1) \right) - \frac{1}{2} F'(1)  \delta A(1)\,.
\end{equation}
Now,
\begin{eqnarray}
    \frac{\delta \Gamma_{tt}}{\Gamma_{tt}} & = & - 2 q - 2 f \frac{F'}{F} + \frac{f ( 2 F - 2 F' - F'' )}{3 p ( 2 F  + F' )}\,, \nl
\frac{\delta \Gamma_{\theta\theta}}{\Gamma_{\theta\theta}} & = & - \frac{f \left( - 2 (1 - 6 p ) F + 2 (1 + 3 p) F' + F'' \right)}{3 p ( 2 F  + F' )}\,.
\end{eqnarray}
where the r.h.s are evaluated at $r = 1$. Given that we must have $f(1) \ne 0$, or we have not done anything, the solution to these is given by,
\begin{equation}
\label{eq:bcs}
q = f(1) \left( 1 - \frac{F'(1)}{F(1)} \right) \; , \quad 0 = F''(1) + 2 (1 + 3 p) F'(1) - 2(1 - 6 p) F(1)\,.
\end{equation}
Note that compared to the `brane' calculation, the second condition is precisely the `brane' linearized condition in equation~\eqref{eq:linear_expo} (for $R_0 = 1$) that $\alpha = 0$. We emphasize that the function $f(r)$ away from the boundary is arbitrary, except that it should preserve the position of the origin of spherical coordinates, or the position of the horizon. However, we could simply choose that it is compactly supported close to the boundary and then these would follow. It is only the fact that it is non-zero at the boundary that is important.

Consider the case of a perturbed flat cavity, where we see the Lorentzian and Euclidean stability precisely agree, with the transition being at $p = 1/6$. The agreement may be understood as the marginal Lorentzian mode at $p=1/6$, being static, can be analytically continued to give a marginal Euclidean mode (a zero mode, so $\tilde{\lambda} = 0$, of section~\ref{sec:EucStaticSpherical}).
Why then is there disagreement for a black hole interior and $p > 1/6$? The Euclidean fluctuations are stable in the spherical sector, but the Lorentzian ones are unstable for sufficiently large black holes. Consider now the Lorentzian marginal perturbation, so $p = p_*$, for a suitably large black hole. This is again a static mode, and so the question is, why can we not continue this to give a Euclidean zero mode, which would separate Euclidean stability from instability? 

Indeed, we can continue the static mode to the Euclidean signature. However, if we consider the perturbation above in equation~\eqref{eq:perturbation} since necessarily $q \ne 0$ it induces a conical deficit at the Euclidean horizon, even if $f(r)$ is compactly supported near the cavity wall, and has no support near this horizon. In the flat space infilling there is no horizon, and so the corresponding marginal mode continues without a problem to the Euclidean setting
\footnote{
One might wonder whether one could consider a perturbation generated by the diffeomorphism, $r \to r ( 1 + \epsilon f(r) )$ and $t \to t ( 1 + \epsilon q(r) )$, where $q(r)$ is chosen to be compactly supported near the cavity boundary as for $f(r)$. However now the metric perturbation will have explicit time dependence in the off-diagonal terms going linearly in $t$, so as $\sim F(r) q'(r) t \,{\rm d}t {\rm d}r$.
After Euclidean continuation the $\tau$ dependence, $\sim \tau\, {\rm d}\tau {\rm d}r$, will not be compatible with having a compact Euclidean time identified as $\tau \sim \tau + \beta$.}.

This suggests that enforcing smoothness of the Euclidean horizon is then the underlying reason that for $p > 1/6$ we appear to have dynamically unstable large black holes that nonetheless are stable Euclidean saddle points. It points to the possibility that perhaps the Euclidean path integral should be taken over metrics that are allowed to have conical deficits.

\subsubsection{End-point of the instabilities}

Let us return to the `brane' calculation of spherical stability, as it allows us to consider the full non-linear evolution in the spherical sector. An obvious question is, where do the dynamical instabilities find above evolve the system to?
Let us start by considering the non-linear evolution of an empty flat spherical cavity, so $\Lambda = 0$ and $F = 1$, where we have seen that for $p < 1/6$ the dynamics of the boundary are unstable. In this case where the only scale is set by the boundary, we may choose units where the static boundary position is $R_0 = 2$ without loss of generality to simplify the equations. Then writing $R(t) = 2 \sqrt{W(t)}$ the dynamical equation for the boundary, equation~\eqref{eq:boundarydynamics}, becomes  \footnote{
Equation~\eqref{eq:WNL} admits an analytic first integral,
\begin{equation}
\dot{W}^2=C_0^2-W^2+2 C_0 \frac{W^{n+1}}{n+1}+\frac{W^{2(n+1)}}{(n+1)^2}
\end{equation}
where $C_0$ is a constant of integration and we defined $n=\frac{1}{2} (1-6 p)$.
We note that for some values of $p$ it can actually be solved analytically. For instance, for $p=-1/6$ it can be solved in terms of some Jacobi ($\rm sn$) function.
}
\begin{equation}
\frac{\ddot{W}}{W} = W^{\frac{1-6p}{2}} \sqrt{1 + \frac{\dot{W}^2}{W^2}} - 1\,.
\label{eq:WNL}
\end{equation}
The static point then corresponds to the solution $W(t) = 1$. In the special case of marginal stability, $p = 1/6$, we see that in fact the cavity wall can be placed anywhere and will remain exactly static. Thus $W = W_0 > 0$ is a solution. {As noted earlier, this is a result of the boundary condition, which fixes $\gamma^{\frac{1}{6}} K$ and the conformal class of $\gamma_{\mu\nu}$, having an exact global scale symmetry. 

However as we see from the general $p=1/6$ solution,
\begin{equation}
W = W_0 \left( 1 + \frac{2 v t + v^2 t^2}{1 - v^2} \right)
\end{equation}
for constants $W_0$ and $v$, at the non-linear level these marginally stable fixed points are in fact unstable. If $W$ is perturbed at $t = 0$ so that $\dot{W} = \frac{2 v W_0}{1 - v^2} \ne 0$, then if $v < 0$ or $v > 1$ the cavity boundary will reach zero size in a finite time,
\begin{equation}
t_{\rm shrink} = \left\{ \begin{array}{cc}
 1 - \frac{1}{v}\,, &  |v| > 1 \\
 \frac{1}{|v|} - 1 \,, & v \in (-1,0) \\

 \end{array}
  \right.\,.
\end{equation}
For $-1 < v < 0$ and $v > 1$ the boundary initially shrinks, but not with the exponential time dependence of the instabilities we have seen in the case $p < 1/6$. Thus this is a non-linear instability. Further, for $v<-1$ initially the cavity expands, before then turning around and shrinking to zero size. For the remaining range $0 < v <1$, the cavity wall initially expands, and continues to expand forever, approaching asymptotically the accelerated expansion $W \sim \frac{v^2 W_0}{1 - v^2} t^2$. The behavior of $W(t)$ for different $v$'s are plotted in the left panel of figure~\ref{fig:WT}.

\begin{figure}
    \centering
    \includegraphics[width=\linewidth]{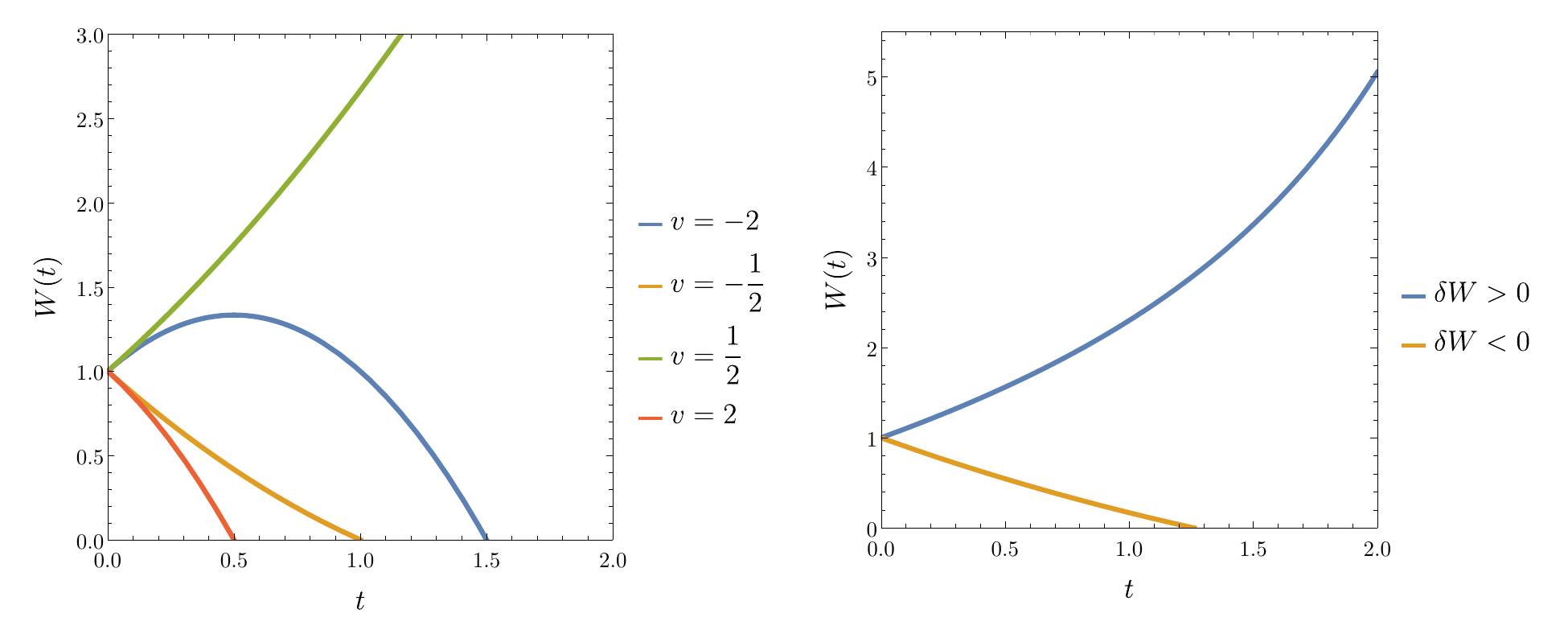}
    \caption{Behavior of the function $W(t)$ under different $p$'s and initial conditions. The left panel has $p=\frac{1}{6}$ and the right panel has $p=\frac{1}{12}<\frac{1}{6}$.}
    \label{fig:WT}
\end{figure}
Now let us consider the unstable case $p < 1/6$ where the fixed point is $R = R_0$, so $W = 1$. Consider that the fixed point is perturbed and then the dynamics takes $W$ away from one. To explore the behaviour let us consider whether we may find $\dot{W} = 0$ at a later time, so that $W$ has an extremum. This condition implies that where $\dot{W} = 0$ then,
\begin{equation}
\frac{\ddot{W}}{W} =  W^{\frac{1-6p}{2}} - 1 \; .
\end{equation}
Now given that $p < 1/6$, this implies that at an extremum, $\ddot{W} > 0$ if $W >1$, and $\ddot{W} < 0$ if $0 < W <1$. Thus $W$ can't have a maximum if $W > 1$, and it can't have a minimum if $0 < W < 1$. As a result, we conclude that if $W$ initially is perturbed in the positive sense from $W = 1$ (so initially $W\ge 1$ and $\dot{W} \ge 0$), it will continue to grow indefinitely. Likewise if it is perturbed in the negative sense (so initially $W\le 1$ and $\dot{W} \le 0$), then $W$ will decrease indefinitely towards $W = 0$. An example of this behavior of $W(t)$ taking  $p=\frac{1}{12}<\frac{1}{6}$ is plotted in the right panel of figure~\ref{fig:WT}.

We may consider the cavity motion in the general background as for the dynamics~\eqref{eq:boundarydynamics}. Then $R = R_0$ is a fixed point for the cavity radius. Let's consider a perturbation of this, and ask again whether we may have $\dot{R} =0$. This implies the condition,
\begin{equation}
\frac{\ddot{R}}{R}   =  \frac{R^{1-6p} F(R)}{R_0^2} \left( Q(R)  -   Q(R_0) \right)\,,
\end{equation}
where we have defined,
\begin{equation}
Q(R) = - \frac{ R F'(R) + 2 F }{R^{1-6p}} \; .
\end{equation}
Now if $R > R_0$ implies $Q(R) > Q(R_0)$, then again $\ddot{R} > 0$ and there can be no maximum in $R$, and hence the cavity will indefinitely grow if initially it is perturbed with $R \ge R_0$ and $\dot{R} \ge 0$. Suppose the minimum size the cavity could be is $R_{\rm min}$ -- this might be zero size, or it might be the horizon size if the cavity contains a black hole. Then if $R_{\rm min} < R < R_0$ implies $Q(R) < Q(R_0)$ then $\ddot{R} < 0$ and there can be no minimum in $R$, and $R$ will decrease to $R_{\rm min}$ if initially it is perturbed with $R \le R_0$ and $\dot{R} \le 0$.
A particular case of such behaviours is found if $Q(R)$ is a monotonically increasing function of $R$ for $R > R_{min}$. 

One example of this is the case of Anderson boundary conditions, so that $p=0$, and an infilling that is Schwarzschild, possibly including a cosmological constant, so,
\begin{equation}
F(r) = \sqrt{ - \frac{r^2 \Lambda}{3} + 1 - \frac{r_0}{r}} \; .
\end{equation}
Then $Q(R)$ is indeed monotonically increasing for all values of $r_0$ and $\Lambda$, and for cavity radii greater than the black hole horizon size. We have previously seen that Anderson boundary conditions are linearly unstable for spherical cavities infilled with black holes. Now we see that this instability leads to unbounded expansion of the cavity, or collapse of the boundary to the horizon.

Let us now consider Schwarzschild with no cosmological constant, so $\Lambda = 0$, and our generalized boundary conditions. We have seen above that for a given $x$ then for $p < p_*(x)$, with $p_*$ given in equation~\eqref{eq:pstar}, the Schwarzschild infilling is linearly unstable.
For small black holes with $0 < x < 2/3$, then $p_* \in (\frac{\sqrt{6}-2}{3} , 1/6)$, and so these mirror the stability of the empty cavity in the sense that for $p > 1/6$ they are stable. However for large black holes with $x > 2/3$ we saw that $p_* > 1/6$, and further that $p_* \sim \frac{1}{12(1-x)}$ as $x \to 1$, and so while an empty cavity is stable for $p > 1/6$, if it is filled with a sufficiently large black hole it then becomes unstable in this spherical dynamical sector. What can we say non-linearly about these dynamics?

Firstly it is easy to show that for $-\frac{\sqrt{6}+2}{3}<p < \frac{\sqrt{6}-2}{3}$ then $Q(R)$ is an increasing monotonic function for any $R$ greater than the horizon size. Small black holes in cavities with such $p$ and are linearly unstable, and this monotonicity of $Q(R)$ shows this instability collapses the cavity to the horizon or expands it forever.

In order to consider the other ranges of $p$ we note that the behaviour of $Q(R)$ near the horizon and asymptotically are,
\begin{eqnarray}
Q(R)=-\frac{1}{2}\frac{r_0^{6p-\frac{1}{2}}}{\sqrt{R-r_0}}+O(\sqrt{R-r_0}) \; , \qquad Q(R) = - 2 R^{6 p-1} \left( 1 + O(R^{-1}) \right)
\end{eqnarray} 
respectively. Hence for $p < 1/6$ we have $Q(R) < Q(R_0)$ for $R$ near the horizon and $Q(R_0) < Q(R)$ for large radius too. Consider the case of small black holes which are unstable with $p$ in the range $p \in (\frac{\sqrt{6}-2}{3} , 1/6)$. Then the function $Q(R)$ is not monontonic, but has both a maximum at the radius $R = R_-$ and a minimum at the radius $R = R_+$, where these are given by,
\begin{eqnarray}
\frac{R_{\pm}}{r_0}=\frac{1 - \frac{21}{4} p}{1 - 6 p}  \pm \frac{\sqrt{9p^2 + 12 p - 2}}{4(1-6p)}\,,
\end{eqnarray}
so we see that $r_0 < R_- < R_+$. Given that the cavity is unstable then either the unperturbed cavity radius $R = R_0$ is at a smaller radius than the maximum, or a larger radius than the minimum; so either $r_0 < R_0 < R_-$ or $R_+ < R_0$. In the first case then $Q(R)$ is monotonic for $r_0 < R < R_0$ and so if the cavity wall is perturbed so that $R \le R_0$ and $\dot{R} \le 0$ then it will continue to fall to the horizon. In the second case it is monotonic for $R_0 < R$ and so if it is perturbed in an increasing sense it will continue to expand forever. 

In the interesting case of $p > 1/6$ the function $Q(R)$ has a single maximum at the value given by $R_-$ above for an unstable cavity, and in particular $R_0 < R_-$. Thus of the cavity is perturbed in the shrinking sense, so initially $R \le R_0$ and $\dot{R} \le 0$ then it will continue to fall to the horizon. If perturbed in the increasing sense is initially grows, but since for large $R$ then $Q(R) \simeq - 2 R^{6 p-1}$, so $Q(R) - Q(R_0) < 0$, then generically it reaches a maximum size and recontracts.

Thus in summary, we appear to find that where a linear dynamical instability exists for a cavity boundary, it tends to act to either shrink the cavity boundary down to a minimal size (either zero radius in an empty cavity or to the black hole horizon) or it acts to expand the cavity radius indefinitely.

\subsection{Stability beyond spherical symmetry}\label{sec:puzzle2}

Having explored dynamical stability for spherical symmetry, we now explore
 the linear stability of our novel boundary conditions concerning perturbations that \emph{break spherical symmetry}, typically examined through the Kodama-Ishibashi formalism \cite{Kodama:2003jz}. We focus on the case of an empty spherical cavity with a vanishing cosmological constant. The background metric thus reads
\begin{equation}
{\rm d}s^2 = -{\rm d}t^2+{\rm d}r^2+r^2 {\rm d}\Omega_2^2\,,
\end{equation}
and we take the cavity to be located at $r=R_0$.

Our initial objective is to understand how our cavity boundary conditions influence the boundary conditions for the Kodama-Ishibashi master variables. These variables govern gravitational perturbations derived from both scalar and vector harmonics on the two-sphere. The vector-derived perturbations can be shown to have the same boundary conditions as in the Dirichlet problem, and were thus studied in \cite{Andrade:2015gja} and no instabilities were found. For this reason, we will focus on scalar-derived perturbations.

The generic form of scalar-derived gravitational perturbations has been explained in section \ref{sec:Fieldtheoretic} in the Euclidean context. The same reasoning applies to the Lorentzian setup so that,
\begin{subequations}
\begin{align}
&h^{\ell_S\,m_S}_{\hat{a}\hat{b}}=\hat{f}_{\hat{a}\hat{b}}^{\ell_S\,m_S}(t,r)\,Y^{\ell_S\,m_S}
\\
& h^{\ell_S\,m_S}_{\hat{a}I}=-r \hat{f}_{\hat{a}}^{\ell_S\,m_S}(t,r)\frac{\mathcal{D}_I Y^{\ell_S\,m_S}}{\sqrt{\ell_S(\ell_S+1)}}
\\
& h^{\ell_S\,m_S}_{IJ}=2r^2\left[\hat{h}^{\ell_S\,m_S}_L(t,r)\,G_{IJ}\,Y^{\ell_S\,m_S}+\hat{h}_T^{\ell_S\,m_S}(t,r)\,\frac{S^{\ell_S\,m_S}_{IJ}}{\ell_S(\ell_S+1)}\right]
\end{align}
\label{eqs:tot}
\end{subequations}\\
where $\mathcal{S}^{\ell_S\,m_S}_{IJ}$ is the traceless symmetric two tensor defined in Eq.~(\ref{eq:SIJ}) and recall that $Y^{\ell_S\,m_S}$ is a spherical harmonic on the two-sphere obeying the earlier equation~\eqref{eq:SphHarm}.
Note that there are a few factors of $2$, $\ell_S(\ell_S+1)$, $r$ and $r^2$ appearing in (\ref{eqs:tot}) that differ from those of the Euclidean analysis detailed in Appendix~\ref{sec:nonsphericalS}. These have been added to make contact with \cite{Kodama:2003jz,Andrade:2015gja}. Since the background is static, \emph{i.e.} has a timelike Killing vector field $\partial/\partial t$, we can decompose all perturbations in Fourier modes as
\begin{multline}
\hat{f}_{\hat{a}\hat{b}}^{\ell_S\,m_S}(t,r)=e^{-i\,\omega\,t}f_{\hat{a}\hat{b}}^{\ell_S\,m_S\,\omega}(r)\,,\quad \hat{f}_{\hat{a}}^{\ell_S\,m_S}(t,r)=e^{-i\,\omega\,t}f_{\hat{a}}^{\ell_S\,m_S\,\omega}(r)\,,
\\
\hat{h}_L^{\ell_S\,m_S}(t,r)=e^{-i\,\omega\,t}h_L^{\ell_S\,m_S\,\omega}(r)\,,\quad\text{and}\quad \hat{h}_T^{\ell_S\,m_S}(t,r)=e^{-i\,\omega\,t}h_T^{\ell_S\,m_S\,\omega}(r)\,.
\end{multline}
For given boundary conditions, we would like to determine
\begin{equation}
\Big\{f_{\hat{a}\hat{b}}^{\ell_S\,m_S\,\omega}(r),f_{\hat{a}}^{\ell_S\,m_S\,\omega}(r),h_L^{\ell_S\,m_S\,\omega}(r),h_T^{\ell_S\,m_S\,\omega}(r),\omega\Big\}\,.
\end{equation}
An instability is then identified by a mode with ${\rm Im}(\omega R_0)>0$.

At this point, we could introduce a particular gauge, and make progress. However, using the procedure outlined by Kodama and Ishibashi in \cite{Kodama:2003jz} we will bypass this altogether and instead work with a master variable that is manifestly gauge invariant. To decrease the clutter, we will drop all the superscripts $\ell_S\,m_S\,\omega$.

We start by presenting what our boundary conditions imply for the metric components. These are simply given by
\begin{multline}
f_{t}(R_0)=0\,,\quad h_T(R_0)=0\,,\quad f_{tt}(R_0)+2h_L(R_0)=0\,,
\\
f^{\prime}_{tt}(R_0)-4 h_L^\prime(R_0)-\frac{24\,p}{R_0}h_L(R_0)+\frac{2}{R_0}f_{rr}(R_0)+\frac{2\sqrt{\ell_S(\ell_S+1)}}{R_0}f_r(R_0)+2 \,i\,\omega\,f_{tr}(R_0)=0\,.
\label{eq:refBCSL}
\end{multline}
These are the boundary conditions that we would like to rewrite in terms of the corresponding Kodama-Ishibashi variable.

Recall that any gauge transformation can also be decomposed in terms of the harmonic decomposition that we used to write the metric perturbations, and in particular, if $\xi_a$ parametrizes such an infinitesimal diffeomorphism we have
\begin{subequations}
\begin{align}
&\xi_{\hat{a}}(t,r,\theta,\phi) = \xi^{\ell_S\,m_S\,\omega}_{\hat{a}}(r) e^{-i\omega t} Y^{\ell_S\,m_S}(\theta,\phi)\,,
\\
&\xi_I(t,r,\theta,\phi) = -\xi^{\ell_S\,m_S\,\omega}(r) e^{-i\omega t} \frac{\mathcal{D}_IY^{\ell_S\,m_S}(\theta,\phi)}{\sqrt{\ell_S(\ell_S+1)}}\,.
\end{align}
\end{subequations}
One can check that the following metric combinations are invariant under infinitesimal transformations generated by $\xi$
\begin{subequations}
\begin{align}
&F=h_L+\frac{1}{2}h_T+\frac{1}{r} (\hat{\nabla}_{\hat{a}}r) 
\hat{g}^{\hat{a}\hat{b}}X_{\hat{a}}
\\
&F_{\hat{a}\hat{b}}=f_{\hat{a}\hat{b}}+\hat{\nabla}_{\hat{a}}X_{\hat{b}}+\hat{\nabla}_{\hat{b}}X_{\hat{a}}
\end{align}
where $\hat{\nabla}$ is the metric preserving connection associated with the two-dimensional Minkowski spacetime
\begin{equation}
\hat{g}_{\hat{a}\hat{b}}{\rm d}x^{\hat{a}}{\rm d}x^{\hat{b}}=-{\rm d}t^2+{\rm d}r^2\,,
\end{equation}
and
\begin{equation}
X_{\hat{a}}\equiv\frac{1}{\sqrt{\ell_S(\ell_S+1)}}\left[f_{\hat{a}}+\frac{r}{\sqrt{\ell_S(\ell_S+1)}}\hat{\nabla}_{\hat{a}}h_T\right]\,.
\end{equation}
\end{subequations}
One can rewrite the boundary conditions (\ref{eq:refBCSL}) in terms of $F$ and $F_{\hat{a}\hat{b}}$. Three of these boundary conditions are manifestly gauge dependent (for instance, one of the boundary conditions demands $h_T(R_0)=0$ and $h_T$ itself is gauge dependent), but can be solved to determine linear combinations of metric functions. However, one can show that a \emph{single} gauge invariant boundary condition emerges that can be expressed in terms of $F$ and $F_{\hat{a}\hat{b}}$
\begin{multline}
F_{tt}^\prime(R_0)-4 F^\prime(R_0)+\frac{1}{R_0}\left[\ell_S(\ell_S+1)-2-\omega^2 R_0^2+12p\right]F_{tt}(R_0)
\\
+\frac{2}{R_0}\left[\ell_S(\ell_S+1)-2-\omega^2 R_0^2\right]F(R_0)+\frac{2}{R_0}F_{rr}(R_0)+2 i \omega F_{tr}(R_0)=0\,.
\label{eq:BCSGAUGE}
\end{multline}
To proceed, we introduce three auxiliary quantities $X$, $Y$ and $Z$ so that
\begin{subequations}
\begin{align}
&F_{tt}(r)=-\frac{X(r)-Y(r)}{2}\,,
\\
&F_{tr}(r)=i\,\omega\,Z(r)\,,
\\
&F_{rr}(r)=-\frac{X(r)-Y(r)}{2}\,,
\\
& F(r) = -\frac{X(r)+Y(r)}{4}\,.
\end{align}
\end{subequations}%
One might wonder about the motivation behind this change of variables. Indeed, using the Einstein equation, it can be demonstrated that both $F_{{\hat{a}}\hat{b}}$ and $F$ adhere to two algebraic constraints. The variables $X$, $Y$, and $Z$ are selected to retain a single algebraic constraint. Consequently, the Einstein equation is simplified to three first-order differential equations for $X$, $Y$, and $Z$, along with a lone algebraic constraint linking them. These equations can be found in \cite{Kodama:2003jz} and turn out to be given by
\begin{subequations}
\begin{align}
&X'(r)-\frac{Z(r) \ell _S \left(\ell _S+1\right)+r Y(r)}{r^2}+\frac{3 Y(r)}{r}+\omega ^2 Z(r)=0\,,
\\
&Y'(r)-\omega ^2 Z(r)=0\,,
\\
&Z'(r)-X(r)=0\,,
\\
&
r \omega ^2 [r X(r)+r Y(r)-2 Z(r)]-Y(r) \left[\ell _S(\ell _S+1)-2\right]=0\,.
\end{align}
\label{eqs:firstorder}
\end{subequations}%

Finally, one can introduce a variable $\Phi$ that is constructed from $X$, $Y$ and $Z$ in terms of which the gravitational perturbations simplify immensely (particularly if we were to study black holes in the interior of the cavity, or higher-dimensional cavities). Defining,
\begin{subequations}
\begin{equation}
\Phi(r)=\frac{2Z(r)-r\,X(r)-r\,Y(r)}{\ell_S(\ell_S+1)-2}\,,
\end{equation}
then we find that it obeys the second-order differential condition,
\begin{equation}
\Phi^{\prime\prime}+\left[\omega^2-\frac{\ell_S(\ell_S+1)}{r^2}\right]\Phi=0\,.
\label{eq:phiKI}
\end{equation}
\end{subequations}%
From the definition of $\Phi$ and from Eqs.~(\ref{eqs:firstorder}), we  reconstruct $X$, $Y$ and $Z$ from $\Phi$ as,
\begin{subequations}
\begin{align}
& X=-\frac{\ell_S(\ell_S+1)-r^2\omega^2}{r}\Phi-2 \Phi^\prime\,,
\\
& Y =-r \omega^2 \Phi\,,
\\
& Z = -\Phi-r \Phi^\prime\,.
\end{align}
\end{subequations}%
Finally, using the above relations, and the definition of $F_{tt}$, $F_{tr}$, $F_{rr}$ and $F$ in terms of $X$, $Y$, $Z$ one can rewrite the boundary condition (\ref{eq:BCSGAUGE}) solely in terms of $\Phi$ and $\Phi^\prime$ finding,
\begin{multline}
\left[4(1-6p)-3\ell_S(\ell_S+1)+2 R_0^2\omega^2\right]R_0 \Phi^\prime(R_0)
\\
-\left\{2R_0^4\omega^4+4[1-6p-\ell_S(\ell_S+1)]R_0^2\omega^2-\ell_S(\ell_S+1)[3(1-4p)-2\ell_S(\ell_S+1)]\right\}\Phi(R_0)=0\,.
\end{multline}
For $p=0$, \emph{i.e.} Anderson boundary conditions, the above boundary condition agrees with the boundary conditions presented in \cite{Anninos:2023epi}, while as $p\to+\infty$ we recover those reported in \cite{Andrade:2015gja}.

The generic solution to Eq.~(\ref{eq:phiKI}) can be readily found in terms of Bessel functions
\begin{equation}
\Phi(r)=\sqrt{r \omega}\left[C_1\,{\rm J}_{\ell_S+\frac{1}{2}}\left(r \omega\right)+C_2\,{\rm Y}_{\ell_S+\frac{1}{2}}\left(r \omega\right)\right]\,.
\end{equation}
Regularity at the origin demands $C_2=0$, and then our boundary condition yields,
\begin{multline}
\left[4(1-6p)-3\ell_S(\ell_S+1)+2 R_0^2\omega^2\right]\,R_0\,\omega\,{\rm J}_{\ell_S+\frac{3}{2}}\left(R_0 \omega\right)
\\
+\Big\{2R_0^4\omega^4+2[2(1-6p)-(2\ell_S+1)(\ell_S+1)]R_0^2\omega^2
\\
-(\ell_S+1)\left[3(1-4p)\ell_S-(2\ell_S+3)(\ell_S+1)\ell_S+4(1-6p)\right]\Big\}{\rm J}_{\ell_S+\frac{1}{2}}\left(R_0 \omega\right)=0\,.
\label{eq:simple}
\end{multline}
The above transcendental equation governs how empty spherical cavities in flat spacetime react to non-spherical gravitational perturbations induced by scalar-derived gravitational deformations.
This is the analog expression to that in equation~\eqref{eq:pexpression} that we analysed earlier for the Euclidean fluctuations.

Before commenting on numerical results obtained from solving Eq.~(\ref{eq:simple}), let us analyse the large $\ell_S$ limit of $\omega R_0$ with $p$ being held fixed. Using \emph{uniform} asymptotic expansions for Bessel functions at large order in terms of Airy functions (see for instance \cite{doi:10.1098/rsta.1954.0021}) one can solve for the large $\ell_S$ limit of $\omega R_0$ appearing in (\ref{eq:simple}), which turns out to be given by
\begin{subequations}%
\begin{equation}
R_0 \omega = \ell_S-\Gamma\left(\frac{\ell_S}{2}\right)^{1/3}+\frac{1}{2}+\frac{1}{16\Gamma}\frac{11-96 p-128 \Gamma^3}{7+16 \Gamma^3}\left(\frac{2}{\ell_S}\right)^{1/3}+\mathcal{O}(\ell_S^{-2/3})
\label{eq:exp}
\end{equation}
where $\Gamma$ can be any solution to the following transcendental equation:
$$
{\rm Ai}^\prime(\Gamma)=4 \Gamma^2 {\rm Ai}(\Gamma)
$$
with ${\rm Ai}$ the Airy function and ${\rm Ai}^\prime$ its first derivative. In particular, we find (at least) \emph{two complex roots}:
\begin{equation}
\Gamma \approx -0.0674243 + 0.427905 i\quad\text{or}\quad \Gamma \approx -0.0674243 - 0.427905 i\,
\end{equation}
\end{subequations}%
showing that an instability exists in the spectrum regardless of $p$, so long as $\ell_S$ is large enough while $p$ is held fixed. We note that this is in broad agreement with the particular case of Anderson boundary conditions studied in~\cite{Anninos:2023epi} which derived similar asymptotics for instabilities for large $\ell_S$, but which differ in detail\footnote{In that work they found instabilities going roughly as $R_0 \omega \simeq \pm \ell_S \pm i \ell_S^{1/3}$ for large $\ell_S$ for the Anderson boundary conditions. Our result above gives the precise asymptotic behaviour which is slightly different in the crucial imaginary subleading term.}.
The point to emphasize here is that for large $\ell_S$ all our family of boundary conditions share the same large $\ell_S$ instabilities originating from the imaginary part of $\Gamma$ above.
This is one of the main results of this paper, since for the case $p > 1/6$ it shows that a spherical cavity endowed with our boundary conditions appears to be free of Euclidean unstable modes (as argued in the previous sections), and yet is dynamically \emph{unstable} in the Lorentzian section with infinitely many unstable non-spherical modes!

Unlike the case of $\ell_S=0$, which can be followed nonlinearly, here we can only speculate as to what the endpoint of this instability might look like. We note also that the instability growth rate grows at large $\ell_S$, indicating that it is going to be hard to shut down the instability in a simple manner. 

This is suggestive that the instability leads to a `turbulent' behaviour with power being transferred to shorter and shorter scales as these large $\ell_S$ instabilities are triggered by non-linear interactions.
The phenomenon of the superradiant instability of Kerr-AdS is a gravitational example where it is conjectured that a linear instability leads to such a non-smooth final state, again with power transferred to arbitrarily short scales \cite{Cardoso:2013pza,Niehoff:2015oga,Chesler:2018txn,Chesler:2021ehz}. It is worth noting that in the case of superradiance the instability timescale becomes slower for modes with increasingly short scale azimuthal dependence, whereas in our case the instability actually grows with $\ell_S$ as the power $\ell_S^{1/3}$, so it may be that the turbulent cascade happens even more rapidly.

We validate the expansion (\ref{eq:exp}) by comparing it with numerical results obtained through the direct solution of (\ref{eq:simple}) using a Newton-Raphson solver. In Fig.\ref{fig:com} we compute the real (left panel) and imaginary (right panel) parts of $\omega$ as functions of $\ell_S$ for the case $p=0$. We successfully tracked the mode up to $\ell_S=2000$. 
Achieving such large values of $\ell_S$ requires resorting to extended precision, keeping at least the first five hundred digits. This is necessary because evaluating the Bessel functions in (\ref{eq:simple}) at such high orders is notoriously challenging. 
On both panels, the solid black line is the asymptotic prediction (\ref{eq:exp}) while the blue disks represent the numerical data, and we see good agreement at large $\ell_S$.
\begin{figure}
    \centering
    \includegraphics[width=\linewidth]{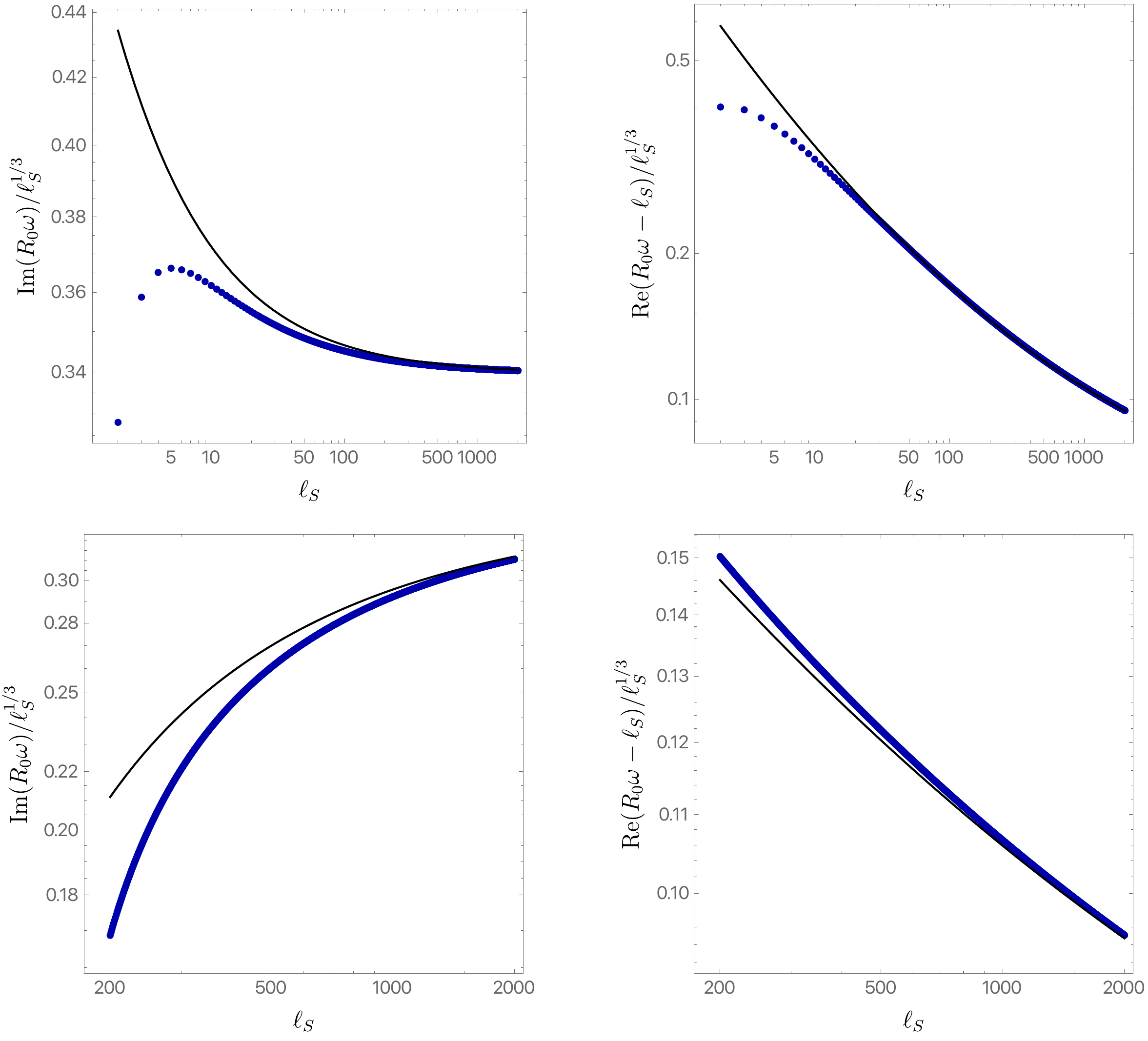}
    \caption{The real part (left panel) and imaginary part (right panel) of $R_0 \omega$ as a function of $\ell_S$ for fixed $p=0$ (top row) and $p=2$ (bottom row). In both panels, the solid black line corresponds to the prediction (\ref{eq:exp}), while the blue disks depict the exact numerical data.}
    \label{fig:com}
\end{figure}

Next, we investigate the spectrum, that is to say $R_0\omega$, for a fixed $\ell_S=2$ while varying $p$. The results are shown in the top row of Fig.~\ref{fig:ell2dyn}, where the real (imaginary) part of $\omega R_0$ is plotted on the left (right) panel. When the mode becomes complex, we represent it with a dashed green curve; for purely real $\omega$, a solid blue line is used. The structure of the spectrum in the complex plane is intricate. 
For sufficiently small $p$, here $p < -2/3$, there 
coexist stable modes with purely real frequency and unstable modes with purely imaginary frequency. For larger $p$ there are intervals of $p$ where the $\ell_S = 2$ modes appear to be stable, with purely real frequency. However these regions are interrupted by `bubbles' where the mode frequencies become complex, and in particular modes with ${\rm Im}(\omega R_0)>0$ exist that correspond to instabilities, as seen in the righthand panel of the figure. The mode tracked in Fig.~\ref{fig:com} connects to the middle `bubble' in Fig.~\ref{fig:ell2dyn} with $p=0$. In fact, for this mode, we recover the same value as in \cite{Anninos:2023epi}, which studied the case of Anderson boundary conditions, so $p=0$. The unstable modes observed in the righthand panel of the figure appear to occur also at larger values of $p$ than those shown, with the existence of further complex `bubbles' at ever greater values of $p$. As $p$ increases, these 'bubbles' become progressively smaller. For instance, there is a complex 'bubble' in the interval $p\in(184.12619,184.15074)$ with ${\rm Re}(\omega R_0)\in(47.08062,47.08219)$ for $\ell_S=2$. We have also explored other values of $\ell_S$ and observed qualitatively similar behavior, as shown in the bottom row of Fig.~\ref{fig:ell2dyn}, where we present the spectrum for $\ell_S=4$.
\begin{figure}
    \centering
    \includegraphics[width=0.9\linewidth]{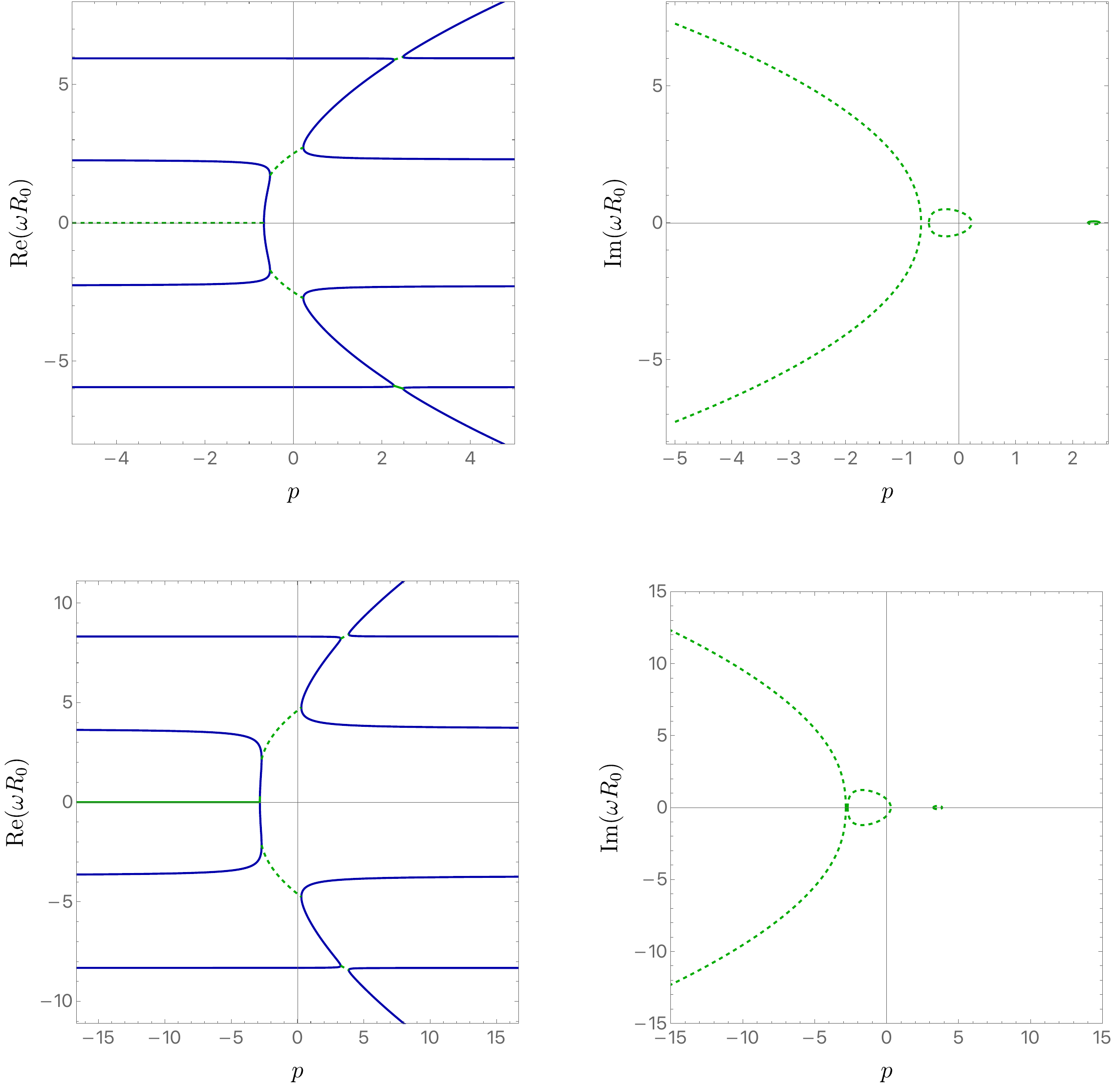}
    \caption{The real (left panel) and imaginary (right panel) parts of $R_0 \omega$ as a function of $p$ for fixed $\ell_S=2$ (top row) and $\ell_S=4$ (bottom row). If the mode turns complex, we represent it with a dashed green curve; when $\omega$ is purely real, we depict it with a solid blue line.}
    \label{fig:ell2dyn}
\end{figure}

\section{Conclusion and Discussion}
\label{sec:discussion}
We have introduced a one-parameter set of elliptic boundary conditions within the framework of Einstein metric infillings in Riemannian geometry. The conventional Dirichlet and Anderson boundary conditions are derived as 
limits in this context. Additionally, we have demonstrated that these boundary conditions lead to a well-defined infilling problem, except in the strict Dirichlet limit. For these boundary conditions, one fixes the conformal structure of the boundary metric $[\gamma_{\mu\nu}]$ and ${\gamma}^p K$, with $p\in\mathbb{R}$ and $K$ being the extrinsic curvature of the boundary metric in the infilling space $(\mathcal{M},g)$. We have referred to these as `generalized conformal boundary conditions'.
In the case that we require the function $K$ to be non-vanishing we may elegantly phrase these boundary conditions in terms of fixing the rescaled boundary metric 
 $\Gamma_{\mu\nu}=K^{\frac{1}{(d-1)p}}\gamma_{\mu\nu}$ on $\partial\mathcal{M}$. 

These boundary conditions naturally emerge when considering the Gibbons-Hawking-York boundary term to have an \emph{arbitrary real coefficient} $\Theta$ which then determines $p$ via the relation~\eqref{eq:finalp}. To the best of our knowledge, this is the first exploration of these boundary conditions.
Perhaps amusingly, these include the case $\Theta = 0$ for which no Gibbons-Hawking-York term is needed, corresponding to the particular case of $p=\frac{1}{2(d-1)}$. 
We have observed that 
rather than fixing ${\gamma}^p K$ and $[\gamma_{\mu\nu}]$, instead one may fix ${\gamma}^p K$ and the traceless part of the Brown-York stress tensor which also results in a well-defined variational problem. We have not explored this second option here, but believe it is interesting to do so.

Given the well-posed nature of these new boundary conditions, it is logical to define an ensemble in reference to them, akin to the traditional concepts introduced by York. We then think of the Euclidean path integral as being a functional of the rescaled boundary metric $\Gamma_{\mu\nu}$.
We have derived a first law of thermodynamics applicable to a broad range of geometries, including those with bolts, such as the Euclidean Schwarzschild black hole. We derived this first law assuming 1) that a hypersurface-orthogonal Killing vector field exists, and extends into the bulk; 2) that $\Gamma_{\mu\nu}$ is ultrastatic. Exploring the implications and consequences of relaxing these assumptions would be an intriguing avenue for further study. For instance, the Euclidean version of Kerr black holes fails to satisfy both. This first law involves the entropy and temperature (measured with respect to $\Gamma_{\mu\nu}$) associated to possible bolts, a work term related to deforming the boundary geometry and a free energy $\mathfrak{F}$ that characterises the ensemble. The free energy can be computed from the Euclidean path integral using the leading saddle point approximation for fixed $\Gamma_{\mu\nu}$.

From this point onward, we specialised to $d=4$ and it would be interesting to explore any dimensionally dependent phenomena. We focused on spherically symmetric infilling solutions with the boundary metric $\gamma_{\mu\nu}$ chosen as $S^1_{\beta}\times S^2_{R_0}$. Identifying up to three \emph{three} saddle points for a given $\Gamma_{\mu\nu}$, we scrutinized both their global and local thermodynamics. Our findings, while differing in some details from those presented in \cite{Anninos:2023epi} for the case of Anderson boundary conditions, so $p=0$, align on the aspects of global and local thermodynamic stability for that case.

We then investigated the Euclidean gravitational path integral beyond the leading saddle point approximation. In particular, we investigated the possible existence of fluctuation instabilities, so Euclidean negative modes of the Lichnerowicz operator, using the methods of \cite{Marolf:2022ntb,Liu:2023jvm} for the case of an empty cavity, \emph{i.e.,} no bolt, with boundary metric $S^1_{\beta}\times S^2_{R_0}$. Using a mixture of numerical and analytic work, we believe we have provided convincing evidence that for $p\geq 1/6$, there are no such Euclidean negative mode
instabilities. However, for $p<1/6$ we explicitly constructed fluctuations whose Lichnerowicz operator eigenvalues have negative real part, yielding  instabilities. Indeed, for certain ranges of $p$ we find Euclidean negative modes even for deformations that break spherical symmetry. The limit $| p | \to \infty$ corresponds to Dirichlet boundary conditions. Thus for Euclidean signature we may regard large positive $p$ as a regulator that deforms the Dirichlet boundary conditions to be well-posed. Then removing the regulator recovers the Dirichlet behaviour. While $p \to - \infty$ also corresponds to the Dirichlet limit, this is not a good regulator, as for large magnitude negative $p$, very negative modes exist that would dominate the behaviour of the system.

We have also explored the existence of negative mode instabilities around the Euclidean-Schwarzschild black hole and found an interesting puzzle that we have not fully solved. In particular, we find that for small black holes an additional negative mode exists beyond any present for the flat spacetime filled cavity, in agreement with the analysis based on local thermodynamic stability. However, we find no evidence for Euclidean negative modes for large  black holes when $p>1/6$, even though we find a negative specific heat for these. Our analysis imposed smoothness at the bolt, \emph{i.e.} at the black hole event horizon, and we suspect that if one relaxes this assumption, a negative mode will exist.

While in this paper we focused mainly on boundary metrics with $S^1\times S^2$ topology, it would be interesting to also investigate boundary metrics with $S^3$ topology. Furthermore, we have only studied the static and spherically symmetric limit in the presence of a cosmological constant (see appendix \ref{app:ads}), but it would be also interesting to perform a more systematic analysis for this more general class of geometries.
It would also be very interesting to understand how RG flow of our boundary term works, following previous work looking at the Gibbons-Hawking-York term~\cite{Jacobson:2013yqa,Neri:2023esr}.

Having studied in some detail the Euclidean problem, we turned our attention to the Lorentzian problem with our generalized conformal boundary conditions. While some boundary conditions have been shown to be well-posed and geometrically unique, much less is known about initial-boundary-value problems.  In particular, the class of boundary conditions for which theorems exist (see for instance \cite{Fournodavlos:2019ckr,Fournodavlos:2020wde,Fournodavlos:2021eye}) do not include Anderson boundary conditions or Dirichlet boundary conditions. In fact, it is strongly believed that Dirichlet boundary conditions will not be consistent with geometric uniqueness (see for instance \cite{An:2020nfw}). In this paper we did not address the issue of geometric uniqueness, and indeed we only investigated the linear stability problem for our new boundary conditions. 
Interestingly for the special case of Anderson boundary conditions the well-posedness of the \emph{linear} Lorentzian evolution was shown in~\cite{Anninos:2022ujl} and it seems likely that this will generalize to our more general class of boundary conditions.

We examined four-dimensional spacetimes with a spherically symmetric infilling and boundary geometry, without cosmological constant. Then Birkhoff's theorem constrains the interior geometry to be flat spacetime, or the Schwarzschild black hole. We then considered dynamics which preserves the spatial spherical symmetry. 
In the case of a cavity filled with Minkowski spacetime
we identified a linear instability for $p<1/6$.
For the black hole case, we have shown that large enough black holes are linearly unstable so long as $p>1/6$, but with the Dirichlet case limit $p \to \infty$ being stable (in accordance to the results reported in \cite{Andrade:2015gja}). Furthermore, the onset of the instability agrees precisely with a change in sign of the specific heat capacity, and hence with the local thermodynamic stability for the large black holes. Because of the spherical symmetry of the dynamics, we were able to determine the fate of the nonlinear evolution. Indeed, we found that, depending on the initial data, either the cavity expands indefinitely or contracts to zero size in the case of flat spacetime, or to the horizon for a black hole interior, in finite boundary time.

Motivated by the analysis of the spherically symmetric sector, we investigated the linear stability of perturbations that disrupt spherical symmetry. 
We took a static spherical cavity filled with flat Minkowski spacetime, and then considered dynamical perturbations to this that break the spherical symmetry.
Our analysis focused on scalar-derived perturbations
which are characterized by the spherical harmonic index $\ell_S$.

Our findings indicate that, for any value of $p$, there exists a critical value of $\ell_S$ above which an instability occurs. Notably, the instability persists unless $p\to+\infty$, aligning with the Dirichlet case as analyzed in \cite{Andrade:2015gja}. Additionally, for the particular case of Anderson boundary conditions, $p=0$, our results agree with those for the values of $\ell_S$ reported in \cite{Anninos:2023epi}. The fate of the linear instability under perturbations that disrupt spherical symmetry remains uncertain but since an instability exists for all $\ell_S$ larger than a critical value, and the instability time scale grows as $\ell_S^{1/3}$ asymptotically, we might expect a `turbulent' behaviour where energy cascades to shorter and shorter scales.

To the best of our knowledge, our results reveal the first example of a rather surprising physical system. Euclidean gravity with our cavity boundary conditions with $p > 1/6$ admits a flat spacetime infilling that appears to be a stable  saddle point. It apparently possesses no Euclidean negative modes, and for small temperatures gives the dominant saddle point contribution to the partition function. However, the corresponding Lorentzian cavity with the same cavity boundary conditions appears to be dynamically unstable, with infinitely many unstable perturbations associated to $\ell_S \to \infty$. Thus we have a system that appears to be thermodynamically stable and well behaved, and yet is dynamically unstable!

It is worth noting that numerous examples exist in the opposite direction, the classic example for gravity being the Schwarzschild black hole which is known to be dynamically stable \cite{Regge:1957td,Vishveshwara:1970cc,Zerilli:1970se,Moncrief:1974am,Bardeen:1973xb,Dafermos:2016uzj,Klainerman:2017nrb,Dafermos:2021cbw}, but  in the Euclidean context has been shown to contain 
negative modes \cite{Gross:1982cv}. However, we know of no examples, in gravity or otherwise, of a system that is thermodynamically stable and yet dynamically unstable.

\acknowledgments
We would like to express gratitude to Don~Marolf for many insightful discussions of this work.  Additionally, J.~E.~S. extends thanks to Aron~Wall for helpful discussions. X.~L. thanks Gary~Horowitz and Zhencheng~Wang for illuminating discussions. J.~E.~S. is partially supported by STFC consolidated grants ST/T000694/1 and ST/X000664/1. X.~L. is supported by NSF grant PHY-2107939, and by funds from the University of California. T.~W. is supported by the STFC consolidated grant ST/T000791/1.

\newpage

\appendix

\section{\label{app:EucPT}Euclidean Fluctuations of the Empty Spherical Cavity}

In this Appendix we give details of the analysis of the Euclidean fluctuations of an empty spherical cavity filled with flat space. We have given a general argument in the main text that for $p > 1/6$ we see no Euclidean instabilities based on the form of the condition in equation~\eqref{eq:pexpression} that governs the modes. In this Appendix we will derive this condition in detail, taking into account various special cases that require separate treatment. In addition we shall explicitly give examples of the behaviour of modes and their instabilities in the case $p < 1/6$.

\subsection{Modes with $n=\ell_S=0$}\label{sec:n0ell0}
These are the simplest modes, and yet, the ones that will impose the most stringent constraints on $p$. A static, \emph{i.e.} $n=0$, and spherically symmetric mode can be written as 
\begin{equation}
\delta {\rm d}s^2 = a(r)\,{\rm d}\tilde{\tau}^2+2\chi(r)\mathrm{d}\tilde{\tau}\mathrm{d}r+b(r){\rm d}r^2+c(r)\,r^2\,{\rm d}\Omega_2^2\,,
\end{equation}
where $a$, $b$ and $c$ parametrize the mode in question and are functions of $r$ only. We introduce an auxiliary variable
\begin{equation}
c(r)=\frac{{\rm f}(r)-a(r)-b(r)}{2}\,,
\end{equation}
so that $h={\rm f}(r)$.
Imposing the de Donder gauge and regularity at the origin further restricts $\chi$ to vanish, and $a$ to take the simple form,
\begin{equation}
a(r)=-3 b(r)+{\rm f}(r)-r\,b^\prime(r)+\frac{1}{2} r {\rm f}^\prime(r)\,.
\end{equation}
The resulting equations for $b$ and ${\rm f}$ can be determined analytically
\begin{subequations}
\begin{align}
&{\rm f}(r)=C_1 \frac{\sin(\sqrt{\lambda}r)}{r\sqrt{\lambda}}+C_2 \frac{\cos(\sqrt{\lambda}r)}{r\sqrt{\lambda}}\,,
\\
&b(r)=\frac{1}{2}{\rm f}(r)+\frac{C_3}{r^2\lambda}\left[\cos(\sqrt{\lambda}r)-\frac{\sin(\sqrt{\lambda}r)}{r\sqrt{\lambda}}\right]+\frac{C_4}{r^2\lambda}\left[\frac{\cos(\sqrt{\lambda}r)}{r\sqrt{\lambda}}+\sin(\sqrt{\lambda}r)\right]\,.
\end{align}
\label{eqs:sphericalmode}
\end{subequations}%
\\Regularity at the origin demands $C_2=C_4=0$, so that we are left with imposing the boundary conditions at the cavity wall. Linearising our boundary conditions around the spherical cavity at radius $r=r_E$ in Euclidean space yields
\begin{equation}
a(r_E)=c(r_E)\quad\text{and}\quad \frac{1}{2} a^\prime(r_{E})+c^\prime(r_{E})-\frac{b(r_{E})}{r_{E}}+\frac{6 p}{r_{E}}c(r_{E})=0\,.
\end{equation}
Substituting the explicit expressions for $a$, $b$ and $c$ in terms of the mode (\ref{eqs:sphericalmode}) yields
\begin{equation}
\left[
\begin{array}{ccc}
 -\frac{1}{\sqrt{\tilde{\lambda}}}\sin (\sqrt{\tilde{\lambda}}) && \frac{1}{2 \tilde{\lambda}^{3/2}}\left[3 \tilde{\lambda} \sin (\sqrt{\tilde{\lambda}})-\sin (\sqrt{\tilde{\lambda
   }})+\sqrt{\tilde{\lambda}} \cos (\sqrt{\tilde{\lambda}})\right]
   \\
   \\
 \frac{1}{4} \cos (\sqrt{\tilde{\lambda}})+\frac{3 (4 p-1)}{4\sqrt{\tilde{\lambda}}} \sin (\sqrt{\tilde{\lambda}}) && \frac{1-6 p}{2 \tilde{\lambda}^{3/2}} \left[\tilde{\lambda} \sin
   (\sqrt{\tilde{\lambda}})-\sin (\sqrt{\tilde{\lambda}})+\sqrt{\tilde{\lambda}} \cos (\sqrt{\tilde{\lambda}})\right] \\
\end{array}
\right]\left[\begin{array}{c}
C_1
\\
\\
\\
C_3
\end{array}\right]=0\,,
\end{equation}
where we defined $\tilde{\lambda}\equiv r_E^2\lambda$. Any nontrivial solution can only exist if the determinant of the above matrix vanishes. This imposes a condition on the possible values of $\tilde{\lambda}$, namely the following must be true 
\begin{multline}
\frac{1}{16\tilde{\lambda}^2}\Bigg\{(1+6 \tilde{\lambda}) \cos (2 \sqrt{\tilde{\lambda}})+3 \tilde{\lambda}^{3/2} \sin (2 \sqrt{\tilde{\lambda}})-1-4 \tilde{\lambda}
\\
+12 p \left[1+\tilde{\lambda
   }-(1+\tilde{\lambda}) \cos (2 \sqrt{\tilde{\lambda}})-\sqrt{\tilde{\lambda}} \sin (2 \sqrt{\tilde{\lambda}})\right]\Bigg\}=0\,.
   \label{eq:flatn0ell0}
\end{multline}
We emphasize that this is the same expression as follows from our one in the main text in section~\ref{sec:Fieldtheoretic}, equation~\eqref{eq:pexpression}, in the special case $n = \ell_S = 0$. As discussed in that section there will be no instabilities for $p > 1/6$. In Fig.~\ref{fig:n0ell0} we plot the real part of $\tilde{\lambda}$ (left and middle panels) as well as its imaginary part (right panel) as a function of $p$. The middle plot shows a zoom in the red region marked on the left panel. It is clear that a single negative mode exists if $p<1/6$. This critical value of $p$ can also be inferred analytically from Eq.(\ref{eq:flatn0ell0}) evaluated at $\tilde{\lambda}=0$. 
\begin{figure}
    \centering
    \includegraphics[width=\linewidth]{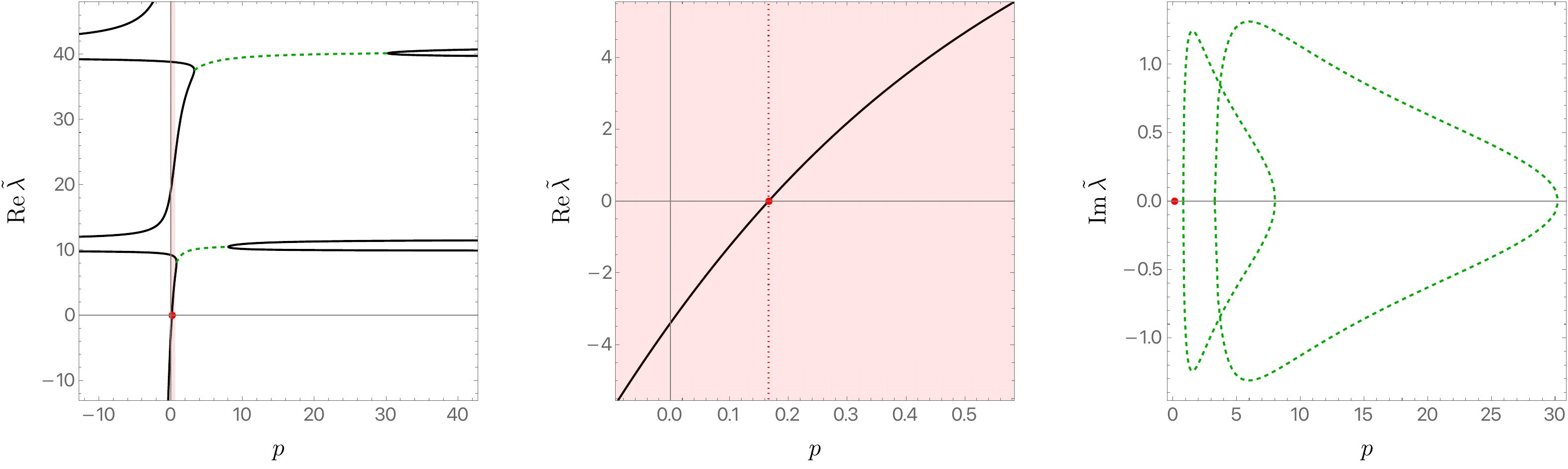}
    \caption{The real part (left and middle panels) and imaginary part (right panel) of $\tilde{\lambda}$ as a function of $p$. The point $(p,\tilde{\lambda})=(1/6,0)$ is marked as a red disk. For $p<1/6$, there exists a single mode with ${\rm Re}\,\tilde{\lambda}<0$, indicating the presence of a field-theoretic negative mode. The green dashed curves indicate that the mode is complex.}
    \label{fig:n0ell0}
\end{figure}

\subsection{Modes with $n\neq0$ and $\ell_S=0$}
The modes in this section are allowed to depend on the thermal circle, {so that $n \ne 0$, but that remain spherically symmetric with $\ell_S=0$. We write these as
\begin{equation}
\delta {\rm d}s^2 = \left[a(r)\,{\rm d}\tilde{\tau}^2+2 \chi(r)\mathrm{d}\tilde{\tau}\mathrm{d}r+b(r){\rm d}r^2+c(r)r^2{\rm d}\Omega_2^2\right]e^{i\tilde{n}\tilde{\tau}}\,.
\end{equation}
Once again, we can solve for all metric functions 
\begin{subequations}
\begin{align}
& c(r)=\frac{{\rm f}(r)-a(r)-b(r)}{2}\,,
\\
&\chi(r) = \frac{i\,\tilde{n}}{\lambda -\tilde{n} ^2}a^\prime(r)-\frac{i\,\tilde{n}}{2 (\lambda -\tilde{n} ^2)}{\rm f}^\prime(r)\,,
\\
&b(r)=\frac{\tilde{n}^2}{\lambda -\tilde{n} ^2}a(r)+\frac{\lambda -2 \tilde{n} ^2}{2 (\lambda -\tilde{n} ^2)}{\rm f}(r)+\frac{\lambda +2 \tilde{n} ^2}{r \left(\lambda -\tilde{n}
   ^2\right)^2} a^\prime(r)+\frac{\lambda -4 \tilde{n} ^2}{2 r \left(\lambda -\tilde{n} ^2\right)^2} {\rm f}^\prime(r)\,,
\\
& a(r)=C_1 \frac{\sin\left(\sqrt{\lambda-\tilde{n}^2}\,r\right)}{\sqrt{\lambda-\tilde{n}^2}\,r}+C_2 \frac{\cos\left(\sqrt{\lambda-\tilde{n}^2}\,r\right)}{\sqrt{\lambda-\tilde{n}^2}\,r}
\\
& {\rm f}(r)=C_3 \frac{\sin\left(\sqrt{\lambda-\tilde{n}^2}\,r\right)}{\sqrt{\lambda-\tilde{n}^2}\,r}+C_4 \frac{\cos\left(\sqrt{\lambda-\tilde{n}^2}\,r\right)}{\sqrt{\lambda-\tilde{n}^2}\,r}\,.
\end{align}
\end{subequations}%
Regularity at the origin demands $C_2=C_4=0$, while our boundary conditions now impose
\begin{equation}
a(r_E)=c(r_E)\quad\text{and}\quad \frac{1}{2} a^\prime(r_{E})+c^\prime(r_{E})-\frac{b(r_{E})}{r_{E}}+\frac{6 p}{r_{E}}c(r_{E})-i \tilde{n} \chi(r_E)=0\,.
\end{equation}
These can be simultaneously solved so long as
\begin{subequations}
\begin{equation}
p=\frac{\tilde{\Lambda}+6 \tilde{\Lambda}^2-\csc ^2\left(\sqrt{\tilde{\Lambda}}\right) \tilde{\Lambda} \left(\varpi^2+\tilde{\Lambda}\right)+\varpi^2 \left(2 \tilde{\Lambda}-7\right)+\cot
   \left(\sqrt{\tilde{\Lambda}}\right) \sqrt{\tilde{\Lambda}} \left[\varpi^2 \left(8+\tilde{\Lambda}\right)-3 \tilde{\Lambda}^2\right]}{12 \left[\varpi^2 \left(\tilde{\Lambda}-3\right)+\tilde{\Lambda}+\tilde{\Lambda}^2-\cot \left(\sqrt{\tilde{\Lambda}}\right) \sqrt{\tilde{\Lambda}} \left(\tilde{\Lambda}-3 \varpi^2\right)\right]}
\end{equation}
where
\begin{equation}
\varpi=\tilde{n}\,r_E\quad\text{and}\quad \tilde{\Lambda}=(\lambda-\tilde{n}^2)r_E^2\,.
\end{equation}
\end{subequations}
Again this is consistent with the expression~\eqref{eq:pexpression} in section~\ref{sec:Fieldtheoretic} and hence will give rise to no unstable modes for $p > 1/6$. We now look at the explicit behaviour of modes, and in particular find negative modes for sufficiently small $p$. We start by noting that $\lambda=0$ now requires $p=(2-\varpi^2)/12<1/6$,  so a zero mode (marking a transition to a negative mode) can only exist in a regime where $p$ is smaller than the case for which $\varpi=0$. Of course, it could still happen that ${\rm Re}\,\tilde{\lambda}={\rm Re}(r_E^2 \lambda)$ becomes negative somewhere on the complex plane. However, we have explicitly checked that this is not the case for $\varpi\in\mathbb{R}$. See, for instance, Fig.~\ref{fig:nneq0ell0}, where we plot the real part of $\tilde{\lambda}$ (left and middle panels) as well as its imaginary part (right panel) as a function of $p$ for $\varpi=1$. The middle panel provides a zoom near the transition point, and shows a negative mode.
\begin{figure}
    \centering
    \includegraphics[width=\linewidth]{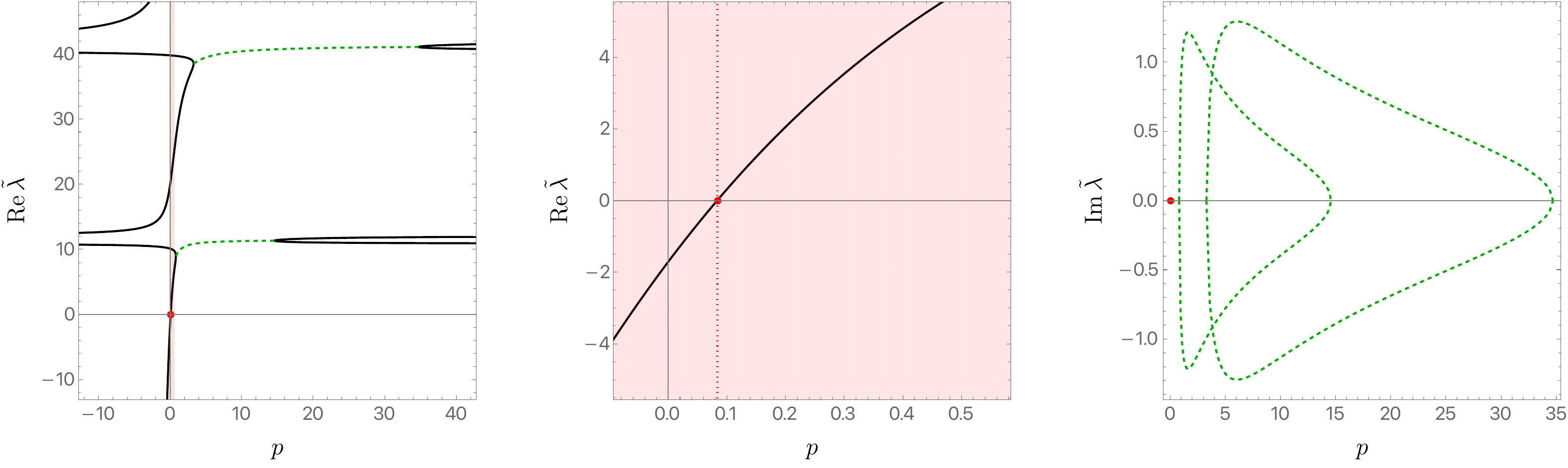}
    \caption{The real part (left and middle panels) and imaginary part (right panel) of $\tilde{\lambda}$ as a function of $p$. The point $(p,\tilde{\lambda})=\left(\frac{2-\varpi^2}{12},0\right)$ is marked as a red disk. For $p<\frac{2-\varpi^2}{12}$, there exists a single mode with ${\rm Re}\,\tilde{\lambda}<0$, indicating the presence of a Euclidean negative mode. This plot was generated with $\varpi=1$ and uses the same colour coding as Fig.~\ref{fig:n0ell0}.}
    \label{fig:nneq0ell0}
\end{figure}

\subsection{Non-spherical scalar-derived gravitational Euclidean modes with $\ell_S \geq 2$}\label{sec:nonsphericalS}

We now consider non-spherical Euclidean fluctuations, so $\ell_S \ne 0$. Our treatment requires $\ell_S \ge 2$ and we later consider the special case $\ell_S = 1$ separately.

\subsubsection{Static modes}\label{sec:staticModesS}

Modes with $\tilde{n}=0$ have to be dealt with separately. Due to the staticity of the modes in question, the components $f_{\tilde{\tau}r}$ and $f_{\tilde{\tau}}$ decouple from the remaining ones, and we shall return to these at the end of this section. There are then a total of five functions to solve for: $\{f_{\tilde{\tau}\tilde{\tau}}(r), f_{rr}(r), f_{r}(r), h_L(r), h_T(r)\}$, which we turn our attention to next.

To simplify our expressions, we define
\begin{equation}
h_L=\frac{r^2}{2}({\rm f}-f_{\tilde{\tau}\tilde{\tau}}-f_{rr})\,,
\end{equation}
so that the trace of the mode is simply $h={\rm f}\,Y^{\ell_S\,m_S}$. Using the de Donder gauge condition, as well as the Lichnerowicz eigenvalue equation (\ref{eq:lichneagain}) we find
\begin{subequations}
\begin{align}
&h_T(r)=-\frac{r^2}{\ell_S \left(\ell_S+1\right)-2}\left[f_{\tilde{\tau}\tilde{\tau}}(r)+f_{rr}(r)-\frac{4\,f_r(r)}{r}-2 f^\prime_r(r)\right]\,,
\label{eq:1}
\\
&f_{rr}(r)=q_1(r)+\frac{{\rm f}(r)}{2}+\frac{f_{\tilde{\tau}\tilde{\tau}}^\prime(r)}{r\,\lambda}+\frac{{\rm f}^\prime(r)}{2\,r\lambda}\,,
\label{eq:2}
\\
&f_r(r)=q_2(r)+\frac{f_{\tilde{\tau}\tilde{\tau}}(r)}{r\,\lambda}+\frac{{\rm f}(r)}{2\,r\,\lambda}\,,
\label{eq:3}
\\
& q_2(r)=\frac{3r}{\ell_S(\ell_S+1)}q_1(r)+\frac{r^2}{\ell_S(\ell_S+1)}q_1^\prime(r)\,,
\label{eq:4}
\end{align}
with
\begin{align}
&\frac{\left(r^2{\rm f}^\prime\right)^\prime}{r^2}+\left[\lambda-\frac{\ell_S(\ell_S+1)}{r^2}\right]{\rm f}=0\,,
\label{eq:5}
\\
&\frac{\left(r^2 f_{\tilde{\tau}\tilde{\tau}}^\prime\right)^\prime}{r^2}+\left[\lambda-\frac{\ell_S(\ell_S+1)}{r^2}\right]f_{\tilde{\tau}\tilde{\tau}}=0\,,
\label{eq:6}
\\
&\frac{\left(r^6 q_1^\prime\right)^\prime}{r^6}+\left[\lambda-\frac{\ell_S(\ell_S+1)-6}{r^2}\right]q_1=0\,.
\label{eq:7}
\end{align}
\end{subequations}
Note that once the solutions for ${\rm f}$, $f_{\tilde{\tau}\tilde{\tau}}$ and $q_1$ are known via Eqs.~(\ref{eq:5})-(\ref{eq:7}), all the remaining functions can be found via Eqs.~(\ref{eq:1})-(\ref{eq:4}). Indeed, one can readily solve in full generality for ${\rm f}$, $f_{\tilde{\tau}\tilde{\tau}}$ and $q_1$
\begin{subequations}
\begin{align}
&{\rm f}(r)=\frac{C_1}{(r\sqrt{\lambda})^{1/2}}{\rm J}_{\ell_S+\frac{1}{2}}\left(r\sqrt{\lambda}\right)+\frac{C_2}{(r\sqrt{\lambda})^{1/2}}{\rm Y}_{\ell_S+\frac{1}{2}}\left(r\sqrt{\lambda}\right)\,,
\\
&f_{\tilde{\tau}\tilde{\tau}}(r)=\frac{C_3}{(r\sqrt{\lambda})^{1/2}}{\rm J}_{\ell_S+\frac{1}{2}}\left(r\sqrt{\lambda}\right)+\frac{C_4}{(r\sqrt{\lambda})^{1/2}}{\rm Y}_{\ell_S+\frac{1}{2}}\left(r\sqrt{\lambda}\right)\,,
\\
&q_1(r)=\frac{C_5}{(r\sqrt{\lambda})^{5/2}}{\rm J}_{\ell_S+\frac{1}{2}}\left(r\sqrt{\lambda}\right)+\frac{C_6}{(r\sqrt{\lambda})^{5/2}}{\rm Y}_{\ell_S+\frac{1}{2}}\left(r\sqrt{\lambda}\right)\,,
\end{align}
\end{subequations}
where ${\rm J}_p$ and ${\rm Y}_p$ are Bessel functions  of the first and second kind of order $p$, respectively. Regularity at the origin requires $C_2=C_4=C_5=0$, and we are left with three unknown constants $\{C_1,C_2,C_3\}$ together with $\lambda$ to determine in what follows.

Imposing our boundary conditions at the cavity wall yields three conditions
\begin{multline}
h_L(r_E)=r_E^2\,f_{\tilde{\tau}\tilde{\tau}}(r_E)\,,\quad h_T(r_E)=0\,,
\\
\text{and}\quad f_{\tilde{\tau}\tilde{\tau}}^\prime(r_E)+\frac{2}{r_E^2} h_L^\prime(r_E)-\frac{4(1-3p)}{r_E}f_{\tilde{\tau}\tilde{\tau}}(r_E)-\frac{2}{r_E}f_{rr}(r_E)+\frac{2\ell_S(\ell_S+1)}{r_E^2}f_r(r_E)=0\,.
\end{multline}
All the above conditions can be simultaneously solved so long as
\begin{subequations}
\begin{multline}
\tilde{\lambda}\,{\rm J}^3_{\ell_S+\frac{3}{2}}\left(\sqrt{\tilde{\lambda}}\right)+\sqrt{\tilde{\lambda}}\,\Theta_1(p,\ell_S,\tilde{\lambda})\,{\rm J}^3_{\ell_S+\frac{1}{2}}\left(\sqrt{\tilde{\lambda}}\right)-\Theta_2(p,\ell_S,\tilde{\lambda})\,{\rm J}^2_{\ell_S+\frac{1}{2}}\left(\sqrt{\tilde{\lambda}}\right){\rm J}_{\ell_S+\frac{3}{2}}\left(\sqrt{\tilde{\lambda}}\right)
\\
-\sqrt{\tilde{\lambda}}\,\Theta_3(p,\ell_S,\tilde{\lambda})\,{\rm J}_{\ell_S+\frac{1}{2}}\left(\sqrt{\tilde{\lambda}}\right){\rm J}^2_{\ell_S+\frac{3}{2}}\left(\sqrt{\tilde{\lambda}}\right)=0\,,
\end{multline}
with $\tilde{\lambda}\equiv\lambda r_E^2$ and
\begin{align}
&\Theta_1(p,\ell_S,\tilde{\lambda})=(12 p-2+3 \ell_S) \tilde{\lambda}-(1+2 \ell_S) (12 p-2+\ell_S+2 \ell_S^2)\,,
\\
&\Theta_2(p,\ell_S,\tilde{\lambda})=(1+2 \ell_S) (12 p-2+\ell_S+2 \ell_S^2)-\left[24 p-1+4 \ell_S (2+\ell_S)\right] \tilde{\lambda}+3 \tilde{\lambda}^2\,,
\\
& \Theta_3(p,\ell_S,\tilde{\lambda})=3-12 p+\ell_S-2 \ell_S^2+2 \tilde{\lambda}\,.
\end{align}
\end{subequations}
Again this is consistent with equation~\eqref{eq:pexpression}, and hence will give no negative modes for $p > 1/6$.
We now give an example of the behaviour, determining allowed values of $\tilde{\lambda}$ for a given value of $p$. The results for $\ell_S=2$ are shown in Fig.~(\ref{fig:ell2}) and similar qualitative results hold for all other values of $\ell_S$ we have investigated. In the region $p>1/6$ we find no modes with ${\rm Re}\,\tilde{\lambda}<0$ consistent with our general argument in Section~\ref{sec:Fieldtheoretic}. Note that negative modes do exist for $p\lesssim-0.64422$, and indeed it is relatively simple to show that as $p$ becomes more and more negative, a single negative mode persists with
\begin{equation}
\tilde{\lambda}=-16 p^2+24 p+\frac{1}{3} \left[11 \ell_S \left(\ell_S+1\right)-3\right]+\frac{29 \ell_S \left(\ell_S+1\right)+40}{12 p}+\mathcal{O}(p^{-2})\,.
\end{equation}
\begin{figure}
    \centering
    \includegraphics[width=\linewidth]{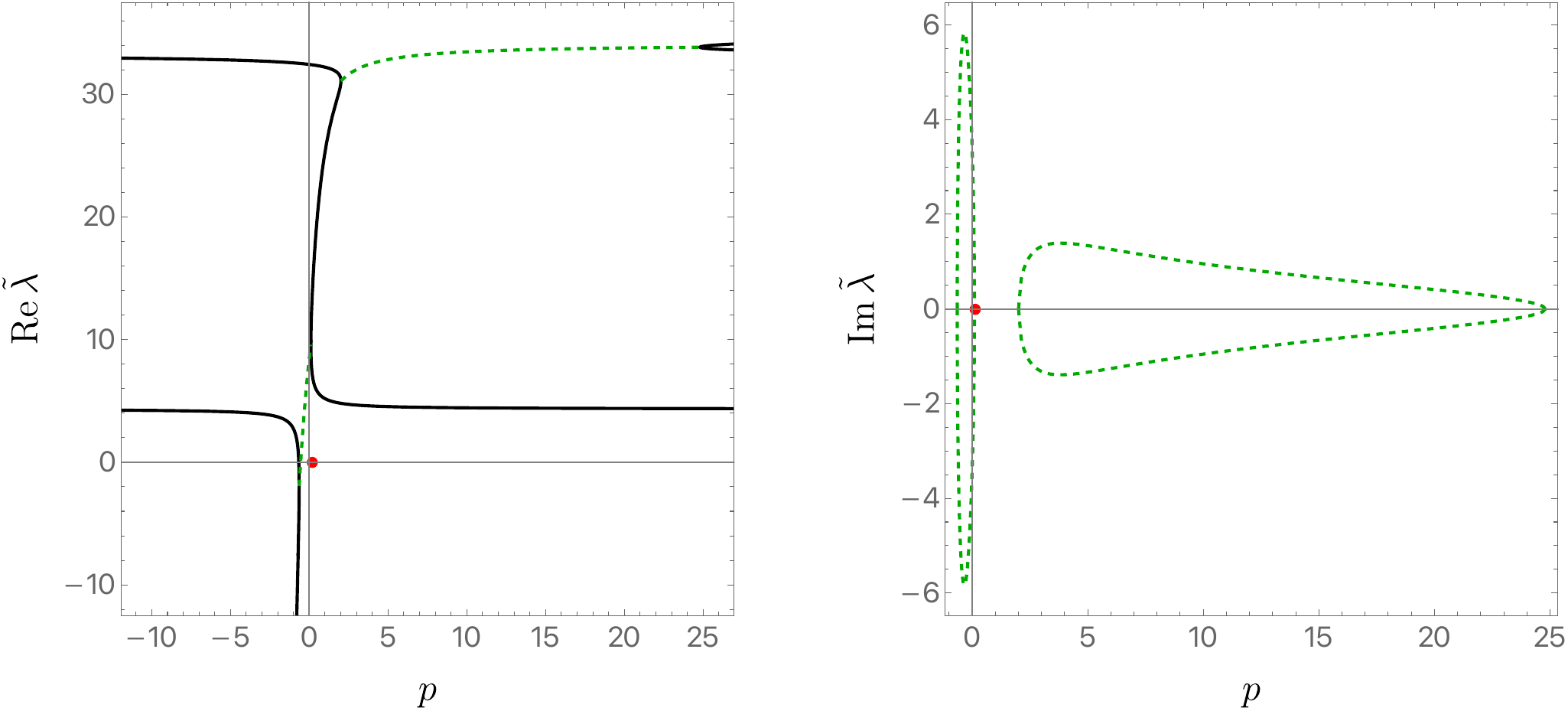}
    \caption{The real part (left panel) and imaginary part (right panel) of $\tilde{\lambda}$ as a function of $p$ for $\ell_S=2$. The point $(p,\tilde{\lambda})=\left(1/6,0\right)$ is marked as a red disk. This plot uses the same colour coding as Fig.~\ref{fig:n0ell0}.}
    \label{fig:ell2}
\end{figure}

We now return to the $f_{\tilde{\tau}r}$ and $f_{\tilde{\tau}}$ components. As anticipated earlier, these metric components decouple from the remaining ones. The de Donder gauge condition demands
\begin{subequations}
\begin{equation}
f_{\tilde{\tau}}(r)=\frac{r}{\ell_S(\ell_S+1)}\left[r\,f^\prime_{\tilde{\tau}r}(r)+2f_{\tilde{\tau}r}(r)\right]
\end{equation}
with
\begin{equation}
\frac{\left(r^4 f_{\tilde{\tau}r}^\prime\right)}{r^4}+\left[\lambda-\frac{\ell_S(\ell_S+1)-2}{r^2}\right]f_{\tilde{\tau}r}=0\,.
\end{equation}
\end{subequations}
The equation for $f_{\tilde{\tau}r}$ can be readily integrated to find
\begin{equation}
f_{\tilde{\tau}r}=\frac{1}{(r\sqrt{\lambda})^{3/2}}\left[C_1{\rm J}_{\ell_S+\frac{1}{2}}\left(r\sqrt{\lambda}\right)+C_2{\rm Y}_{\ell_S+\frac{1}{2}}\left(r\sqrt{\lambda}\right)\right]\,.
\end{equation}
Regularity at the origin once again demands $C_2=0$, while our boundary conditions at the cavity wall demand
\begin{equation}
f_{\tilde{\tau}}(r_E)=0\Rightarrow {\rm J}_{\ell _S+\frac{3}{2}}\left(\sqrt{\tilde{\lambda}}\right) \sqrt{\tilde{\lambda}}-\left(\ell _S+1\right) {\rm J}_{\ell _S+\frac{1}{2}}\left(\sqrt{\tilde{\lambda}}\right)=0\,.
\end{equation}
This condition is independent of $p$ and thus will yield the same results as for the Dirichlet case where stability was seen in~\cite{Marolf:2022ntb}. It is easy to show that the above equation only has real roots, and that for $\ell_S=2$ the first four modes are approximately given by
\begin{equation}
\tilde{\lambda}=\{14.97874(7), 55.39954(5),114.76860(4), 193.78091(7),\ldots\}\,,
\label{eq:afew0}
\end{equation}
thus showing that no negative modes exist in this sector as well.

\subsubsection{Generic modes}\label{sec:GenericModesS}
This is the most complicated sector, since both $\tilde{n}\neq0$ and $\ell_S\geq2$. Just as in previous sections, we begin by introducing an auxiliary quantity ${\rm f}$
\begin{equation}
h_L=\frac{r^2}{2}({\rm f}-f_{\tilde{\tau}\tilde{\tau}}-f_{rr})\,.
\end{equation}
The de Donder gauge condition, together with the Lichnerowicz eigenvalue equation (\ref{eq:lichneagain}), now imply the following relations
\begin{subequations}
\begin{multline}
f_{\tilde{\tau}}(r)=i\frac{\tilde{n}}{\lambda -\tilde{n}^2}f_{\tilde{\tau}\tilde{\tau}}(r)-\frac{i}{2}\frac{\tilde{n}}{\lambda -\tilde{n}^2} {\rm f}(r)+i\frac{\ell_S \left(\ell_S+1\right)+9-r^2 \left(\lambda
   -\tilde{n}^2\right) }{\ell_S \left(\ell_S+1\right) \tilde{n}}q_1(r)
   \\
   -\frac{4 }{r \tilde{n}}\,i\,q_2(r)+\frac{3 r}{\ell_S \left(\ell_S+1\right) \tilde{n}}\,i\,q_1^\prime(r)-\frac{i}{\tilde{n}} q_2^\prime(r)\,,
   \label{eq:sin1}
\end{multline}
\begin{equation}
f_{\tilde{\tau}r}=i\frac{3}{r \tilde{n}} q_1(r)-i\frac{\ell_S \left(\ell_S+1\right) }{r^2 \tilde{n}}q_2(r)+i\frac{\tilde{n}}{\lambda -\tilde{n}^2} f_{\tilde{\tau}\tilde{\tau}}^\prime(r)-\frac{i}{2}\frac{\tilde{n}
  }{\lambda -\tilde{n}^2} {\rm f}^\prime(r)+\frac{i}{\tilde{n}} q_1^\prime(r)
  \label{eq:sin2}
\end{equation}
\begin{multline}
h_T(r)=\frac{1}{\ell_S \left(\ell_S+1\right)-2}\Bigg\{\frac{\left(\lambda +2 \tilde{n}^2\right) \left[2-r^2 \left(\lambda -\tilde{n}^2\right)\right]}{\left(\lambda -\tilde{n}^2\right)^2}f_{\tilde{\tau}\tilde{\tau}}(r)+\frac{\left(\lambda -4 \tilde{n}^2\right) \left[2-r^2
   \left(\lambda -\tilde{n}^2\right)\right]}{2 \left(\lambda -\tilde{n}^2\right)^2} {\rm f}(r)
\\
   -\frac{r^2 \left[3 \ell_S \left(\ell_S+1\right)+18-2 r^2 \left(\lambda -\tilde{n}^2\right)\right]
   }{\ell_S \left(\ell_S+1\right)}q_1(r)+12\,r\,q_2(r)+\frac{\lambda +2 \tilde{n}^2}{\left(\lambda -\tilde{n}^2\right)^2} \,r\,f_{\tilde{\tau}\tilde{\tau}}^\prime(r)
   \\
   +\frac{\lambda -4 \tilde{n}^2}{2
   \left(\lambda -\tilde{n}^2\right)^2}\,r\,{\rm f}^\prime(r)-\frac{6 r^3 }{\ell_S \left(\ell_S+1\right)}q_1^\prime(r)+4 r^2 q_2^\prime(r)\Bigg\}\,,
   \label{eq:sin3}
\end{multline}
\begin{equation}
f_{rr}(r)=q_1(r)+\frac{1}{2}\frac{\lambda -2 \tilde{n}^2}{\lambda -\tilde{n}^2}{\rm f}(r)+\frac{\tilde{n}^2}{\lambda -\tilde{n}^2}f_{\tilde{\tau}\tilde{\tau}}(r)+\frac{\lambda +2 \tilde{n}^2}{\left(\lambda -\tilde{n}^2\right)^2 r}f^\prime_{\tilde{\tau}\tilde{\tau}}(r)+\frac{\lambda-4 \tilde{n}^2}{2 \left(\lambda-\tilde{n} ^2\right)^2 r}{\rm f}^\prime(r)\,,
\label{eq:sin4}
\end{equation}
\begin{equation}
f_r(r)=q_2(r)+\frac{\lambda +2 \tilde{n}^2}{\left(\lambda -\tilde{n}^2\right)^2 r}f_{\tilde{\tau}\tilde{\tau}}(r)+\frac{\lambda -4 \tilde{n}^2}{2 \left(\lambda -\tilde{n}^2\right)^2 r}{\rm f}(r)\,
\label{eq:sin5}
\end{equation}
where
\begin{equation}
\frac{\left(r^2{\rm f}^\prime\right)^\prime}{r^2}+\left[\lambda-\frac{\ell_S(\ell_S+1)}{r^2}-\tilde{n}^2\right]{\rm f}=0\,,
\label{eq:sin6}
\end{equation}
\begin{equation}
\frac{\left(r^2 f_{\tilde{\tau}\tilde{\tau}}^\prime\right)^\prime}{r^2}+\left[\lambda-\frac{\ell_S(\ell_S+1)}{r^2}-\tilde{n}^2\right]f_{\tilde{\tau}\tilde{\tau}}=0\,,
\label{eq:sin7}
\end{equation}
\begin{equation}
\frac{\left(r^2 q_1^\prime\right)^\prime}{r^2}+\left[\lambda-\frac{\ell_S(\ell_S+1)+6}{r^2}-\tilde{n}^2\right]q_1+\frac{4\ell_S(\ell_S+1)}{r^3}q_2(r)=0\,,
\label{eq:sin8}
\end{equation}
\begin{equation}
\frac{\left(r^4 q_2^\prime\right)^\prime}{r^4}+\left[\lambda-\frac{\ell_S(\ell_S+1)-8}{r^2}-\tilde{n}^2\right]q_2-\frac{6}{\ell_S(\ell_S+1)}\frac{(r^3q_1)^\prime}{r^3}+\frac{2\,r}{\ell_S(\ell_S+1)}(\lambda-\tilde{n}^2)q_1=0\,.
\label{eq:sin9}
\end{equation}
\end{subequations}
Note that once the solutions for ${\rm f}$, $f_{\tilde{\tau}\tilde{\tau}}$, $q_1$ and $q_2$ are known via Eqs.~(\ref{eq:sin6})-(\ref{eq:sin9}), all the remainder functions can be found via Eqs.~(\ref{eq:sin1})-(\ref{eq:sin5}). Indeed, all the above four equations can be integrated in full generality
\begin{subequations}
\begin{equation}
{\rm f}(r)=\frac{C_1}{(r\sqrt{\lambda-\tilde{n}^2})^{1/2}}{\rm J}_{\ell_S+\frac{1}{2}}\left(r\sqrt{\lambda-\tilde{n}^2}\right)+\frac{C_2}{(r\sqrt{\lambda-\tilde{n}^2})^{1/2}}{\rm Y}_{\ell_S+\frac{1}{2}}\left(r\sqrt{\lambda-\tilde{n}^2}\right)
\end{equation}
\begin{equation}
f_{\tilde{\tau}\tilde{\tau}}(r)=\frac{C_3}{(r\sqrt{\lambda-\tilde{n}^2})^{1/2}}{\rm J}_{\ell_S+\frac{1}{2}}\left(r\sqrt{\lambda-\tilde{n}^2}\right)+\frac{C_4}{(r\sqrt{\lambda-\tilde{n}^2})^{1/2}}{\rm Y}_{\ell_S+\frac{1}{2}}\left(r\sqrt{\lambda-\tilde{n}^2}\right)
\end{equation}
\begin{multline}
q_1(r)=\frac{C_5}{(r\sqrt{\lambda-\tilde{n}^2})^{3/2}}{\rm J}_{\ell_S+\frac{3}{2}}\left(r\sqrt{\lambda-\tilde{n}^2}\right)+\frac{C_6}{(r\sqrt{\lambda-\tilde{n}^2})^{3/2}}{\rm Y}_{\ell_S+\frac{3}{2}}\left(r\sqrt{\lambda-\tilde{n}^2}\right)
\\
+\frac{C_7}{(r\sqrt{\lambda-\tilde{n}^2})^{3/2}}{\rm J}_{\ell_S-\frac{1}{2}}\left(r\sqrt{\lambda-\tilde{n}^2}\right)+\frac{C_8}{(r\sqrt{\lambda-\tilde{n}^2})^{3/2}}{\rm Y}_{\ell_S-\frac{1}{2}}\left(r\sqrt{\lambda-\tilde{n}^2}\right)\,,
\end{multline}
and
\begin{equation}
q_2(r)=\frac{1}{2 \ell_S \left(\ell_S+1\right)}\Bigg\{\frac{1}{2} r \left[\ell_S \left(\ell_S+1\right)+6-r^2 \left(\lambda-\tilde{n} ^2\right)\right] q_1(r)-r^2 q_1^\prime(r)-\frac{1}{2} r^3 q_1^{\prime\prime}(r)\Bigg\}\,.
\end{equation}
\end{subequations}
Imposing regularity at the origin in the above expressions demands $C_2=C_4=C_6=C_8=0$. At this stage we impose our boundary conditions, which yield
\begin{multline}
h_L(r_E)=r_E^2f_{\tilde{\tau}\tilde{\tau}}(r_E)\,,\quad f_{\tilde{\tau}}(r_E)=0\,,\quad h_T(r_E)=0\,,
\\
\quad\text{and}\quad f_{\tilde{\tau}\tilde{\tau}}^\prime(r_E)+\frac{2}{r_E^2} h_L^\prime(r_E)-\frac{4(1-3p)}{r_E}f_{\tilde{\tau}\tilde{\tau}}(r_E)-\frac{2}{r_E}f_{rr}(r_E)+\frac{2\ell_S(\ell_S+1)}{r_E^2}f_r(r_E)-2\,i\,\tilde{n}f_{\tilde{\tau}r}(r_E)=0\,.
\end{multline}
Imposing all of these conditions simultaneously constrains the possible values of $\lambda$ to obey the following complicated relation summarized in equation~\eqref{eq:pexpression} in the main text in Section~\ref{sec:Fieldtheoretic},
\begin{subequations}\label{eq:generalLambda}
\begin{equation}
\sum_{i=0}^4\left[\eta_i(\varpi,\tilde{\Lambda},\ell_S)+\iota_i(\varpi,\tilde{\Lambda},\ell_S) p\right]\cdot\left[{\rm J}_{\ell_S+\frac{1}{2}}\left(\sqrt{\tilde{\Lambda}}\right)\right]^{4-i}\left[{\rm J}_{\ell_S+\frac{3}{2}}\left(\sqrt{\tilde{\Lambda}}\right)\right]^{i}=0
\end{equation}
where $\tilde{\Lambda}\equiv (\lambda-\tilde{n}^2)r_E^2$ and $\varpi\equiv\tilde{n}\,r_E$ and
\begin{equation}
\eta_0(\varpi,\tilde{\Lambda},\ell_S)=\sqrt{\tilde{\Lambda}} (\ell_S+1) (\tilde{\Lambda}+\varpi^2) \left[2-2 \tilde{\Lambda}+3 (1+\tilde{\Lambda}) \ell_S-4\ell_S^3-4 \ell_S^2\right]\,,
\end{equation}
\begin{multline}
\eta_1(\varpi,\tilde{\Lambda},\ell_S)=3 \varpi^2 \tilde{\Lambda}^2+\varpi^2 \tilde{\Lambda}-\tilde{\Lambda}^3-3 \tilde{\Lambda}^2+2 \tilde{\Lambda}-8 \varpi^2 \tilde{\Lambda} \ell_S^3-16 \varpi^2 \tilde{\Lambda} \ell_S^2+2 \varpi^2 \tilde{\Lambda}^2 \ell_S
\\
-4 \tilde{\Lambda} \ell_S^4+8 \tilde{\Lambda}^2 \ell_S^3-8 \tilde{\Lambda} \ell_S^3+16 \tilde{\Lambda}^2 \ell_S^2-\tilde{\Lambda} \ell_S^2-6 \tilde{\Lambda}^3 \ell_S+4 \tilde{\Lambda}^2 \ell_S+5 \tilde{\Lambda}\ell_S+8 \varpi^2 \ell_S^5+28 \varpi^2 \ell_S^4
\\
+26 \varpi^2 \ell_S^3-7 \varpi^2 \ell_S^2-19 \varpi^2 \ell_S-6 \varpi^2\,,
\end{multline}
\begin{multline}
\eta_2(\varpi,\tilde{\Lambda},\ell_S)=\tilde{\Lambda}^{1/2}\Bigg[3 \tilde{\Lambda}^3-\varpi^2 \tilde{\Lambda}^2-\varpi^2 \tilde{\Lambda}-\tilde{\Lambda}^2-5 \tilde{\Lambda}+4 \varpi^2 \tilde{\Lambda} \ell_S^2+10 \varpi^2 \tilde{\Lambda} \ell_S+6 \tilde{\Lambda} \ell_S^3
\\
-4\tilde{\Lambda}^2 \ell_S^2+5 \tilde{\Lambda} \ell_S^2-10 \tilde{\Lambda}^2 \ell_S-7 \tilde{\Lambda} \ell_S-4 \varpi^2 \ell_S^4-20 \varpi^2 \ell_S^3-23 \varpi^2 \ell_S^2+11 \varpi^2 \ell_S+
11 \varpi^2\Bigg]\,,
\end{multline}
\begin{multline}
\eta_3(\varpi,\tilde{\Lambda},\ell_S)=2 \tilde{\Lambda}^3-2 \varpi^2 \tilde{\Lambda}^2-4 \varpi^2 \tilde{\Lambda}+4 \tilde{\Lambda}^2+4 \varpi^2 \tilde{\Lambda} \ell_S^2
+8 \varpi^2 \tilde{\Lambda} \ell_S-2 \tilde{\Lambda}^2 \ell_S^2+2 \tilde{\Lambda}^2 \ell_S
\,,
\end{multline}
\begin{equation}
\eta_4(\varpi,\tilde{\Lambda},\ell_S)=-\tilde{\Lambda}^{3/2} (\tilde{\Lambda}+\varpi^2)\,,
\end{equation}
and
\begin{equation}
\iota_0(\varpi,\tilde{\Lambda},\ell_S)=-12 \sqrt{\tilde{\Lambda}} (\ell_S+1) (\tilde{\Lambda}+\varpi^2)(1-\tilde{\Lambda}+2 \ell_S)\,,
\end{equation}
\begin{multline}
\iota_1(\varpi,\tilde{\Lambda},\ell_S)=36 \tilde{\Lambda}^2-12 \varpi^2 \tilde{\Lambda}^2-12 \varpi^2 \tilde{\Lambda}-12 \tilde{\Lambda}^3-12 \tilde{\Lambda}
\\
-48 \varpi^2 \tilde{\Lambda} \ell_S+48 \tilde{\Lambda}^2 \ell_S-24 \tilde{\Lambda} \ell_S^2-36 \tilde{\Lambda} \ell_S+48 \varpi^2 \ell_S^3+144\varpi^2 \ell_S^2+132 \varpi^2 \ell_S+36 \varpi^2\,,
\end{multline}
\begin{equation}
\iota_2(\varpi,\tilde{\Lambda},\ell_S)=\tilde{\Lambda}^{1/2} \left(24 \varpi^2 \tilde{\Lambda}-24 \tilde{\Lambda}^2+24 \tilde{\Lambda}+36
   \tilde{\Lambda} \ell_S-24 \varpi^2 \ell_S^2-132 \varpi^2 \ell_S-72\varpi^2\right)\,,
\end{equation}
\begin{equation}
\iota_3(\varpi,\tilde{\Lambda},\ell_S)= 36 \varpi^2 \tilde{\Lambda}-12 \tilde{\Lambda}^2\,,
\end{equation}
\begin{equation}
\iota_4(\varpi,\tilde{\Lambda},\ell_S)=0\,.
\end{equation}
\end{subequations}
Naturally, this transcendental equation has no analytic solutions, but we can proceed numerically, just as in previous sections. 
{Our general argument implies no negative modes can exist for $p > 1/6$. 
We display explicit results for the example mode with $\ell_S=2$ and $\varpi=1$ in Fig.~(\ref{fig:ell2ntnot0}) and similar qualitative results hold for all other values of $\ell_S$ and/ or $\varpi$ we have investigated. Consistent with our general argument in Section~\ref{sec:Fieldtheoretic}, in the region $p>1/6$, there are no modes with ${\rm Re},\tilde{\lambda}<0$.
\begin{figure}
    \centering
    \includegraphics[width=\linewidth]{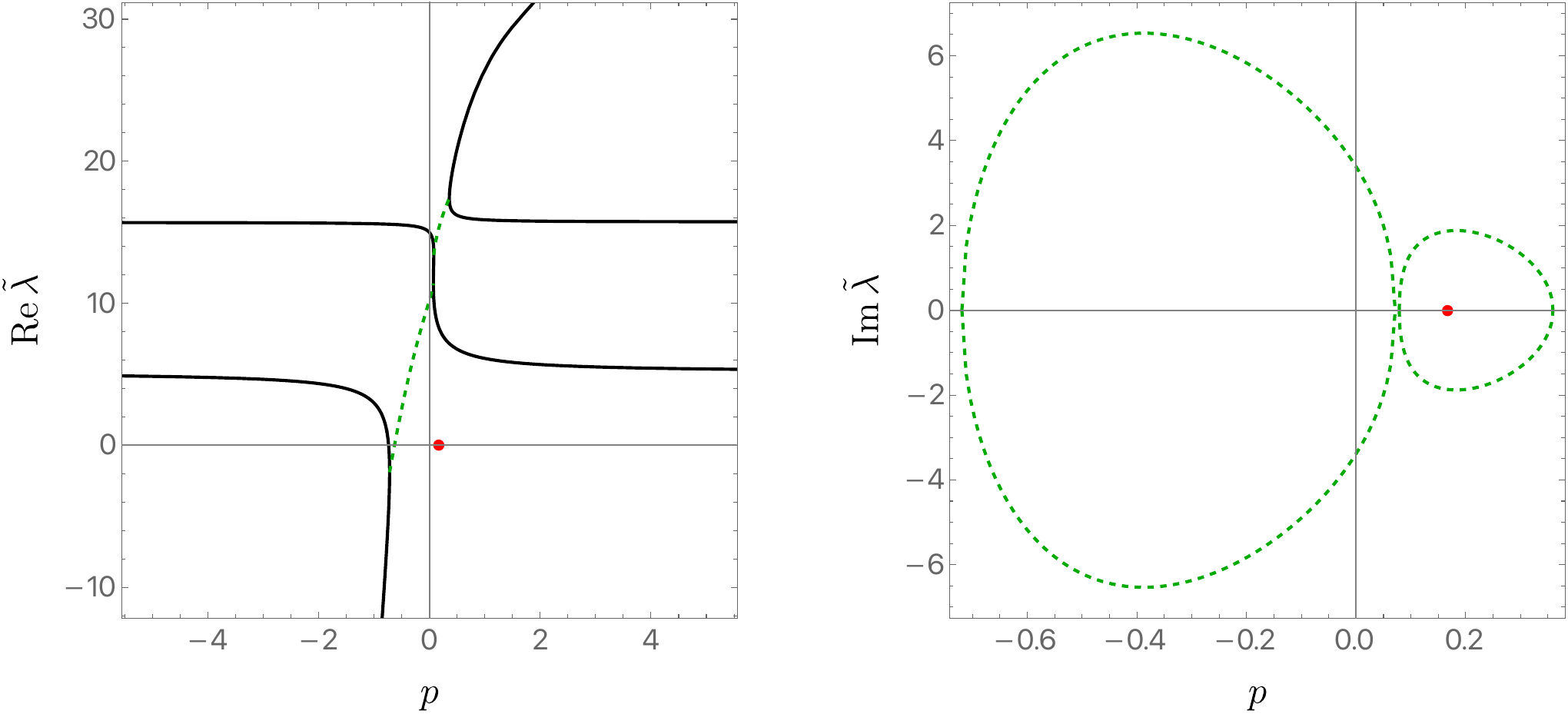}
    \caption{The real part (left panel) and imaginary part (right panel) of $\tilde{\lambda}=\lambda r_E^2$ as a function of $p$ for $\ell_S=2$ and $\varpi=1$. The point $(p,\tilde{\lambda})=\left(1/6,0\right)$ is marked as a red disk. This plot uses the same colour coding as Fig.~\ref{fig:n0ell0}.}
    \label{fig:ell2ntnot0}
\end{figure}
Finally, we note that at large negative values of $p$ we obtain a single negative mode which obeys
\begin{equation}
\tilde{\Lambda}=-16 p^2+24 p+\frac{1}{3} \left[11 \ell_S \left(\ell_S+1\right)-3+8\varpi^2\right]+\frac{29 \ell_S \left(\ell_S+1\right)+40+40 \varpi^2}{12 p}+\mathcal{O}(p^{-2})\,.
\end{equation}

\subsection{Modes with $n=0$ and $\ell_S=1$\label{eq:special0ellS1}}
Next we consider scalar-derived gravitational modes with $\ell_S=1$. On-shell, these modes represent infinitesimal translations. For these modes, $S^{\ell_S\,m_S}_{IJ}=0$ and the equivalent of $h^{\ell_S\,m_S}_T$ in Eq.~(\ref{eq:SIJ}) is altogether absent. This is the reason why these modes have to be analysed separately.
Due to staticity, the metric components $f_{\tilde{\tau}}$ and $f_{\tilde{\tau}r}$ decouple from the remaining and shall be analysed at the end of this section.
The analysis is very similar to the previous sections. We first introduce the trace variable ${\rm f}$ via
\begin{equation}
h_L=\frac{r^2}{2}({\rm f}-f_{\tilde{\tau}\tilde{\tau}}-f_{rr})\,.
\end{equation}
Since we are interested in modes with $n=0$, the metric components $f_{\tilde{\tau}}$ and $f_{\tilde{\tau}r}$ decouple and shall be discussed at the end. After some algebra, the de Donder condition together with the Lichnerowicz eigenvalue equation (\ref{eq:lichneagain}) yield
\begin{subequations}
\begin{align}
&f_{rr}(r)=-\frac{1}{r^2 \lambda }f_{\tilde{\tau}\tilde{\tau}}(r)+\left(\frac{1}{2}-\frac{1}{2 r^2 \lambda }\right) {\rm f}(r)+\frac{1}{r \lambda}f^\prime_{\tilde{\tau}\tilde{\tau}}(r)+\frac{1}{2 r \lambda }{\rm f}^\prime(r)
\label{eq:sina}
\\
&f_r(r)=\frac{1}{2 r \lambda }f_{\tilde{\tau}\tilde{\tau}}(r)+\frac{1}{4 r \lambda }{\rm f}(r)-\frac{1}{2 \lambda }f^\prime_{\tilde{\tau}\tilde{\tau}}(r)-\frac{1}{4 \lambda }{\rm f}^\prime(r)\,,
\label{eq:sinb}
\end{align}
together with
\begin{equation}
\frac{(r^2 f^\prime_{\tilde{\tau}\tilde{\tau}})^\prime}{r^2}+\left(\lambda-\frac{2}{r^2}\right)f_{\tilde{\tau}\tilde{\tau}}=0\,,
\label{eq:sinc}
\end{equation}
and
\begin{equation}
\frac{(r^2 {\rm f}^\prime)^\prime}{r^2}+\left(\lambda-\frac{2}{r^2}\right){\rm f}=0\,.
\label{eq:sind}
\end{equation}
\end{subequations}
Once $f_{\tilde{\tau}\tilde{\tau}}$ and ${\rm f}$ are known, one can determine $f_{rr}$ and $f_r$ via Eq.~(\ref{eq:sina}) and Eq.~(\ref{eq:sinb}), respectively. Eq.~(\ref{eq:sinc}) and Eq.~(\ref{eq:sind}) can be readily solved in full generality to yield
\begin{align}
&{\rm f}(r)=\frac{C_1}{r \sqrt{\lambda}}\left[\cos(\sqrt{\lambda}r)-\frac{\sin(\sqrt{\lambda}r)}{\sqrt{\lambda}r}\right]+\frac{C_2}{r \sqrt{\lambda}}\left[\frac{\cos(\sqrt{\lambda}r)}{\sqrt{\lambda}r}+\sin(\sqrt{\lambda}r)\right]\,,
\\
&f_{\tilde{\tau}\tilde{\tau}}(r)=\frac{C_3}{r \sqrt{\lambda}}\left[\cos(\sqrt{\lambda}r)-\frac{\sin(\sqrt{\lambda}r)}{\sqrt{\lambda}r}\right]+\frac{C_4}{r \sqrt{\lambda}}\left[\frac{\cos(\sqrt{\lambda}r)}{\sqrt{\lambda}r}+\sin(\sqrt{\lambda}r)\right]\,.
\end{align}
Once again, regularity demands $C_2=C_4=0$. On the other hand, our boundary conditions at the cavity wall require
\begin{subequations}
\begin{equation}
h_L(r_E)=r_E^2f_{\tilde{\tau}\tilde{\tau}}(r_E)
\end{equation}
and
\begin{equation}
f_{\tilde{\tau}\tilde{\tau}}^\prime(r_E)+\frac{2}{r_E^2} h_L^\prime(r_E)-\frac{4(1-3p)}{r_E}f_{\tilde{\tau}\tilde{\tau}}(r_E)-\frac{2}{r_E}f_{rr}(r_E)+\frac{4}{r_E^2}f_r(r_E)=0\,.
\end{equation}
\end{subequations}
Both conditions can be satisfied so long as $\tilde{\lambda}=\lambda r_E^2$ obeys
\begin{multline}
4 \tilde{\lambda}^2+3 \tilde{\lambda}-12+3 \cos \left(2 \sqrt{\tilde{\lambda}}\right) \left(4-9 \tilde{\lambda}+4 \tilde{\lambda}^2\right)+\sqrt{\tilde{\lambda}} \left(24-23 \tilde{\lambda}+3
   \tilde{\lambda}^2\right) \sin \left(2 \sqrt{\tilde{\lambda}}\right)
   \\
   -12 p \left[3+3 \tilde{\lambda}+\tilde{\lambda}^2+\cos \left(2 \sqrt{\tilde{\lambda}}\right) \left(\tilde{\lambda}^2+3
   \tilde{\lambda}-3\right)-\sqrt{\tilde{\lambda}} \left(6+\tilde{\lambda}\right) \sin \left(2 \sqrt{\tilde{\lambda}}\right)\right]=0\,,
\end{multline}
again consistent with equation~\eqref{eq:pexpression} showing there are no negative modes for $p > 1/6$. The point $p=-1/12$ marks the location with $\tilde{\lambda}=0$, so that the modes develop a positive real part for $p>-1/12$, and there is a single negative mode for $p<-1/12$. Away from this special point and at large values of $-p$ we find that the negative mode approaches
\begin{equation}
\lambda =-16 p^2+24 p+\frac{19}{3}+\frac{49}{6 p}+\mathcal{O}(p^{-2})\,.
\end{equation}
Once again, consistent with our general argument in the range $p>1/6$, we find no negative modes. In Fig.~\ref{fig:ellS1nt0} we plot ${\rm Re}\,\tilde{\lambda}$ (left and middle panels) and ${\rm Im}\,\tilde{\lambda}$ (right panel) as a function of $p$. The solid black lines indicate regions of moduli space where the mode is purely real, while the dashed green line shows regions where the modes become complex. The point $(p,\tilde{\lambda})=(-1/12,0)$ is marked as a red disk, and the middle panel provides a zoom of the pink-shaded region of the left panel.
\begin{figure}
    \centering
    \includegraphics[width=\linewidth]{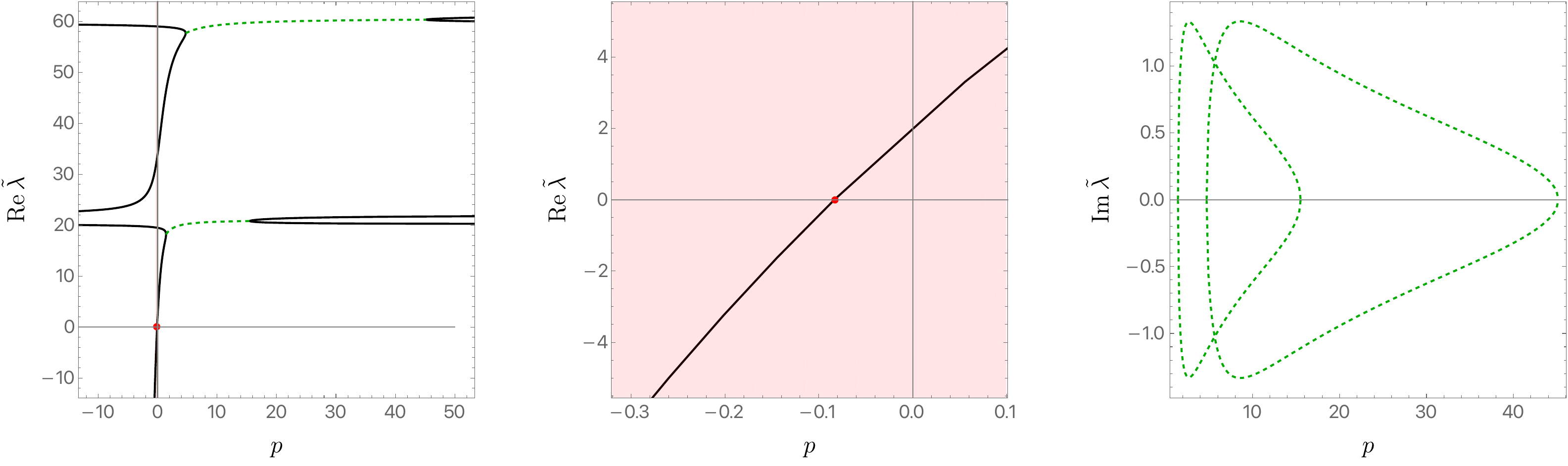}
    \caption{The real part (left and middle panels) and imaginary part (right panel) of $\tilde{\lambda}$ as a function of $p$. The point $(p,\tilde{\lambda})=(-1/12,0)$ is marked as a red disk. For $p<-1/12$, there exists a single mode with ${\rm Re}\,\tilde{\lambda}<0$, indicating the presence of a field-theoretic negative mode. The green dashed curves indicate that the mode is complex.}
    \label{fig:ellS1nt0}
\end{figure}

We now return to the metric components $f_{\tilde{\tau}}$ and $f_{\tilde{\tau}r}$. The de Donder gauge condition imposes
\begin{equation}
f_{\tilde{\tau}}(r)=\frac{r}{2}[r f_{\tilde{\tau}r}^\prime(r)+2 f_{\tilde{\tau}r}(r)]\,,
\end{equation}
while from the Lichnerowicz eigenvalue equation (\ref{eq:lichneagain}) we find
\begin{equation}
\frac{\left(r^4 f_{\tilde{\tau}r}^\prime\right)^\prime}{r^4}+\lambda f_{\tilde{\tau}r}=0\,.
\end{equation}
The equation for $f_{\tilde{\tau}r}$ can be easily integrated in full generality and one finds
\begin{equation}
f_{\tilde{\tau}r}(r)=\frac{C_1}{(r \sqrt{\lambda})^3}\left[r\sqrt{\lambda} \cos\left(r\sqrt{\lambda}\right)-\sin\left(r\sqrt{\lambda}\right)\right]+\frac{C_2}{(r \sqrt{\lambda})^3}\left[\cos\left(r\sqrt{\lambda}\right)+r\sqrt{\lambda}\sin\left(r\sqrt{\lambda}\right)\right]\,.
\end{equation}
Regularity at the origin demands $C_2=0$, whereas our boundary conditions require
\begin{equation}
f_{\tilde{\tau}}(r_E)=0\Rightarrow \frac{1}{\tilde{\lambda}^{3/2}}\left[\sqrt{\tilde{\lambda}} \cos \left(\sqrt{\tilde{\lambda}}\right)-\left(1-\tilde{\lambda}\right) \sin \left(\sqrt{\tilde{\lambda}}\right)\right]=0\,,
\end{equation}
with $\tilde{\lambda}=\lambda r_E^2$.  We see this is independent of $p$ and hence yields the same results as in the Dirichlet case studied in~\cite{Marolf:2022ntb} where stability was found. It is a relatively simple exercise to show that no complex zeroes of the above transcendental equation exist, and that the real zeroes all like on the positive axis. The first few are approximately given by
\begin{equation}
\tilde{\lambda}=\{7.52793(0),37.41480(5),86.79932(7),155.89863(2),\ldots\}\,.
\end{equation}
\subsection{Modes with $n\neq0$ and $\ell_S=1$\label{eq:special1ellS1}}
We finally reach the last scalar mode to consider. Again $S^{\ell_S\,m_S}_{IJ}$ vanishes identically and the equivalent of $h^{\ell_S\,m_S}_T$ in Eq.~(\ref{eq:SIJ}) is altogether absent. However, in general, $f_{\tilde{\tau}}$ and $f_{\tilde{\tau}r}$ will not decouple from the remaining metric components. We follow a familiar procedure, mirroring the preceding sections. We commence by introducing the trace variable ${\rm f}$ through the equation:
\begin{equation}
h_L=\frac{r^2}{2}({\rm f}-f_{\tilde{\tau}\tilde{\tau}}-f_{rr})\,.
\end{equation}
Following some algebraic manipulations, the conjunction of the de Donder condition and the Lichnerowicz eigenvalue equation (\ref{eq:lichneagain}) produces
\begin{subequations}
\begin{multline}
f_{\tilde{\tau}}(r)=\frac{i}{4 \tilde{n}}\Bigg\{\left(2+r^2 \tilde{n}^2-\frac{\lambda}{\lambda -\tilde{n}^2}\right) f_{\tilde{\tau}\tilde{\tau}}(r)+ \left[8-r^2 \left(\lambda -\tilde{n}^2\right)\right] f_{rr}(r)
\\
-\frac{1}{2} \left[8-r^2 \left(\lambda -2\tilde{n}^2\right)-\frac{\lambda}{\lambda -\tilde{n}^2}\right] {\rm f}(r)+\frac{r \lambda }{\lambda -\tilde{n}^2}f_{\tilde{\tau}\tilde{\tau}}^\prime(r)+2 r f_{rr}^\prime(r)-\frac{r \lambda }{2 \left(\lambda-\tilde{n}^2\right)}{\rm f}^\prime(r)\Bigg\}\,,
\label{eq:din1}
\end{multline}
\begin{multline}
f_{\tilde{\tau}r}(r)=\frac{i}{2 \tilde{n}} \Bigg[\frac{4}{r} f_{rr}(r)-\frac{1}{2 r}\left(4-\frac{\lambda}{\lambda -\tilde{n}^2}\right) {\rm f}(r)+\frac{1}{r}\left(2-\frac{\lambda}{\lambda-\tilde{n}^2}\right)f_{\tilde{\tau}\tilde{\tau}}(r)
\\
+\frac{\tilde{n}^2}{\lambda -\tilde{n}^2}f^\prime_{\tilde{\tau}\tilde{\tau}}(r)+f_{rr}'(r)-\frac{\lambda}{2 \left(\lambda -\tilde{n}^2\right)}{\rm f}^\prime(r)\Bigg]\,,
\label{eq:din2}
\end{multline}
\begin{multline}
f_{r}(r)=\frac{r}{2} \Bigg[f_{rr}(r)+\frac{1}{2}\frac{\lambda}{\lambda -\tilde{n}^2}f_{\tilde{\tau}\tilde{\tau}}(r)-\frac{1}{4}\frac{\lambda}{\lambda -\tilde{n}^2}  {\rm f}(r)-\frac{r}{2}\frac{\tilde{n}^2}{\lambda
   -\tilde{n}^2} f_{\tilde{\tau}\tilde{\tau}}^\prime(r)+\frac{r}{2} f_{rr}'(r)-\frac{r}{4}\frac{\lambda -2 \tilde{n}^2}{\lambda -\tilde{n}^2} {\rm f}^\prime(r)\Bigg]
   \label{eq:din3}
\end{multline}
\begin{equation}
h_{rr}(r)=q_1(r)+\frac{\tilde{n}^2}{\lambda -\tilde{n}^2}f_{\tilde{\tau}\tilde{\tau}}(r)+\frac{1}{2}\frac{\lambda -2 \tilde{n}^2}{\lambda -\tilde{n}^2}{\rm f}(r)\,,
\label{eq:din4}
\end{equation}
together with
\begin{equation}
\frac{(r^2 f^\prime_{\tilde{\tau}\tilde{\tau}})^\prime}{r^2}+\left(\lambda-\frac{2}{r^2}-\tilde{n}^2\right)f_{\tilde{\tau}\tilde{\tau}}=0\,,
\label{eq:din5}
\end{equation}
\begin{equation}
\frac{(r^2 {\rm f}^\prime)^\prime}{r^2}+\left(\lambda-\frac{2}{r^2}-\tilde{n}^2\right){\rm f}=0\,.
\label{eq:din6}
\end{equation}
and
\begin{equation}
\frac{(r^4 q_1^\prime)^\prime}{r^4}+\left(\lambda-\frac{4}{r^2}-\tilde{n}^2\right)q_1=0\,.
\label{eq:din7}
\end{equation}
\end{subequations}
Solutions to $f_{\tilde{\tau}\tilde{\tau}}$, ${\rm f}$ and $q_1$ are found via Eqs.~(\ref{eq:din5})-(\ref{eq:din7}) and one can then determine all remaining variables using Eqs.~(\ref{eq:din1})-(\ref{eq:din4}). For $f_{\tilde{\tau}\tilde{\tau}}$, ${\rm f}$ and $q_1$ we find
\begin{equation}
{\rm f}(r)=\frac{C_1}{r \sqrt{\Lambda}}\left[\cos(\sqrt{\Lambda}r)-\frac{\sin(\sqrt{\Lambda}r)}{\sqrt{\Lambda}r}\right]+\frac{C_2}{r \sqrt{\Lambda}}\left[\frac{\cos(\sqrt{\Lambda}r)}{\sqrt{\Lambda}r}+\sin(\sqrt{\Lambda}r)\right]\,,
\end{equation}
\begin{equation}
f_{\tilde{\tau}\tilde{\tau}}(r)=\frac{C_3}{r \sqrt{\Lambda}}\left[\cos(\sqrt{\Lambda}r)-\frac{\sin(\sqrt{\Lambda}r)}{\sqrt{\Lambda}r}\right]+\frac{C_4}{r \sqrt{\Lambda}}\left[\frac{\cos(\sqrt{\Lambda}r)}{\sqrt{\Lambda}r}+\sin(\sqrt{\Lambda}r)\right]\,,
\end{equation}
\begin{multline}
q_1(r)=\frac{C_5}{r^2\Lambda}\left[\frac{3}{r \sqrt{\lambda}} \cos \left(r \sqrt{\lambda}\right)-\frac{3}{r^2 \Lambda } \sin \left(r \sqrt{\lambda}\right)+\sin \left(r \sqrt{\lambda}\right)\right]
\\
+\frac{C_6}{r^2\Lambda}\left[\frac{3}{r \sqrt{\lambda}} \sin \left(r \sqrt{\lambda}\right)+\frac{3}{r^2 \Lambda } \cos \left(r \sqrt{\lambda}\right)-\cos \left(r \sqrt{\lambda}\right)\right]
\end{multline}
where we defined $\Lambda\equiv\lambda-\tilde{n}^2$. Regularity at the origin demands $C_2=C_4=C_6=0$, while our boundary conditions impose
\begin{subequations}
\begin{equation}
h_L(r_E)=r_E^2f_{\tilde{\tau}\tilde{\tau}}(r_E)\,,\quad f_{\tilde{\tau}}(r_E)=0\,,
\end{equation}
and
\begin{equation}
f_{\tilde{\tau}\tilde{\tau}}^\prime(r_E)+\frac{2}{r_E^2} h_L^\prime(r_E)-\frac{4(1-3p)}{r_E}f_{\tilde{\tau}\tilde{\tau}}(r_E)-\frac{2}{r_E}f_{rr}(r_E)+\frac{4}{r_E^2}f_r(r_E)-2 i \tilde{n}f_{\tilde{\tau}r}(r_E)=0\,.
\end{equation}
\end{subequations}
All the last three conditions are satisfied if the eigenvalue $\lambda$ (or equivalently $\Lambda$) satisfies
\begin{multline}
\sqrt{\tilde{\Lambda}} \sin \left(\sqrt{\tilde{\Lambda}}\right) \Bigg[324 \varpi^2 \tilde{\Lambda}^2-9 \varpi^2 \tilde{\Lambda}^5-198 \varpi^2 \tilde{\Lambda}^4-81 \varpi^2 \tilde{\Lambda}^3-9
   \tilde{\Lambda}^6+126 \tilde{\Lambda}^5-405 \tilde{\Lambda}^4+324\tilde{\Lambda}^3
   \\
   +p \left(3888 \varpi^2 \tilde{\Lambda}^3-108 \varpi^2 \tilde{\Lambda}^5-108 \varpi^2 \tilde{\Lambda}^4+972
   \varpi^2 \tilde{\Lambda}^2-108 \tilde{\Lambda}^6-108 \tilde{\Lambda}^5+972 \tilde{\Lambda}^3\right)\Bigg]
   \\
   +\sqrt{\tilde{\Lambda}} \sin \left(3 \sqrt{\tilde{\Lambda}}\right) \Bigg[459 \varpi^2 \tilde{\Lambda}^3-9 \varpi^2
   \tilde{\Lambda}^5-306 \varpi^2 \tilde{\Lambda}^4-108 \varpi^2 \tilde{\Lambda}^2+135 \tilde{\Lambda}^6-558 \tilde{\Lambda}^5+567 \tilde{\Lambda}^4-108
   \tilde{\Lambda}^3
   \\
   +p \left(1620 \varpi^2-108 \varpi^2 \tilde{\Lambda}^5 \tilde{\Lambda}^4-324 \varpi^2 \tilde{\Lambda}^2-108 \tilde{\Lambda}^6-108 \tilde{\Lambda}^5+1296 \tilde{\Lambda}^4-324
   \tilde{\Lambda}^3\right)\Bigg]
   \\
   +\cos \left(3 \sqrt{\tilde{\Lambda}}\right) \Bigg[9 \varpi^2 \tilde{\Lambda}^6+90 \varpi^2 \tilde{\Lambda}^5-450 \varpi^2 \tilde{\Lambda}^4+324 \varpi^2
   \tilde{\Lambda}^3-27 \tilde{\Lambda}^7+342 \tilde{\Lambda}^6-666 \tilde{\Lambda}^5+324 \tilde{\Lambda}^4
   \\
   +p \left(1728 \varpi^2 \tilde{\Lambda}^4-648 \varpi^2 \tilde{\Lambda}^5+972 \varpi^2
   \tilde{\Lambda}^3-216 \tilde{\Lambda}^6-864 \tilde{\Lambda}^5+972 \tilde{\Lambda}^4\right)\Bigg]
   \\
   +\cos \left(\sqrt{\tilde{\Lambda}}\right) \Bigg[198 \varpi^2
   \tilde{\Lambda}^5-9 \varpi^2 \tilde{\Lambda}^6+18 \varpi^2 \tilde{\Lambda}^4-324 \varpi^2 \tilde{\Lambda}^3+27 \tilde{\Lambda}^7-54 \tilde{\Lambda}^6+234 \tilde{\Lambda}^5-324 \tilde{\Lambda}^4
   \\
   +p \left(216 \varpi^2
   \tilde{\Lambda}^5-3024 \varpi^2 \tilde{\Lambda}^4-972 \varpi^2 \tilde{\Lambda}^3-216 \tilde{\Lambda}^6-432 \tilde{\Lambda}^5-972 \tilde{\Lambda}^4\right)\Bigg]=0\,,
\end{multline}
where we defined $\tilde{\Lambda}=\Lambda r_E^2$. {Again this expression is consistent with~\eqref{eq:pexpression} in the main text, showing there are no negative modes for $p > 1/6$.

Using the above expression we can verify that a zero mode exists if $p=-(1+\tilde{n}^2)/12$, becoming a negative mode in the region $p<-(1+\tilde{n}^2)/12$ and positive otherwise. Away from this special point, we have to proceed numerically and we find that for $p<-(1+\tilde{n}^2)/12$ a single negative mode exists, and that at large values of $-p$ it approaches
\begin{equation}
\lambda =-16 p^2+24 p+\frac{1}{3} \left(19+8 \tilde{n}^2\right)+\frac{49+20 \tilde{n}^2}{6 p}+\mathcal{O}(p^{-2})\,.
\end{equation}
Consistent with our general argument, within the range $p>1/6$, we observe the absence of negative modes. To conclusively affirm the non-problematic nature of this mode, numerical exploration is once again undertaken. In Fig.~\ref{fig:ellS1ntnot0} we depict ${\rm Re}\,\tilde{\lambda}$ (left and middle panels) and ${\rm Im}\,\tilde{\lambda}$ (right panel) against $p$. Solid black lines delineate regions in moduli space where the mode is exclusively real, while dashed green lines demarcate areas where the modes become complex. A red disk at $(p,\tilde{\lambda})=(-(1+\tilde{n}^2)/12,0)$ serves as a reference point, and the middle panel offers an amplified view of the pink-shaded region in the left panel. In generating this figure we took $\tilde{n}=1$, but we found similar results for different values of $\tilde{n}$.
\begin{figure}
    \centering
    \includegraphics[width=\linewidth]{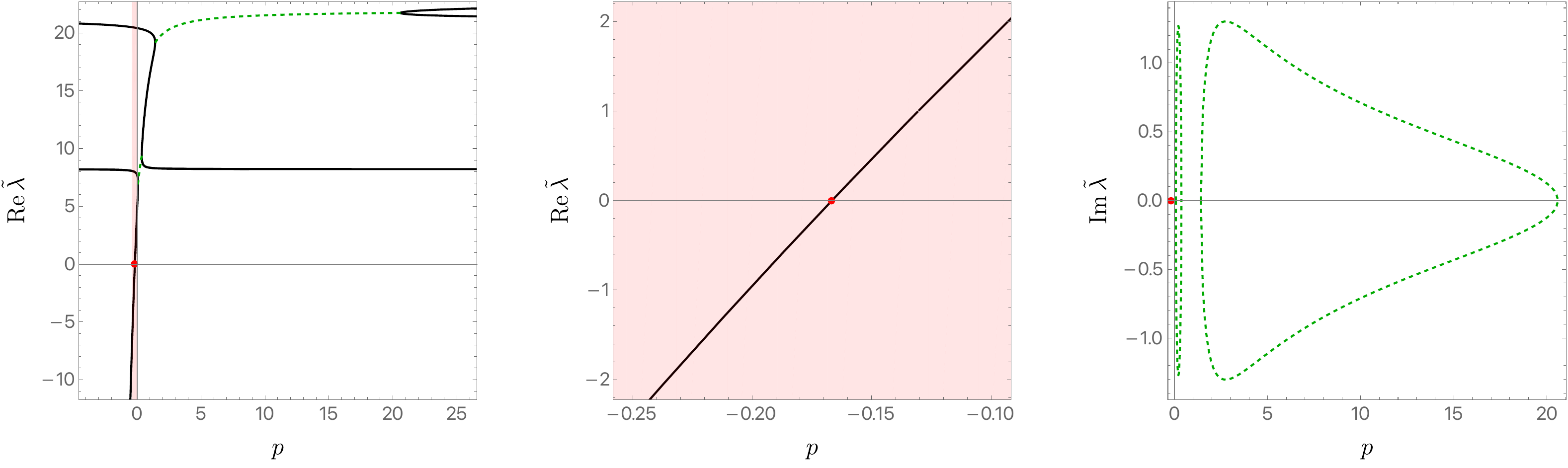}
    \caption{The real part (left and middle panels) and imaginary part (right panel) of $\tilde{\lambda}$ as a function of $p$. The point $(p,\tilde{\lambda})=(-(1+\tilde{n}^2)/12,0)$ is marked as a red disk. For $p<-(1+\tilde{n}^2)/12$, there exists a single mode with ${\rm Re}\,\tilde{\lambda}<0$, indicating the presence of a field-theoretic negative mode. The green dashed curves indicate that the mode is complex. In generating this plot we took $\tilde{n}=1$.}
    \label{fig:ellS1ntnot0}
\end{figure}
\subsection{Non-spherical Euclidean modes with $\ell_V \geq 2$}\label{sec:nonsphericalV}
We now turn our attention to vector-derived gravitational perturbations. These are easier to present than the scalars, owing to the fact that less components of the metric are non-zero. Importantly we find that the behaviour of the vector modes is precisely the same as that in the Dirichlet case. Following the stability of the Dirichlet case studied in~\cite{Marolf:2022ntb} then all vector modes will be stable for our general conformal boundary condition for any $p$.

These modes are built from harmonic vectors on the two-sphere, \emph{i.e.}
\begin{equation}
\mathcal{D}^I Y_I^{\ell_V\,m_V}=0\quad \text{and}\quad \mathcal{D}_I\mathcal{D}^I Y^{\ell_V\,m_V}_J+\left[\ell_V(\ell_V+1)-1\right]Y^{\ell_V\,m_V}_J=0\,.
\end{equation}
with $\ell_V\geq1$. For the case of a two-sphere, it turns out to be relatively simple to write these modes in terms of \emph{scalar} spherical harmonics. Namely,
\begin{equation}
\mathbb{Y}^{\ell_V\,m_V}\equiv Y_I^{\ell_V\,m_V}\mathrm{d}x^I=\star_{S^2}\left(\mathrm{d}Y^{\ell_S\,m_S}\right)\,
\end{equation}
where $\star_{S^2}$ is the Hodge dual on the two-sphere with $\ell_V=\ell_S$, and non-zero modes start with $\ell_V\geq1$. A general vector-derived mode can now be written as
\begin{subequations}
\begin{align}
&h^{\ell_V\,m_V}_{\hat{a}\hat{b}}=0
\\
& h^{\ell_V\,m_V}_{\hat{a}I}=\hat{f}^{\ell_V\,m_V}_{\hat{a}}(\tilde{\tau},r)\,Y_I^{\ell_V\,m_V}
\\
& h^{\ell_V\,m_V}_{IJ}=\hat{h}_T^{\ell_V\,m_V}(\tilde{\tau},r)\,S^{\ell_V\,m_V}_{IJ}
\end{align}
where $\mathcal{S}^{\ell_S\,m_S}_{IJ}$ is a traceless symmetric two tensor defined as
\begin{equation}
\mathcal{S}^{\ell_V\,m_V}_{IJ}=\mathcal{D}_IY_J^{\ell_V\,m_V}+\mathcal{D}_JY_I^{\ell_V\,m_V}\,.
\label{eq:StIJ}
\end{equation}
\end{subequations}
For each mode, there are a total of (at most) three functions to be determined. Owing to the background $\partial/\partial \tau$ Killing vector field, we can further decompose the scalars $\hat{f}^{\ell_V\,m_V}_{\hat{a}}(\tilde{\tau},r)$ and $\hat{h}_T^{\ell_V\,m_V}(\tilde{\tau},r)$ as
\begin{equation}
\hat{f}^{\ell_V\,m_V}_{\hat{a}}(\tilde{\tau},r)=e^{i\,\tilde{n}\,\tilde{\tau}}f^{\ell_V\,m_V\,\tilde{n}}_{\hat{a}}(r)\quad\text{and}\quad \hat{h}_T^{\ell_V\,m_V}(\tilde{\tau},r)=e^{i\,\tilde{n}\tilde{\tau}}\,h_T^{\ell_V\,m_V\,\tilde{n}}(r)\,.
\end{equation}
Just like with scalars, modes featuring distinct values of $\ell_V$, $m_V$, and/or $\tilde{n}$ decouple from one another at the quadratic level in the action. Consequently, we will omit the subscripts $\ell_V,m_V,\tilde{n}$ in the subsequent discussion. It is worth noting that modes with $\ell_V=1$ hold a unique status, as for these modes, $S_{IJ}^{1,m_V}=0$. The treatment of these specific modes will be addressed in sections \ref{sec:special0ellV1} and \ref{sec:special1ellV1}. Finally, we also note that for the vectors, our boundary conditions reduce to those of the standard Dirichlet problem, being independent of $p$, and thus there are no negative modes.
\subsubsection{Static modes\label{sec:staticModesV}}
In section we analyse vector-derived perturbations with $\tilde{n}=0$. Due to staticity, the metric component $f_{\tilde{\tau}}$ decouples from the remaining ones, and we shall discuss it at the end of this section. We are left with determining $f_{r}$ and $h_T$. Imposing the de Donder gauge condition and the Lichnerowicz eigenvalue equation (\ref{eq:lichneagain}) yields
\begin{subequations}
\begin{equation}
h_T=\frac{r}{\ell_V(\ell_V+1)-2}\left[r\,f_{r}^\prime(r)+2\,f_{r}(r)\right]
\end{equation}
and
\begin{equation}
\frac{(r^2f_r^\prime)^\prime}{r^2}+\left[\lambda-\frac{\ell_V(\ell_V+1)}{r^2}\right]f_r=0\,.
\end{equation}
\end{subequations}
The last equation for $f_r$ can be readily solved as
\begin{equation}
f_r(r)=\frac{C_1}{(r\sqrt{\lambda})^{1/2}}{\rm J}_{\ell_V+\frac{1}{2}}\left(r\sqrt{\lambda}\right)+\frac{C_2}{(r\sqrt{\lambda})^{1/2}}{\rm Y}_{\ell_V+\frac{1}{2}}\left(r\sqrt{\lambda}\right)\,.
\end{equation}
Regularity at the origin demands $C_2=0$ while our boundary conditions require
\begin{equation}
h_T(r_E)=0\Rightarrow{\rm J}_{\ell _V+\frac{3}{2}}\left(\sqrt{\tilde{\lambda}}\right) \sqrt{\tilde{\lambda}}-\left(\ell _V+2\right) {\rm J}_{\ell _V+\frac{1}{2}}\left(\sqrt{\tilde{\lambda}}\right)=0\,,
\end{equation}
again showing the mode behaviour is independent of $p$, and thus the same as for the Dirichlet case. It is easy to show that the above equation only has real roots, and that for $\ell_V=2$ the first four modes are approximately given by
\begin{equation}
\tilde{\lambda}=\{17.91715(3), 57.60145(7),116.86211(3), 195.83544(5),\ldots\}\,.
\label{eq:afew}
\end{equation}
To sum up, there are no negative modes for this sector of modes.
We now return to analysing $f_{\tilde{\tau}}$. This component turns out to be gauge invariant. The Lichnerowicz eigenvalue equation (\ref{eq:lichneagain}) yields
\begin{equation}
f^{\prime\prime}_{\tilde{\tau}}+\left[\lambda-\frac{\ell_V(\ell_V+1)}{r^2}\right]f_{\tilde{\tau}}=0\,.
\end{equation}
The generic solution to $f_{\tilde{\tau}}$ reads
\begin{equation}
f_{\tilde{\tau}}=(r\sqrt{\lambda})^{1/2}\left[C_1 {\rm J}_{\frac{\ell_V}{2}+1}\left(r\sqrt{\lambda}\right)+C_2 {\rm Y}_{\frac{\ell_V}{2}+1}\left(r\sqrt{\lambda}\right)\right]\,.
\end{equation}
Once again, regularity at the origin demands $C_2=0$, while our boundary conditions give
\begin{equation}
f_{\tilde{\tau}}(r_E)=0\Rightarrow {\rm J}_{\frac{\ell_V}{2}+1}\left(\sqrt{\tilde{\lambda}}\right)\,,
\end{equation}
with $\tilde{\lambda}=\lambda r_E^2$, again showing no dependence on $p$. There are no Euclidean modes lying on the complex plane, and in fact $\lambda$ is positive definite. For the first four modes we find
\begin{equation}
\tilde{\lambda}=\{26.37461(7),70.84999(9),135.02070(9),218.92018(9),\ldots\}\,.
\end{equation}
\subsubsection{Generic modes\label{sec:GenericModesV}}
Having discussed the case with $\tilde{n}=0$, we now turn to the more complicated case with $\tilde{n}\neq0$. In this case all three functions $f_{\tilde{\tau}}(r)$, $f_r(r)$ and $h_T(r)$ can in principle be non-vanishing. Just as for the scalar-derived modes, these can also be solved in terms of Bessel functions as follows
\begin{subequations}
\begin{equation}
h_{r}(r)=q_1(r)+\frac{\ell_V-1}{r}q_2(r)\,,
\end{equation}
\begin{equation}
h_{T}(r)=-\frac{r}{\ell_V+2}q_1(r)+q_2(r)\,,
\end{equation}
and
\begin{equation}
h_{\tilde{\tau}}(r)=\frac{i}{\tilde{n} r} \left[\left(1+\ell _V\right) q_1(r)+\frac{1-\ell _V^2}{r} q_2(r)+r q_1^\prime(r)-\left(1-\ell _V\right) q_2^\prime(r)\right]\,,
\end{equation}
with
\begin{equation}
q_1^{\prime\prime}+\left[\lambda-\frac{(\ell_V+1)(\ell_V+2)}{r^2}-\tilde{n}^2\right]q_1=0\,,
\end{equation}
and
\begin{equation}
r^2\left(\frac{q^\prime_2}{r^2}\right)^{\prime}+\left[\lambda-\frac{\ell_V(\ell_V-1)-2}{r^2}-\tilde{n}^2\right]q_2=0\,.
\end{equation}
\end{subequations}
The equations for $q_1$ and $q_2$ can be readily solved as,
\begin{align}
&q_1(r)=(r\Lambda)^{1/2}\left[C_1\,{\rm J}_{\ell_V+\frac{3}{2}}\left(r\sqrt{\Lambda}\right)+C_2\,{\rm Y}_{\ell_V+\frac{3}{2}}\left(r\sqrt{\Lambda}\right)\right]
\\
&q_2(r)=(r\Lambda)^{3/2}\left[C_3\,{\rm J}_{\ell_V-\frac{1}{2}}\left(r\sqrt{\Lambda}\right)+C_4\,{\rm Y}_{\ell_V-\frac{1}{2}}\left(r\sqrt{\Lambda}\right)\right]\,,
\end{align}
where we defined $\Lambda\equiv \lambda-\tilde{n}^2$. Regularity at the origin demands $C_2=C_4=0$, while our boundary conditions demand
\begin{equation}
f_{\tilde{\tau}}(r_E)=h_T(r_E)=0\,,
\end{equation}
which translates into
\begin{equation}
{\rm J}_{\ell _V+\frac{1}{2}}\left(\sqrt{\tilde{\Lambda}}\right) \left[\sqrt{\tilde{\Lambda}} {\rm J}_{\ell _V+\frac{3}{2}}\left(\sqrt{\tilde{\Lambda}}\right)-\left(\ell _V+2\right) {\rm J}_{\ell
   _V+\frac{1}{2}}\left(\sqrt{\tilde{\Lambda}}\right)\right]=0
\end{equation}
where we have defined $\tilde{\Lambda}=\Lambda r_E^2$. Since each factor can be separately vanishing we see that we recover the same values eigenvalues as in the previous section, except that we should make the replacement $\tilde{\lambda}$ by $\tilde{\Lambda}$.
\subsection{Modes with $n=0$ and $\ell_V=1$\label{sec:special0ellV1}}
This mode is perhaps the easiest to analyse. Here $h_T=0$ in Eq.~(\ref{eq:StIJ}) and additionally, since the metric is static, $f_r$ and $f_{\tilde{\tau}}$ decouple from each other. Indeed, the de Donder gauge fixes $f_r$ to be 
\begin{equation}
f_{r}(r)=\frac{C_1}{r^2}\,,
\end{equation}
and then regularity at the origin demands $C_1=0$, establishing that no modes exist with $f_r\neq0$ in this sector.
For $f_{\tilde{\tau}}$ we find 
\begin{equation}
f_{\tilde{\tau}}(r)=C_1\left[\cos\left(r\sqrt{\lambda}\right)-\frac{\sin \left(r\sqrt{\lambda}\right)}{r\sqrt{\lambda}}\right]+C_2\left[\frac{\cos \left(r\sqrt{\lambda}\right)}{r\sqrt{\lambda}}+\sin\left(r\sqrt{\lambda}\right)\right]
\end{equation}
and regularity at the origin demands $C_2=0$, while our boundary conditions at the cavity wall impose
\begin{equation}
f_{\tilde{\tau}}(r_E)=0\Rightarrow \cos\sqrt{\tilde{\lambda}}-\frac{\sin \sqrt{\tilde{\lambda}}}{\sqrt{\tilde{\lambda}}}\,
\end{equation}
with $\tilde{\lambda}=\lambda r_E^2$. Again there is no dependence on $p$ so the behaviour is as for Dirichlet boundary conditions. There are no zeroes on the complex plane, and the first four real zeroes are positive (note that $\tilde{\lambda}=0$ is not a valid mode, since then the perturbation vanishes identically) and read
\begin{equation}
\tilde{\lambda}=\{4.49340(9),7.72525(2),10.90412(2),14.06619(4)\}\,.
\end{equation}
\subsection{Modes with $n\neq 0$ and $\ell_V=1$\label{sec:special1ellV1}}
This is the last case to analyse. The de Donder gauge condition imposes
\begin{equation}
f_{\tilde{\tau}}(r)=\frac{i}{r\tilde{n}}\left[r f^\prime_r(r)+f_r(r)\right]\,,
\end{equation}
while the Lichnerowicz eigenvalue equation (\ref{eq:lichneagain}) yields
\begin{equation}
f_r^{\prime\prime}+\left(\lambda-\tilde{n}^2-\frac{6}{r^2}\right)f_r=0\,.
\end{equation}
The equation for $h_r$ can be found in full generality
\begin{multline}
f_r(r)=C_1 \left[\frac{3 \cos \left(r \sqrt{\Lambda }\right)}{r \sqrt{\Lambda }}-\frac{3 \sin \left(r \sqrt{\Lambda }\right)}{r^2 \Lambda }+\sin \left(r \sqrt{\Lambda }\right)\right]
\\
+C_2 \left[\cos \left(r
   \sqrt{\Lambda }\right)-\frac{3 \cos \left(r \sqrt{\Lambda }\right)}{r^2 \Lambda }-\frac{3 \sin \left(r \sqrt{\Lambda }\right)}{r \sqrt{\Lambda }}\right]\,,
\end{multline}
with $\Lambda\equiv \lambda-\tilde{n}^2$. Regularity at the origin demands $C_2=0$, while our boundary condition imposes
\begin{equation}
f_{\tilde{\tau}}(r_E)=0\Rightarrow \sqrt{\tilde{\Lambda}}\cos \sqrt{\tilde{\Lambda}}-\sin \sqrt{\tilde{\Lambda}}=0\,,
\end{equation}
with $\tilde{\Lambda}=\Lambda r_E^2$. Once again, the behaviour is independent of $p$, and so is the same as in the Dirichlet case. Furthermore, there are no zeroes of the above equation on the real axis. Furthermore, the mode with $\tilde{\Lambda}=0$ corresponds to a vanishing mode function $f_{\tilde{\tau}}$ and on the positive real axis we find
\begin{equation}
\tilde{\Lambda}=\{20.19072(9),59.67951(6), 118.89986(9), 197.85781(1),\ldots\}\,.
\end{equation}

\section{\label{app:analyticity_p}
Analyticity of $p$ in the Unstable Complex $\tilde{\lambda}$ Plane}

We now argue that ${\rm Re}\,p$ is analytic in the complex half plane ${\rm Re}\,\tilde{\lambda} \le 0$. 
While the expression for $p$ given in~\eqref{eq:pexpression} is given in terms of Bessel functions of $\sqrt{\tilde{\Lambda}}$, the form of the expressions ensures that in fact $p$ is analytic at $\tilde{\Lambda} = 0$. 
As noted in the main text, the numerator in this expression has no poles for ${\rm Re}\,\tilde{\Lambda} \le 0$, and other than a branch cut at $\tilde{\Lambda} = 0$ is analytic, and so any non-analyticity in the expression for $p$ would have to derive from the denominator $B_{\ell_S}(\tilde{\Lambda}) + \varpi^2 D_{\ell_S}(\tilde{\Lambda})$. Since the explicit form of $B_{\ell_S}(\tilde{\Lambda})$ and $D_{\ell_S}(\tilde{\Lambda})$ in terms of Bessel functions again guarantee analyticity for ${\rm Re}\,\tilde{\Lambda} \le 0$, except for branch cuts at $\tilde{\Lambda}$ (which cancel with the denominator) then non-analyticity of $p$ may only derive from zeros of this denominator. 
Now an important point is that $\tilde{\Lambda} = \tilde{\lambda} - \varpi^2$, and $\varpi^2 > 0$ and should be real.
Thus to show that ${\rm Re}\,p$ is analytic for ${\rm Re}\,\tilde{\lambda} \le 0$ we should show that the denominator, $B_{\ell_S}(\tilde{\Lambda}) + \varpi^2 D_{\ell_S}(\tilde{\Lambda})$, has no zeros for ${\rm Re}\,\tilde{\Lambda} \le - \varpi^2$.

Firstly we will consider the special case $\varpi = 0$ where we can make an analytic argument. Then we will consider the general case $\varpi \ne 0$ where we will use a combination of analytic and numerical methods to make the argument. 

\subsection{The static case $\varpi = 0$}

In the static case the denominator is simply the function $B_{\ell_S}(\tilde{\Lambda})$. We note that in this case $\tilde{\Lambda} = \tilde{\lambda}$. Looking at $B_{\ell_S}(\tilde{\Lambda})$ we find it factorizes as,
\be
B_{\ell_S}(\tilde{\Lambda}) & = & 12 \, \tilde{\Lambda} \, \mathrm{J}_{\frac{1}{2} + \ell_{S}}(\sqrt{\tilde{\Lambda}}) F_1 F_2 F_3 
\ee
with factors,
\be
F_1(\tilde{\Lambda}) &=& \sqrt{\tilde{\Lambda}} \mathrm{J}_{\frac{1}{2} + \ell_{S}}(\sqrt{\tilde{\Lambda}}) +
\mathrm{J}_{\frac{3}{2} + \ell_{S}}(\sqrt{\tilde{\Lambda}}) \nl
F_2(\tilde{\Lambda}) &=&  (1 + \ell_S) \mathrm{J}_{\frac{1}{2} + \ell_{S}}(\sqrt{\tilde{\Lambda}}) - \sqrt{\tilde{\Lambda}} 
\mathrm{J}_{\frac{3}{2} + \ell_{S}}(\sqrt{\tilde{\Lambda}})
 \nl
F_3(\tilde{\Lambda}) &=&  (1 + 2 \ell_S - \tilde{\Lambda}) \mathrm{J}_{\frac{1}{2} + \ell_{S}}(\sqrt{\tilde{\Lambda}}) - \sqrt{\tilde{\Lambda}} 
\mathrm{J}_{\frac{3}{2} + \ell_{S}}(\sqrt{\tilde{\Lambda}}) \; .
\label{eq:C2}
\ee
We now provide a simple proof, adapted from \cite{watson1922}, that all the functions $F_i(\tilde{\Lambda})$ only admit zeroes on the \emph{positive} real axis, and hence the denominator in this static case, $B_{\ell_S}(\tilde{\Lambda}) $, has no zeros for ${\rm Re}\,\tilde{\lambda} \le 0$.

For the first two conditions, we note that the statement we are after is equivalent to proving that $f_i(x)\equiv F_i(x^2)$ only admits zeroes $x_i$ for real values of $x$. Indeed, the first two conditions can be written in the following form
\begin{subequations}
\begin{equation}
A_i {\rm J}_{\nu_i}(x)+\,x\,{\rm J}^\prime_{\nu_i}(x)=0\,
\label{eq:Jscond}
\end{equation}
with
\begin{align}
&A_1=\ell_S+\frac{5}{2}\,,\qquad \nu_1=\ell_S+\frac{3}{2}\,.
\\
& A_2=\frac{1}{2}\,,\qquad \nu_2=\ell_S+\frac{1}{2}\,.
\end{align}
\end{subequations}
Next, we note that a Bessel function $J_\nu$ satisfies the following differential equation
\begin{equation}
\partial_x( x \partial_x {\rm J}_\nu)-\frac{\nu^2}{x}{\rm J}_\nu=-x {\rm J}_{\nu}\,.
\label{eq:besselJ}
\end{equation}
We now perform a change of variable, and set $x = y \alpha$, and take $y\in(0,1)$. Later on, $\alpha$ will be related to the zeroes of the $f_i$, but for now it is an arbitrary complex number. Under this change of variable, Eq.~(\ref{eq:besselJ}) is transformed into
\begin{equation}
\partial_y( y \partial_y j^{\alpha}_\nu)-\frac{\nu^2}{y}j^{\alpha}_\nu=-\alpha^2 y j^{\alpha}_{\nu}\,,
\label{eq:besselJy}
\end{equation}
with $j^{\alpha}_\nu(y)\equiv {\rm J}_{\nu}(\alpha y)$. We recognise the above as a St\"urm-Liouville problem (if appropriate boundary conditions are chosen) with $\alpha^2$ playing the role of eigenvalue and $j^{\alpha}_\nu$ the corresponding eigenfunction. Consider now two distinct eigenfunctions $j^{\beta}_\nu$ and $j^{\alpha}_\nu$. Consider the following combination
\begin{align}
-(\alpha^2-\beta^2)\int_0^{y} \tilde{y} j^{\beta}_\nu j^{\alpha}_\nu {\rm}d \tilde{y}& =\int_0^{y}\left\{j^{\beta}_\nu\left[\partial_{\tilde{y}}( {\tilde{y}} \partial_{\tilde{y}} j^{\alpha}_\nu)-\frac{\nu^2}{{\tilde{y}}}j^{\alpha}_\nu\right]-j^{\alpha}_\nu\left[\partial_{\tilde{y}}( {\tilde{y}} \partial_{\tilde{y}} j^{\beta}_\nu)-\frac{\nu^2}{{\tilde{y}}}j^{\beta}_\nu\right]\right\}{\rm d}\tilde{y} \nonumber
\\
& =y \left(j^{\beta}_{\nu}\partial_y j^{\alpha}_{\nu}-j^{\alpha}_{\nu}\partial_y j^{\beta}_{\nu}\right)\,,
\end{align}
where we assumed that $\nu>-1$, so that the boundary term evaluated at $y=0$ vanishes. We have thus concluded that
\begin{equation}
\int_0^{y} \tilde{y} j^{\alpha}_\nu j^{\beta}_\nu {\rm}d \tilde{y}=\frac{y}{\alpha^2-\beta^2}\left(j^{\alpha}_{\nu}\partial_y j^{\beta}_{\nu}-j^{\beta}_{\nu}\partial_y j^{\alpha}_{\nu}\right)\,.
\label{eq:eval}
\end{equation}
Let us take $\alpha$ to be a complex zero of $f_i$ and $\beta=\bar{\alpha}$ to be its complex conjugate. We will assume that $\alpha$ is \emph{not} purely imaginary for the moment, and return to this case later. Note that if $\alpha$ were purely imaginary, we would have $\alpha^2=\bar{\alpha}^2$, and the argument below would not work. Note that if $\alpha$ is a zero of $f_i$, so will $\bar{\alpha}$ because $f_i$ is a real function of its arguments. In terms of the $j^\alpha_{\nu}$, the condition (\ref{eq:Jscond}) can be written as
\begin{equation}
A_i\,j^{\alpha}_{\nu_i}(1)+{j^{\alpha}_{\nu_i}}^\prime(1)=0\,.
\label{eq:BCjalphanu}
\end{equation}
Evaluating (\ref{eq:eval}) with $y=1$ (and noting that $\alpha^2\neq \bar{\alpha}^2$) yields
\begin{equation}
\int_0^{y} \tilde{y} j^{\alpha}_{\nu_i} j^{\bar{\alpha}}_{\nu_i} {\rm}d \tilde{y}=0
\label{eq:c9}
\end{equation}
with Eq.~(\ref{eq:BCjalphanu}) regarded as the relevant boundary condition. But the left hand side is positive definite, since $j^{\bar{\alpha}}_{\nu_i}=\bar{j^{\alpha}_{{\nu_i}}}$. This is a contradiction, and as such complex zeroes of (\ref{eq:Jscond}) cannot exist.

We are left with analysing the case where $\alpha$ can potentially be purely imaginary. For this, we recall the series definition of the Bessel function
\begin{equation}
{\rm J}_\nu(\alpha)=\sum_{m=0}^{+\infty}\frac{(-1)^m}{m!\Gamma(\nu+m+1)}\left(\frac{\alpha}{2}\right)^{2m+\nu}\,,
\label{eq:series}
\end{equation}
where $\Gamma$ is a Gamma function. Our condition (\ref{eq:Jscond}) can be written as
\begin{equation}
\frac{{\rm d}}{{\rm d}\alpha}\left[\left(\frac{\alpha}{2}\right)^{A_i}{\rm J}_{\nu_i}(\alpha)\right]=0\Rightarrow \left(\frac{\alpha}{2}\right)^{\nu_i+A_i}\sum_{m=0}^{+\infty}(2m+\nu_i+A_i)\frac{(-1)^m}{m!\Gamma(\nu+m+1)}\left(\frac{\alpha}{2}\right)^{2m}=0\,.
\end{equation}
So long as $\nu_i+A_i>0$, all the terms on the right hand side of the above expression that are being summed over are positive definite for $\alpha$ being purely imaginary, thus excluding the possibility of purely imaginary zeroes. For our choice of $f_i$ then $\nu_i>0$ and $A_i>0$, thus establishing our proof. In fact, one can show with some effort that for $\nu_i+A_i<0$ two purely imaginary zeroes do exist.

The last condition in (\ref{eq:C2}) is a little more subtle, because it cannot be written as in (\ref{eq:Jscond}). However, it can be written as
\begin{equation}
(\nu-x^2){\rm J}_\nu(x)+x\,{\rm J}_\nu^\prime(x)=0\quad\text{with}\quad \nu=\ell_S+\frac{1}{2}\,.
\end{equation}
One could repeat the above argument replacing (\ref{eq:BCjalphanu}) with
\begin{equation}
(\nu-\alpha^2)j^\alpha_{\nu}(1)+{j^\alpha_\nu}^\prime(1)=0
\end{equation}
as the boundary condition to use on the St\"urm-Liouville problem. The right hand side in Eq.~(\ref{eq:c9}) no longer vanishes, but it is purely imaginary, and thus we still reach a contradiction for complex modes that are not purely imaginary or purely real. Finally, the case with purely imaginary zeroes can be dealt with in a slightly different manner. We note that this final boundary condition can also be written as
\begin{equation}
\nu(1-\nu)\,{\rm J}_\nu(\alpha)+2 \alpha\,{\rm J}_\nu^\prime(\alpha)+\alpha^2\,{\rm J}_\nu^{\prime\prime}(\alpha)=0\quad\text{with}\quad \nu=\ell_S+\frac{1}{2}\,.
\end{equation}
Using the series definition for ${\rm J}_{\nu}$ in (\ref{eq:series}) yields
\begin{equation}
\left(\frac{\alpha}{2}\right)^{\nu}\sum_{m=0}^{+\infty}2(2m+1)(m+\nu)\frac{(-1)^m}{m!\Gamma(\nu+m+1)}\left(\frac{\alpha}{2}\right)^{2m}=0\,
\end{equation}
which again is positive definite so long as $\nu>0$ and $\alpha$ is purely imaginary, establishing that $F_3$ only has zeroes on the positive real axis, just like $F_1$ and $F_2$.

Thus we conclude that $B_{\ell_S}(\tilde{\Lambda})$ has no zeros for ${\rm Re}\,\tilde{\Lambda} \le 0$, which implies that in the static case $\varpi = 0$ then the denominator in the expression for $p$ has no zeros for ${\rm Re}\,\tilde{\lambda} \le 0$, and hence $p$ is analytic in that region.

\subsection{The general case $\varpi \ne 0$}

Now let us consider the general non-static case where the denominator for $p$ is $B_{\ell_S}({\tilde{\Lambda}}) + \varpi^2 D_{\ell_S}({\tilde{\Lambda}})$. Now $D_{\ell_S}$ takes a more complicated form than $B_{\ell_S}$ and doesn't have a simple factorization. We proceed by noting that at a zero of the denominator,
\be
 \varpi^2 = - \frac{B_{\ell_S}({\tilde{\Lambda}})}{D_{\ell_S}({\tilde{\Lambda}})}
\ee
and $\varpi$ should be positive and real. We now aim to show that such a zero cannot occur for ${\rm Re}\,\tilde{\lambda} \le 0$ by showing that $\varpi$ cannot be positive real in the above expression there.

From their explicit forms both $B_{\ell_S}$ and $D_{\ell_S}$ are analytic for ${\rm Re}\,\tilde{\Lambda} < 0$, except for the branch cut at $\tilde{\Lambda} =0$ along the negative real axis. Hence, away from this negative real axis, $\varpi^2$ is an analytic function for ${\rm Re}\,\tilde{\Lambda} < 0$, and thus for ${\rm Re}\,\tilde{\lambda} < 0$, provided that $D_{\ell_S}$ has no zeros.
To analyze this it is convenient to normalize $D_{\ell_S}$ by the function
\be
F_{\ell_S}(\tilde{\Lambda}) = - \tilde{\Lambda}^3 \mathrm{J}_{\frac{1}{2} + \ell_{S}}\left(\sqrt{\tilde{\Lambda}}\right)^3  \mathrm{J}_{\frac{3}{2} + \ell_{S}}\left(\sqrt{\tilde{\Lambda}}\right) \; .
\ee
and consider the ratio,
\be
I_{\ell_S}(\tilde{\Lambda}) \equiv \frac{D_{\ell_S}(\tilde{\Lambda}) }{F_{\ell_S}(\tilde{\Lambda}) } \; .
\ee
Writing $\tilde{\Lambda} = x + i y$, then due to the properties of Bessel functions and our choice of normalizing factor, then $I_{\ell_S}$ is analytic for $x < 0$ and $y > 0$, and zeros of $D_{\ell_S}$ correspond to zeros of $I_{\ell_S}$ in that region. Now asymptotically, $I_{\ell_S}(\tilde{\Lambda}) = - 12/\tilde{\Lambda} + O\left(\tilde{\Lambda}^{-3/2}\right)$ for $x \le 0$ and $y \ge 0$. Hence ${\rm Im}\,I_{\ell_S}(\tilde{\Lambda}) \to 0$ asymptotically.
We also note that ${\rm Im}\,I_{\ell_S}$ vanishes on the real negative $\tilde{\lambda}$ axis. Thus on all boundaries or asymptotic regions of the quadrant $x \le 0$ and $y \ge 0$ except the imaginary axis then $I_{\ell_S}(\tilde{\Lambda})  \to 0$. So now we may plot this on the imaginary axis. This is shown in figure~\ref{fig:I_imagandrealaxis} for a range of $\ell_S = 0, \ldots , 6$. The lowest curve is the one for $\ell_S = 0$, and increasing $\ell_S$ gives the curves with greater value. This is true for the values of $\ell_S$ shown, but is true for all other values of $\ell_S$ we have tested. Thus it is positive on the imaginary axis. We note that it has a singularity at $\tilde{\Lambda} = 0$ and hence is discontinuous between the real and imaginary axes at the origin.
Since $I_{\ell_S}$ is analytic for $x < 0$ and $y>0$ then ${\rm Im}\,I_{\ell_S}$ is harmonic in the complex plane, and hence its minimum is found on a boundary or asymptotically. Hence we conclude that ${\rm Im}\,I_{\ell_S}$ is strictly positive for $x < 0$ and $y>0$, vanishing only on the negative real axis, $y =0$, or asymptotically, and hence ${\rm Im}\,I_{\ell_S}$ cannot have a zero except on the negative real axis. Then $I_{\ell_S}$ can have a zero only on the negative real axis. We may then check whether this occurs by considering ${\rm Re}\,I_{\ell_S}$ on the negative real axis. This is also plotted in the same figure~\ref{fig:I_imagandrealaxis} for $\ell_S = 0, \ldots, 6$ and again the lower curve is $\ell_S = 0$, bounding the others from below, and is everywhere greater than zero. Again this is true for the values of $\ell_S$ shown and all other values of $\ell_S$ we have tested. Thus we conclude that $I_{\ell_S}$ and hence $D_{\ell_S}$ has no zeros for ${\rm Re}\,\tilde{\Lambda} < 0$, and hence no zeros in the unstable complex plane ${\rm Re}\,\tilde{\lambda} < 0$.

\begin{figure}[ht]
    \centering
    \includegraphics[width=\textwidth]{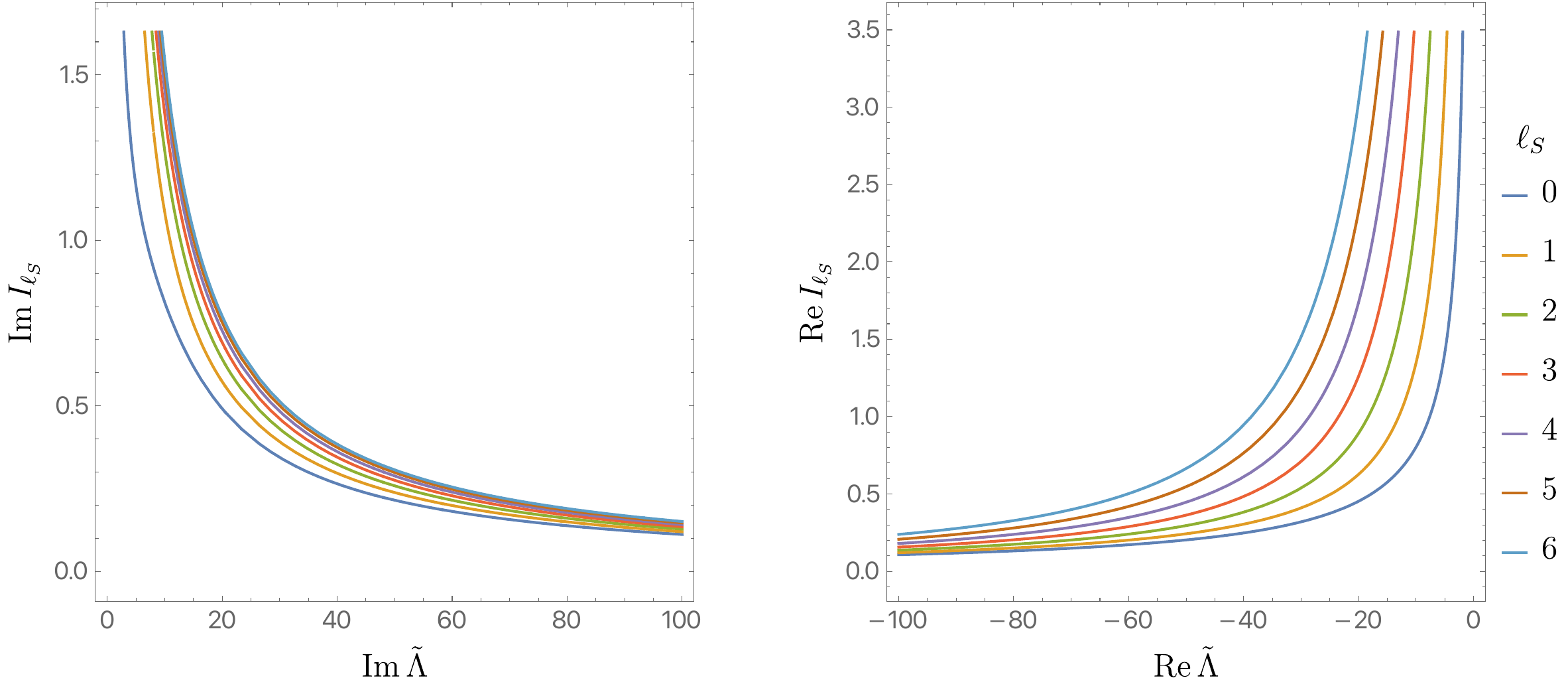}
    \caption{The lefthand figure shows a plot of ${\rm Im}\,I_{\ell_S}$ as a function of $\tilde{\Lambda}$ on the positive imaginary axis for various $\ell_S = 0,1,\ldots, 6$. We see that the curve $\ell_S = 0$ which is positive bounds those for the higher $\ell_S$ shown from below, so that the ${\rm Im}\,I_{\ell_S}$ are strictly positive on the positive imaginary axis. This holds for all other $\ell_S$ tested.
    The righthand figure shows ${\rm Re}\,I_{\ell_S}$ on the negative real $\tilde{\Lambda}$ axis for the same $\ell_S$. Again we see the curve for $\ell_S = 0$ bounds the others from below, and being positive implies ${\rm Re}\,I_{\ell_S} > 0$ on this negative real $\tilde{\Lambda}$ axis. Again this holds for all other $\ell_S$ tested.}
    \label{fig:I_imagandrealaxis}
\end{figure}

As a consequence we have that at a zero of the denominator $\varpi^2$ is analytic as a function of $\tilde{\Lambda}$ in the half plane ${\rm Re}\,\tilde{\Lambda} < 0$. While the numerator and denominator are singular at $\tilde{\Lambda} = 0$, their ratio which gives $\varpi^2$ is analytic there. Asymptotically it goes as,
\be
\varpi^2 & = & - \tilde{\Lambda} - 4 i \sqrt{\tilde{\Lambda}} + O(1) 
\ee
for $x < 0$ and $y > 0$ with the $\ell_S$ dependence being subleading.
Now ${\rm Im}\,\varpi^2 = 0$ on the negative real axis and its value on the positive imaginary axis is shown in figure~\ref{fig:varpi_imagandrealaxis} for $\ell_S = 0, \ldots , 6$, together with the asymptotic behaviour shown as a black dashed curve. We see it is everywhere negative on the positive imaginary axis, except at the origin. Its value asymptotically in the $x < 0$ and $y > 0$ quadrant away from the axes is ${\rm Im}\,\varpi^2 \to - {\rm Im}\,\tilde{\Lambda}$ for large $|\tilde{\Lambda}|$, and so is negative for $x < 0$ and $y > 0$. This holds for all other values of $\ell_S$ tested. Thus we conclude that since $\varpi^2$ is analytic for $x < 0$ and $y > 0$, then ${\rm Im}\,\varpi^2$ is harmonic in $x$ and $y$, so its minimum value lies on a boundaries or asymptotically, and hence it vanishes only on the negative real axis, and otherwise has non-zero imaginary part.

\begin{figure}[ht]
    \centering
    \includegraphics[width=\textwidth]{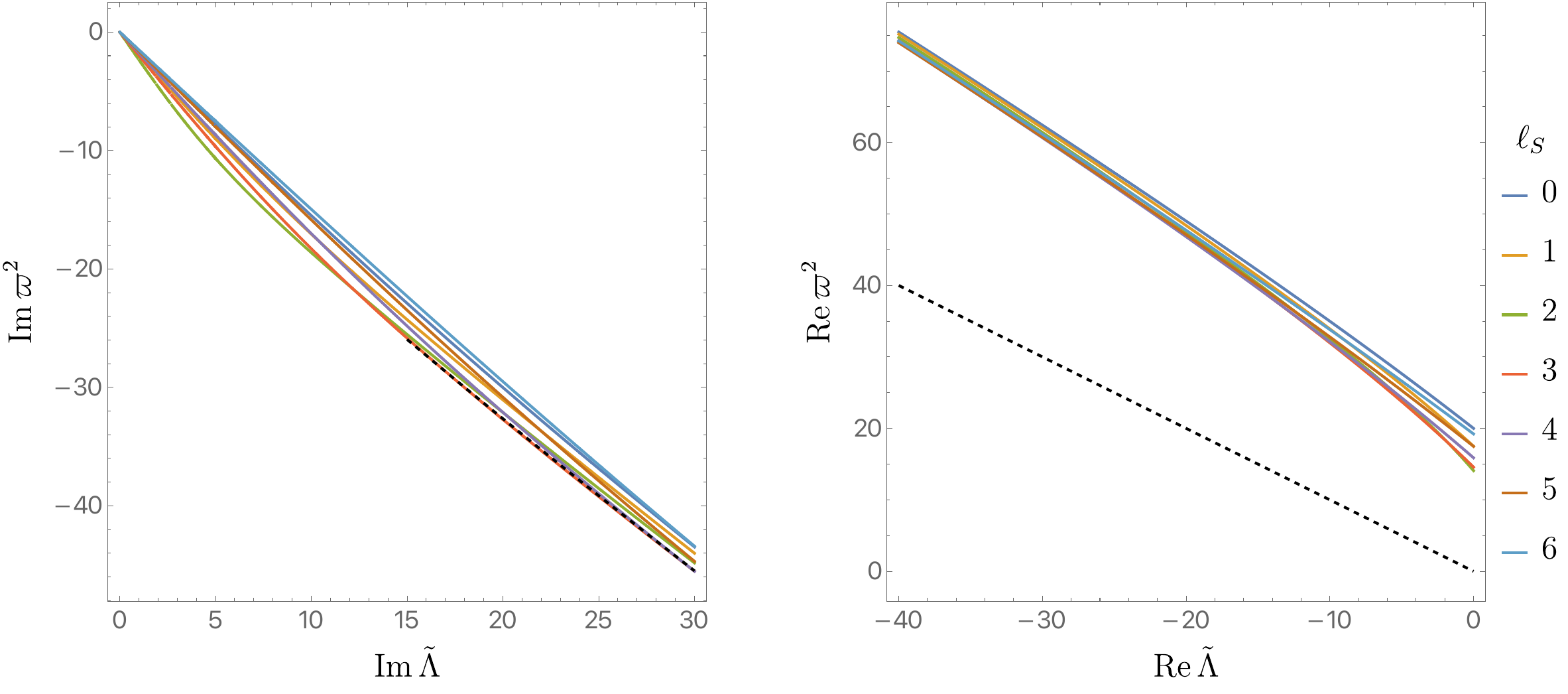}
    \caption{ The lefthand figure shows a plot of ${\rm Im}\,\varpi^2$ on the positive imaginary $\tilde{\Lambda}$ axis for various $\ell_S = 0,1,\ldots, 6$. The black dotted curve is the asymptotic behaviour given in the text. We find that ${\rm Im}\,\varpi^2$ is negative away from the origin on this positive imaginary axis for these $\ell_S$ and all other values examined.
    The righthand plot shows ${\rm Re}\,\varpi^2$ on the negative real $\tilde{\Lambda}$ axis again for the same $\ell_S$. Also plotted as a black dotted line is the condition $\varpi^2 = - \tilde{\Lambda}$, which lies below all the curves. We see that for these $\ell_S$ shown that while zeros of the denominator of $p$ can occur for ${\rm Re}\,\tilde{\Lambda} < 0$, they do not occur for ${\rm Re}\,\tilde{\Lambda} = - \varpi^2$, and hence for ${\rm Re}\,\tilde{\lambda} < 0$ where the corresponding eigenvalue would be unstable. This holds for all other values of $\ell_S$ we have tested.}
    \label{fig:varpi_imagandrealaxis}
\end{figure}

Since $\varpi^2$ must be real and positive, then for ${\rm Re}\,\tilde{\Lambda} < 0$ a zero of the denominator can only occur on the negative real $\tilde{\Lambda}$ axis, as $\varpi^2$ is otherwise complex. The question is then the behaviour of $\varpi^2$ on this negative real axis.
Thus finally in figure~\ref{fig:varpi_imagandrealaxis} we also plot $\varpi^2$ on the negative real $\tilde{\Lambda}$ axis for $\ell_S = 0, \ldots , 6$.  
For these values shown, and all others tested, we see $\varpi^2$ is positive there, and hence zeros are potentially allowed for ${\rm Re}\,\tilde{\Lambda} < 0$. However as we now argue, they are not allowed for ${\rm Re}\,\tilde{\lambda} < 0$, which implies the tighter condition ${\rm Re}\,\tilde{\Lambda} < - \varpi^2$. In order to see this on the plot we show a black dashed curve the condition $\tilde{\Lambda} = - \varpi^2$. 
Thus for all these $\ell_S$ shown, and all others checked, we find that for any given $\tilde{\Lambda} < 0$ we see that for a zero of the $p$ denominator, $\varpi^2$ is greater than $- \tilde{\Lambda}$, and hence corresponds to real strictly positive $\tilde{\lambda} = \varpi^2 + \tilde{\Lambda}$, rather than negative.

Thus finally we conclude that for ${\rm Re}\,\tilde{\lambda} < 0$ there are no zeros of the denominator of $p$. Hence $p$ is analytic in the half plane ${\rm Re}\,\tilde{\lambda} < 0$ corresponding to unstable eigenmodes.
\section{\label{app:ads}Static and Spherical Euclidean Modes for Vacuum Filled Cavity with $\Lambda\neq0$}

In this section we will restrict to four spacetime dimensions and study the eigenspectrum of~(\ref{eq:lichne}) in the static spherically symmetric sector for a cavity filled with the Euclidean vacuum (A)dS geometry. We write this background as,
\begin{subequations}
\begin{equation}
\widehat{{\rm d}s^2}\equiv \hat{g}_{ab}\mathrm{d}x^a\mathrm{d}x^b=F(r)^2{\rm d}\tau^2+\frac{{\rm d}r^2}{F(r)^2}+r^2 {\rm d}\Omega_2^2,
\end{equation}
with $\tau\sim \tau+\beta$ and
\begin{equation}
F(r)^2=1-3 \Lambda r^2\,.
\end{equation}
\end{subequations}
We will take the boundary to be the surface $r=R_0$ endowed with the metric on $S^1_{\beta}\times S^2_{R_0}$. For positive $\Lambda$, demanding $F(r)^2>0$ yields the bound $R_0\leq 1/\sqrt{3\Lambda}$.

We restrict our attention to modes that preserve spherical symmetry and are $\tau$ independent. Such modes can always be written as
\begin{equation}
\delta{\rm d}s^2\equiv h_{ab}\mathrm{d}x^a\mathrm{d}x^b=F(r)^2\,a(r)\,{\rm d}\tau^2+b(r)\,\frac{{\rm d}r^2}{F(r)^2}+r^2\,c(r)\,{\rm d}\Omega_2^2,
\end{equation}
where $a$, $b$ and $c$ are to be determined in what follows. The boundary condition $\delta \Gamma_{\mu\nu}=0$ discussed in section~\ref{sec:newbc} implies that the following relations must be obeyed at the cavity wall $r=R_0$:
\begin{subequations}\label{eq:bc}
\begin{equation}
a(R_0)=c(R_0)
\end{equation}
and
\begin{equation}
2\,F(r_0)\,R_0\,\left[a'(R_0)+2 c'(R_0)\right]-b(R_0) \left[4 F(r_0)+2R_0 F^\prime(R_0)\right]+6\,p \,a(R_0) \left[4 F(r_0)+2R_0 F^\prime(R_0)\right]=0\,.
\label{eq:bcpr}
\end{equation}
\end{subequations}
with $^\prime$ denoting differentiation with respect to $r$.
To proceed, it is convenient to introduce the scalar 
\begin{equation}
{\rm f}\equiv \hat{g}^{ab} h_{ab}=a(r)+b(r)+2c(r)\,.
\end{equation}
Exchanging $c$ by ${\rm f}$ and solving the de Donder gauge condition yields
\begin{subequations}\label{eq:dedonder}
\begin{align}
& a(r)=F(r)^2\, {\rm f}(r)+\frac{1}{2} r F(r)^2\,{\rm f}^\prime(r)-r F(r)^2\,b^\prime(r)+[1-4 F(r)^2] b(r)\,,
\\
& c(r)=\frac{1}{2}[{\rm f}(r)-a(r)-b(r)]\,.
\end{align}
\end{subequations}
The eigenvalue equation (\ref{eq:lichne}) is then solved with respect to ${\rm f}$ and ${\rm b}$, yielding
\begin{subequations}
\begin{align}
&\frac{1}{r^2}\left[r^2 F(r)^2 {\rm f}^\prime(r)\right]^\prime+2\,\Lambda\,{\rm f}(r)=-\lambda \,{\rm f}(r)
\\
& \frac{1}{r^4 F(r)^2}\left[r^4 F(r)^4 b^\prime(r)\right]^\prime+\frac{4}{3}\,\Lambda \, {\rm f}(r)-\frac{2 F(r)^2-1}{r} {\rm f}^\prime(r)-\frac{10}{3}\,\Lambda\,b(r)=-\lambda 
  \, b(r)\,.
\end{align}
\end{subequations}
For $\Lambda\neq0$, these equations can be solved in terms of hypergeometric functions $_2 F_1(a,b;c;z)$
\begin{subequations}
\begin{multline}
b(r)= \frac{2-\Lambda  r^2}{3 \lambda+2 \Lambda}\frac{{\rm f}'(r)}{r}+\frac{3 \lambda +4 \Lambda}{6 \lambda +4 \Lambda }{\rm f}(r) +\frac{C_1}{r^3} \, _2F_1\left(\frac{1}{4}-\eta _b,\frac{1}{4}+\eta _b;-\frac{1}{2};\frac{r^2 \Lambda }{3}\right)
\\
+C_2 \, _2F_1\left(\frac{7}{4}-\eta _b,\frac{7}{4}+\eta _b;\frac{5}{2};\frac{r^2 \Lambda
   }{3}\right)
\end{multline}
\begin{equation}
{\rm f}(r)=\frac{C_3}{r} \, _2F_1\left(\frac{1}{4}-\eta_{\rm f} ,\frac{1}{4}+\eta_{\rm f} ;\frac{1}{2};\frac{r^2 \Lambda }{3}\right)+C_4 \, _2F_1\left(\frac{3}{4}-\eta_{\rm f} ,\frac{3}{4}+\eta_{\rm f} ;\frac{3}{2};\frac{r^2 \Lambda }{3}\right)
\end{equation}
where we have defined
\begin{equation}
\eta_{b}=\frac{1}{4} \sqrt{9+\frac{12 \lambda }{\Lambda}}\quad\text{and}\quad \eta_{\rm f}=\frac{1}{4} \sqrt{33+\frac{12 \lambda }{\Lambda}}\,.
\end{equation}
\label{eqs:solp}
\end{subequations}\\
Regularity at the origin demands $C_1=C_3=0$, while our boundary conditions (\ref{eq:bcpr}) can be schematically written as
\begin{equation}
{\rm A}.\left[\begin{array}{c}
C_2
\\
C_4\end{array}\right]=0
\label{eq:matrixeq}
\end{equation}
with ${\rm A}$ a $2\times 2$ matrix that is a complicated function of the hypergeometric functions appearing in Eqs.~(\ref{eqs:solp}), and their derivatives, evaluated at $r=R_0$. We provide the explicit form at the end of this Appendix. To determine the eingenvalue $\lambda$ we simply demand $\det \rm A=0$, and solve numerically for $\lambda$ using a simple Newton-Raphson algorithm.

Before proceeding, let us note that for $\lambda=0$ the solutions we have determined coincide with those of the \emph{dynamical} linear problem that we have investigated in section \ref{sec:Lorentzian}. Indeed, it is a tedious but simple exercise to show that $\det \rm A=0$ when $\lambda=0$ and
\begin{equation}
p\equiv p^\star(R_0^2\Lambda)=\frac{6-R_0^2 \Lambda }{6 (3-R_0^2 \Lambda ) (2-R_0^2 \Lambda )}\,,
\label{eq:pcritical}
\end{equation}
in agreement with Eq.~(\ref{eq:growthads}) with $\alpha=0$. 

In the case $\Lambda =0$ this reduces to the case analysed in Section~\ref{sec:EucStaticSpherical}. Here we are interested in the case where $\Lambda\neq0$ where we can repeat the analysis but the calculations quickly become rather complicated. However, with some effort, one can find the equivalent of Eq.~(\ref{eq:linearp}) which we write as
\begin{subequations}
\begin{equation}
p=p^\star(R_0^2\Lambda)\left[1+\Xi(R_0^2\Lambda)R_0^2\lambda+\mathcal{O}(\lambda^2)\right]\Leftrightarrow R_0^2\lambda=\frac{p-p^*(R_0^2\Lambda)}{p^*(R_0^2\Lambda) \Xi \left(\Lambda  R_0^2\right)}
\label{eq:lambdaL}
\end{equation}
where
\begin{multline}
\label{eq:Xi}
\Xi(\tilde{\Lambda})=\frac{1}{(\tilde{\Lambda}-6) (\tilde{\Lambda}-3)}\Bigg\{\frac{9}{4}\frac{(\tilde{\Lambda}-4)^2}{(\tilde{\Lambda}-3)}\chi(\tilde{\Lambda})-\frac{9 \sqrt{3}}{2 \tilde{\Lambda
   }^{3/2}} {\rm arctanh}\left(\frac{\sqrt{\tilde{\Lambda}}}{\sqrt{3}}\right)
   \\
   -\frac{1}{(\tilde{\Lambda}-3) \tilde{\Lambda}}\left[\frac{27}{2}+51 \tilde{\Lambda}-44 \tilde{\Lambda}^2+12 \tilde{\Lambda}^3-\frac{9 \tilde{\Lambda}^4}{8}+\frac{1}{8} \sqrt{33} (\tilde{\Lambda}-4)^2 \tilde{\Lambda
   }^2\right]\Bigg\}\,,
\end{multline}
and
\begin{equation}
\chi(\tilde{\Lambda})=\frac{\, _2F_1\left(\frac{1}{4} \left(7-\sqrt{33}\right),\frac{1}{4} \left(3+\sqrt{33}\right);\frac{3}{2};\frac{\tilde{\Lambda}}{3}\right)}{\, _2F_1\left(\frac{1}{4} \left(7-\sqrt{33}\right),\frac{1}{4}
   \left(7+\sqrt{33}\right);\frac{5}{2};\frac{\tilde{\Lambda}}{3}\right)}\,.
   \label{eq:chi}
\end{equation}
\end{subequations}
In Fig.~\ref{fig:theta} we plot $\Xi(\tilde{\Lambda})$ as a function of $\tilde{\Lambda}$ and find that it is positive definite, diverging as $\Xi(\tilde{\Lambda})\propto (\tilde{\Lambda}-3)^{-1}$ near $\tilde{\Lambda}=3$.
\begin{figure}
    \centering
    \includegraphics[width=0.48\linewidth]{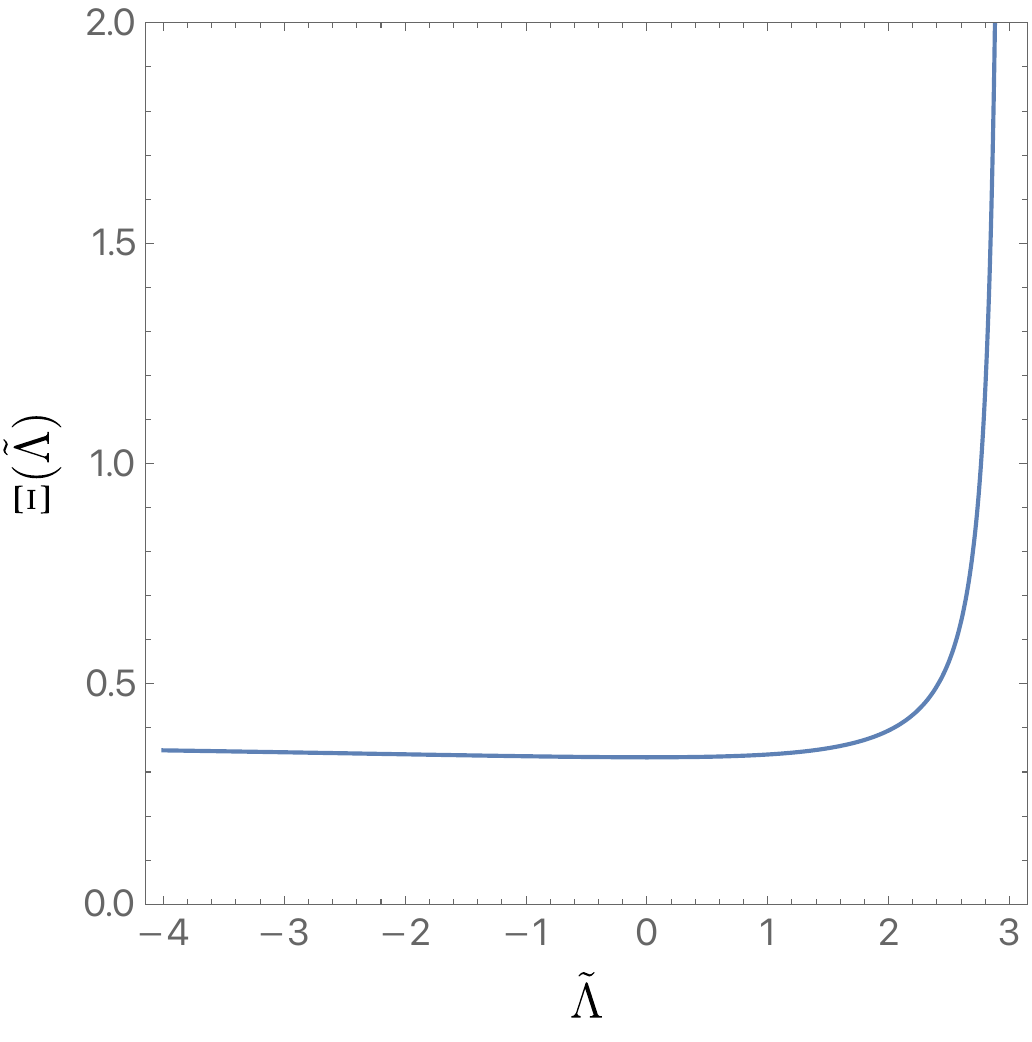}
    \caption{The function $\Xi(\tilde{\Lambda})$ defined in Eq.~(\ref{eq:Xi}) as a function of its argument.}
    \label{fig:theta}
\end{figure}

Since we have established positivity of $\Xi(\tilde{\Lambda})$ for $\tilde{\Lambda}\leq3$, we can go back to Eq.~(\ref{eq:lambdaL}) and read off the consequences for $\lambda$. We see that if $p^\star(R_0^2\Lambda)>0$, then we must require $p>p^\star(R_0^2\Lambda)$ in order to have stability. This will certainly happen for $\Lambda R_0^2<2$. However, if $p^\star(R_0^2\Lambda)<0$ stability now requires $p<p^\star(R_0^2\Lambda)$. Indeed, in the range $2< R_0^2\Lambda<3$ this is precisely what happens. Note that for $R_0^2\Lambda=2$ one can argue that only Dirichlet boundary conditions yield stability, since $\lim_{R_0^2 \Lambda\to 2^-}p^\star(R_0^2\Lambda)=+\infty$.

Finally, as promised above, we give the explicit form for the $2\times2$ matrix $\rm A$ appearing in Eq.~(\ref{eq:matrixeq})
\begin{multline}
4 {\rm A}_{11}=\left[R_0^2 \left(1+4 \eta _b\right) \Lambda +5-12 \eta _b\right] \, _2F_1\left(\frac{7}{4}-\eta _b,\frac{7}{4}+\eta _b;\frac{5}{2};\frac{R_0^2 \Lambda }{3}\right)
\\
-\left(4 \eta _b-7\right) \left(R_0^2
   \Lambda -3\right) \, _2F_1\left(\frac{11}{4}-\eta _b,\frac{7}{4}+\eta _b;\frac{5}{2};\frac{R_0^2 \Lambda }{3}\right)\,,
\end{multline}
\begin{multline}
8 (3 \lambda +2 \Lambda ) R_0^2 {\rm A}_{12}=4 \left(3-4 \eta_{\rm f}\right) \, _2F_1\left(\frac{7}{4}-\eta_{\rm f},\frac{3}{4}+\eta_{\rm f};\frac{3}{2};\frac{\Lambda  R_0^2}{3}\right)
\\
+ \left[16 \eta_{\rm f}-12-16 \eta_{\rm f}^2 \Lambda  R_0^2 \left(\Lambda  R_0^2-2\right)+3 (4 \lambda +11 \Lambda ) R_0^2 \left(\Lambda  R_0^2-2\right)\right]\, _2F_1\left(\frac{3}{4}-\eta_{\rm f},\frac{3}{4}+\eta_{\rm f};\frac{3}{2};\frac{\Lambda 
   R_0^2}{3}\right)\,,
\end{multline}
\begin{multline}
{\rm A}_{21}=2 (1+2 p) \, _2F_1\left(\frac{7}{4}-\eta _b,\frac{7}{4}+\eta _b;\frac{5}{2};\frac{\Lambda  R_0^2}{3}\right) \left(\Lambda  R_0^2-2\right)
\\
-\frac{1}{90} \left(16 \eta _b^2-49\right) \Lambda \,R_0^2 \left(\Lambda  R_0^2-3\right) \,
   _2F_1\left(\frac{11}{4}-\eta _b,\frac{11}{4}+\eta _b;\frac{7}{2};\frac{\Lambda  R_0^2}{3}\right)\,,
\end{multline}
and
\begin{multline}
3240 (3 \lambda +2 \Lambda )\,{\rm A}_{22}=-3240 [(6 p-3) \lambda -4 \Lambda ] \left(\Lambda  R_0^2-2\right) \, _2F_1\left(\frac{3}{4}-\eta_{\rm f},\frac{3}{4}+\eta_{\rm f};\frac{3}{2};\frac{\Lambda R_0^2}{3}\right)
\\
-\Lambda  \left(16 \eta_{\rm f}^2-9\right) \Bigg\{\Lambda  R_0^2 \left[6+\Lambda  R_0^2 \left(\Lambda  R_0^2-5\right)\right] \left(16 \eta_{\rm f}^2-49\right) \, _2F_1\left(\frac{11}{4}-\eta
   _h,\frac{11}{4}+\eta_{\rm f};\frac{7}{2};\frac{\Lambda  R_0^2}{3}\right)-
   \\
   30 \left[24-9 (\lambda +4 \Lambda ) R_0^2+\Lambda  (3 \lambda +10 \Lambda ) R_0^4+12 p \left(\Lambda  R_0^2-2\right){}^2\right] \, _2F_1\left(\frac{7}{4}-\eta
   _h,\frac{7}{4}+\eta_{\rm f};\frac{5}{2};\frac{\Lambda  R_0^2}{3}\right)
   \Bigg\}\,.
\end{multline}

\section{Euclidean Spherical Static Fluctuations of (A)dS Black Holes}\label{sec:ccthermo}

In this section, we numerically compute the spectrum of the Lichnerowicz operator of the Euclidean Schwarzschild black hole background with a non-vanishing cosmological constant.

The black hole solution to the bulk equations of motion inside a cavity at radius $r=r_B$ has the same form  as that in equation~\eqref{eq:BHsol}, 
with the blackening factor now given by 
    \begin{equation}\label{eq:BHf_A.dS}
    F(r)^2=-\kappa\frac{r^2}{\ell^2}+1-\frac{\rp}{r}\left(-\kappa\frac{\rp^2}{\ell^2}+1\right)\,,
    \end{equation}
where $\ell$ is the (A)dS length and is related to the cosmological constant via 
\begin{equation}
    \ell=\sqrt{\frac{3}{\kappa \Lambda}}\,.
\end{equation}
Here $\kappa=1$ and $\kappa=-1$ correspond to dS and AdS spacetimes, respectively. In addition to $x\equiv r_+/r_B$, we have another length scale $y\equiv r_+/\ell>0$\footnote{For dS black holes, we further require that the cavity is inside the cosmological horizon, which puts another constraint on the range of $x$ and $y$.}. 

The action is now given by
\begin{equation}
    S=S_{\rm EH}+S_{\rm bdy}=\frac{2\pi\ell^2}{G}\cdot\frac{(1-x^3)y^4}{x^3(-\kappa+3y^2)}-\frac{\Theta\beta_B\ell y}{4Gx^2}\cdot\frac{[(4-3x)x^2+3(-2+x^3)y^2\kappa]}{\sqrt{x^2(1-x)+\kappa y^2(x^3-1)}} \, ,
\end{equation}
and we have
\begin{equation}
    \mathcal{B}=\frac{4\pi}{1-3\kappa y^2}\sqrt{x^2(1-x)+(x^3-1)\kappa y^2}\,.
\end{equation}
The expression for $\mathcal{K}$ is rather tedious and not very illuminating, so we do not write it down explicitly here. Similar to the asymptotically flat case, there are two families of saddles with the same $\mathcal{B}$ and $\mathcal{K}$, parametrized by their own $x$ and $y$. We refer to the family with larger/smaller $x$ at a fixed $\mathcal{B}$ and $\mathcal{K}$ as the large/small black hole. One can verify that the small black hole always has a larger action than the large black hole, thus never dominates the ensemble.

Unlike the Dirichlet case\footnote{See equation (B.11) in \cite{Marolf:2022ntb} for the expression for Euclidean Schwarzschild AdS black holes. Our $x$ and $y$ are called $y_0^{-1}$ and $\yp$ in \cite{Marolf:2022ntb}, respectively.}, there is no analytical relationship between $y$ and $x$ at the place where $\mathcal{B}$ and the action are both extremized. This is also the place where we expect the black hole to be marginally stable. In figure~\ref{fig:critical_y_A.dS} we plot the numerical relationship between $y$ and $x$ at the critical place for $p=0$ (Anderson boundary condition). We also include the difference between critical $y$ for $p=0$ and that for the Dirichlet case $p\to+\infty$ in figure \ref{fig:critical_y_diff} to show explicitly that the critical points are $p$-dependent. The calculations become numerically challenging for general $p$ but in principle, one can get similar figures using a high enough precision.

\begin{figure}
    \centering
    \begin{subfigure}{0.3\textwidth}
    \includegraphics[width=\textwidth]{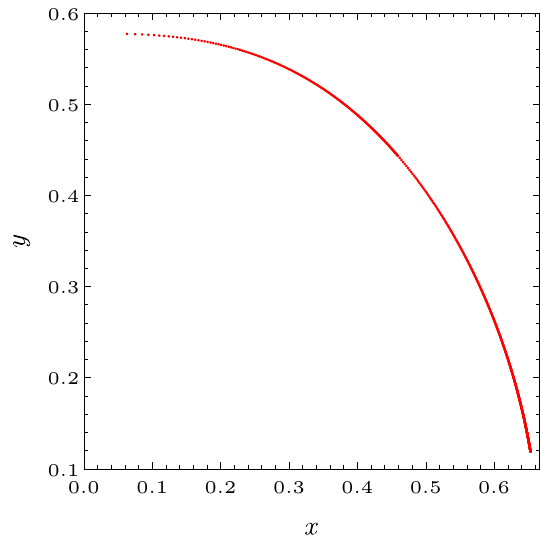}
    \caption{}
    \label{fig:critical_y}
    \end{subfigure}
     \begin{subfigure}{0.32\textwidth}
        \includegraphics[width=\textwidth]{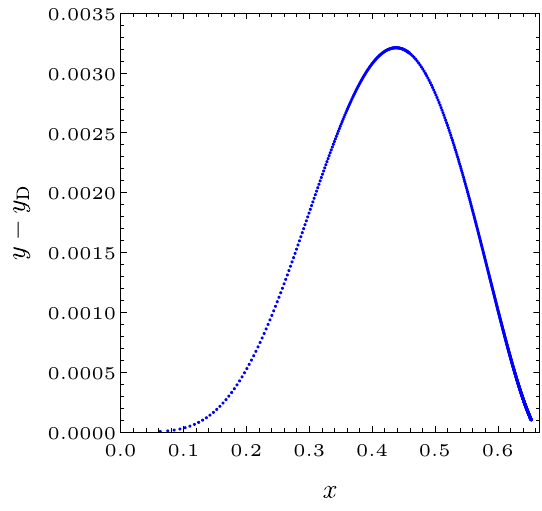}
    \caption{}
    \label{fig:critical_y_diff}
    \end{subfigure}
    \begin{subfigure}{0.307\textwidth}
        \includegraphics[width=\textwidth]{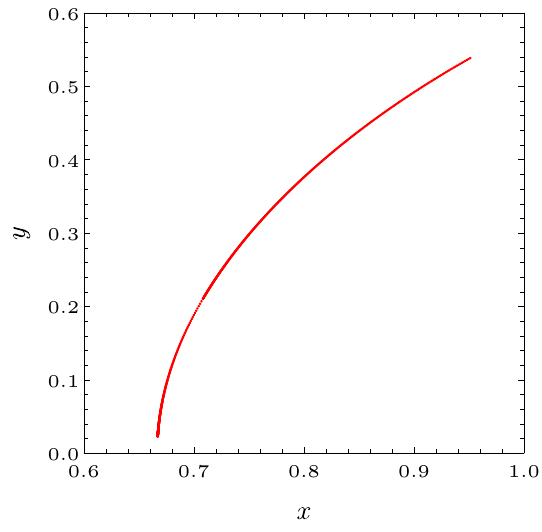}
    \caption{}
    \label{fig:critical_y_dS}
    \end{subfigure}
    \caption{The left/right panel gives the relationship between $y$ and $x$ at the place where $\mathcal{B}$ is extremized for Schwarzschild AdS/dS black holes. The middle panel gives the difference between the critical $y$ for the Anderson boundary condition and that for the Dirichlet boundary condition for Schwarzschild AdS black holes. At the place where $\mathcal{B}$ is extremized, we expect a marginal stability of the black hole.}
    \label{fig:critical_y_A.dS}
\end{figure}

Figures \ref{fig:AdS_BH_modes} and \ref{fig:dS_BH_modes} give the lowest four modes of the Lichnerowicz operator in the background of a Schwarzschild black hole in AdS and dS spacetimes inside a cavity, respectively. Our numerical data matches precisely (to numerical precision) with the prediction of marginal stability in section \ref{sec:ccthermo}. Again, the existence of another negative mode reflects the instability of the vacuum solution under Anderson boundary conditions.
\begin{figure}
    \centering
    \includegraphics[width=0.67\linewidth]{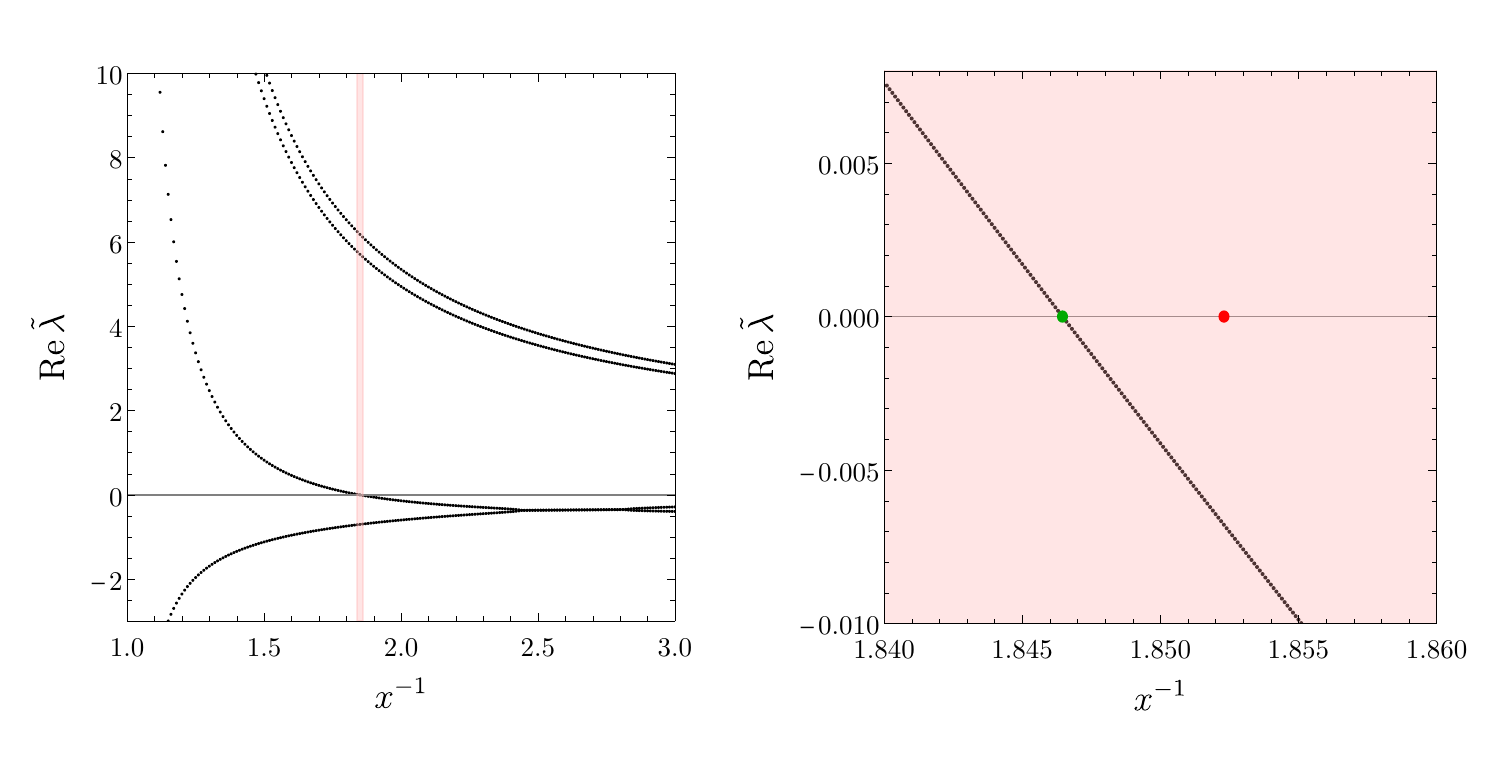}
    \caption[Caption for LOF]{The left panel shows the real part of the four lowest lying eigenvalues of $\hat{\Delta}_L$ as a function of $x^{-1}$ for the case with $y\equiv\rp/\ell\sim 0.3551$ and $\Lambda<0$. The right panel is a zoom-in version of the left panel at the place where there is a marginal stable mode. The green dots indicate the place of critical $x^{-1}$ predicted by $p=0$, as in figure \ref{fig:critical_y}. The red dots indicates the place of critical $x^{-1}$ predicted by the Dirichlet formula, equation (B.11) in \cite{Marolf:2022ntb}. The gray line is $\mathrm{Re}\,\tilde\lambda=0$.}
    \label{fig:AdS_BH_modes}
\end{figure}

\begin{figure}
    \centering
    \includegraphics[width=0.35\linewidth]{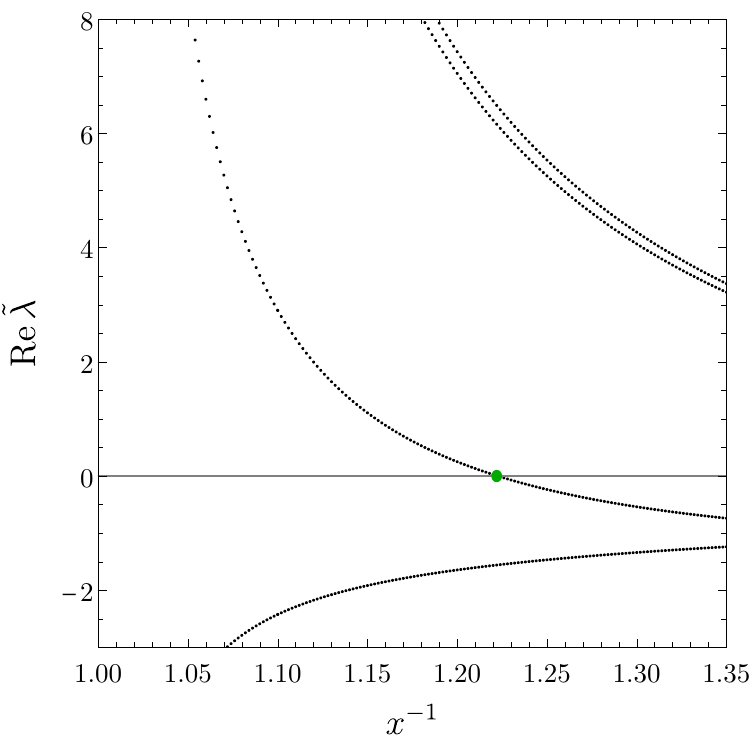}
    \caption[Caption for LOF]{The left panel shows the real part of the four lowest lying eigenvalues of $\hat{\Delta}_L$ as a function of $x^{-1}$ for the case with $y\equiv\rp/\ell= 0.4$ and $\Lambda>0$. The green dots indicate the place of critical $x^{-1}$ predicted by $p=0$, as in figure \ref{fig:critical_y_dS}. The gray line is $\rm Re\,\tilde\lambda=0$.}
    \label{fig:dS_BH_modes}
\end{figure}

~~~~~~~~~~~~
\addcontentsline{toc}{section}{References} 
\bibliographystyle{JHEP} 
\bibliography{references}

\providecommand{\href}[2]{#2}\begingroup\raggedright\begin{thebibliography}{10}

\bibitem{York:1986it}
J.~W. York, Jr., \emph{{Black hole thermodynamics and the Euclidean Einstein
  action}}, \href{https://doi.org/10.1103/PhysRevD.33.2092}{\emph{Phys. Rev. D}
  {\bfseries 33} (1986) 2092}.

\bibitem{Gross:1982cv}
D.~J. Gross, M.~J. Perry and L.~G. Yaffe, \emph{{Instability of Flat Space at
  Finite Temperature}},
  \href{https://doi.org/10.1103/PhysRevD.25.330}{\emph{Phys. Rev. D} {\bfseries
  25} (1982) 330}.

\bibitem{Whiting:1988qr}
B.~F. Whiting and J.~W. York, Jr., \emph{{Action Principle and Partition
  Function for the Gravitational Field in Black Hole Topologies}},
  \href{https://doi.org/10.1103/PhysRevLett.61.1336}{\emph{Phys. Rev. Lett.}
  {\bfseries 61} (1988) 1336}.

\bibitem{Prestidge:1999uq}
T.~Prestidge, \emph{{Dynamic and thermodynamic stability and negative modes in
  Schwarzschild-anti-de Sitter}},
  \href{https://doi.org/10.1103/PhysRevD.61.084002}{\emph{Phys. Rev. D}
  {\bfseries 61} (2000) 084002}
  [\href{https://arxiv.org/abs/hep-th/9907163}{{\ttfamily hep-th/9907163}}].

\bibitem{Reall:2001ag}
H.~S. Reall, \emph{{Classical and thermodynamic stability of black branes}},
  \href{https://doi.org/10.1103/PhysRevD.64.044005}{\emph{Phys. Rev. D}
  {\bfseries 64} (2001) 044005}
  [\href{https://arxiv.org/abs/hep-th/0104071}{{\ttfamily hep-th/0104071}}].

\bibitem{Maldacena:1997re}
J.~M. Maldacena, \emph{{The Large N limit of superconformal field theories and
  supergravity}}, \href{https://doi.org/10.4310/ATMP.1998.v2.n2.a1}{\emph{Adv.
  Theor. Math. Phys.} {\bfseries 2} (1998) 231}
  [\href{https://arxiv.org/abs/hep-th/9711200}{{\ttfamily hep-th/9711200}}].

\bibitem{Gubser:1998bc}
S.~S. Gubser, I.~R. Klebanov and A.~M. Polyakov, \emph{{Gauge theory
  correlators from noncritical string theory}},
  \href{https://doi.org/10.1016/S0370-2693(98)00377-3}{\emph{Phys. Lett. B}
  {\bfseries 428} (1998) 105}
  [\href{https://arxiv.org/abs/hep-th/9802109}{{\ttfamily hep-th/9802109}}].

\bibitem{Witten:1998qj}
E.~Witten, \emph{{Anti-de Sitter space and holography}},
  \href{https://doi.org/10.4310/ATMP.1998.v2.n2.a2}{\emph{Adv. Theor. Math.
  Phys.} {\bfseries 2} (1998) 253}
  [\href{https://arxiv.org/abs/hep-th/9802150}{{\ttfamily hep-th/9802150}}].

\bibitem{Witten:1998zw}
E.~Witten, \emph{{Anti-de Sitter space, thermal phase transition, and
  confinement in gauge theories}},
  \href{https://doi.org/10.4310/ATMP.1998.v2.n3.a3}{\emph{Adv. Theor. Math.
  Phys.} {\bfseries 2} (1998) 505}
  [\href{https://arxiv.org/abs/hep-th/9803131}{{\ttfamily hep-th/9803131}}].

\bibitem{Avramidi:1997sh}
I.~G. Avramidi and G.~Esposito, \emph{{Lack of strong ellipticity in Euclidean
  quantum gravity}},
  \href{https://doi.org/10.1088/0264-9381/15/5/006}{\emph{Class. Quant. Grav.}
  {\bfseries 15} (1998) 1141}
  [\href{https://arxiv.org/abs/hep-th/9708163}{{\ttfamily hep-th/9708163}}].

\bibitem{Avramidi:1997hy}
I.~G. Avramidi and G.~Esposito, \emph{{Gauge theories on manifolds with
  boundary}}, \href{https://doi.org/10.1007/s002200050539}{\emph{Commun. Math.
  Phys.} {\bfseries 200} (1999) 495}
  [\href{https://arxiv.org/abs/hep-th/9710048}{{\ttfamily hep-th/9710048}}].

\bibitem{Anderson:2006lqb}
M.~T. Anderson, \emph{{On boundary value problems for Einstein metrics}},
  \href{https://doi.org/10.2140/gt.2008.12.2009}{\emph{Geom. Topol.} {\bfseries
  12} (2008) 2009} [\href{https://arxiv.org/abs/math/0612647}{{\ttfamily
  math/0612647}}].

\bibitem{Figueras:2011va}
P.~Figueras, J.~Lucietti and T.~Wiseman, \emph{{Ricci solitons, Ricci flow, and
  strongly coupled CFT in the Schwarzschild Unruh or Boulware vacua}},
  \href{https://doi.org/10.1088/0264-9381/28/21/215018}{\emph{Class. Quant.
  Grav.} {\bfseries 28} (2011) 215018}
  [\href{https://arxiv.org/abs/1104.4489}{{\ttfamily 1104.4489}}].

\bibitem{Witten:2018lgb}
E.~Witten, \emph{{A note on boundary conditions in Euclidean gravity}},
  \href{https://doi.org/10.1142/S0129055X21400043}{\emph{Rev. Math. Phys.}
  {\bfseries 33} (2021) 2140004}
  [\href{https://arxiv.org/abs/1805.11559}{{\ttfamily 1805.11559}}].

\bibitem{Anninos:2023epi}
D.~Anninos, D.~A. Galante and C.~Maneerat, \emph{{Gravitational
  Observatories}},  \href{https://arxiv.org/abs/2310.08648}{{\ttfamily
  2310.08648}}.

\bibitem{Adam:2011dn}
A.~Adam, S.~Kitchen and T.~Wiseman, \emph{{A numerical approach to finding
  general stationary vacuum black holes}},
  \href{https://doi.org/10.1088/0264-9381/29/16/165002}{\emph{Class. Quant.
  Grav.} {\bfseries 29} (2012) 165002}
  [\href{https://arxiv.org/abs/1105.6347}{{\ttfamily 1105.6347}}].

\bibitem{Gibbons:1976ue}
G.~W. Gibbons and S.~W. Hawking, \emph{{Action Integrals and Partition
  Functions in Quantum Gravity}},
  \href{https://doi.org/10.1103/PhysRevD.15.2752}{\emph{Phys. Rev. D}
  {\bfseries 15} (1977) 2752}.

\bibitem{Randall:1999vf}
L.~Randall and R.~Sundrum, \emph{{An Alternative to compactification}},
  \href{https://doi.org/10.1103/PhysRevLett.83.4690}{\emph{Phys. Rev. Lett.}
  {\bfseries 83} (1999) 4690}
  [\href{https://arxiv.org/abs/hep-th/9906064}{{\ttfamily hep-th/9906064}}].

\bibitem{Brown:1992br}
J.~D. Brown and J.~W. York, Jr., \emph{{Quasilocal energy and conserved charges
  derived from the gravitational action}},
  \href{https://doi.org/10.1103/PhysRevD.47.1407}{\emph{Phys. Rev. D}
  {\bfseries 47} (1993) 1407}
  [\href{https://arxiv.org/abs/gr-qc/9209012}{{\ttfamily gr-qc/9209012}}].

\bibitem{Marolf:2022ybi}
D.~Marolf, \emph{{Gravitational thermodynamics without the conformal factor
  problem: partition functions and Euclidean saddles from Lorentzian path
  integrals}}, \href{https://doi.org/10.1007/JHEP07(2022)108}{\emph{JHEP}
  {\bfseries 07} (2022) 108}
  [\href{https://arxiv.org/abs/2203.07421}{{\ttfamily 2203.07421}}].

\bibitem{Headrick:2006ti}
M.~Headrick and T.~Wiseman, \emph{{Ricci flow and black holes}},
  \href{https://doi.org/10.1088/0264-9381/23/23/006}{\emph{Class. Quant. Grav.}
  {\bfseries 23} (2006) 6683}
  [\href{https://arxiv.org/abs/hep-th/0606086}{{\ttfamily hep-th/0606086}}].

\bibitem{Marolf:2022ntb}
D.~Marolf and J.~E. Santos, \emph{{The canonical ensemble reloaded: the
  complex-stability of Euclidean quantum gravity for black holes in a box}},
  \href{https://doi.org/10.1007/JHEP08(2022)215}{\emph{JHEP} {\bfseries 08}
  (2022) 215} [\href{https://arxiv.org/abs/2202.11786}{{\ttfamily
  2202.11786}}].

\bibitem{Gibbons:1978ac}
G.~W. Gibbons, S.~W. Hawking and M.~J. Perry, \emph{{Path Integrals and the
  Indefiniteness of the Gravitational Action}},
  \href{https://doi.org/10.1016/0550-3213(78)90161-X}{\emph{Nucl. Phys. B}
  {\bfseries 138} (1978) 141}.

\bibitem{Marolf:2022jra}
D.~Marolf and J.~E. Santos, \emph{{Stability of the microcanonical ensemble in
  Euclidean Quantum Gravity}},
  \href{https://doi.org/10.1007/JHEP11(2022)046}{\emph{JHEP} {\bfseries 11}
  (2022) 046} [\href{https://arxiv.org/abs/2202.12360}{{\ttfamily
  2202.12360}}].

\bibitem{Liu:2023jvm}
X.~Liu, D.~Marolf and J.~E. Santos, \emph{{Stability of saddles and choices of
  contour in the Euclidean path integral for linearized gravity: Dependence on
  the DeWitt Parameter}},  \href{https://arxiv.org/abs/2310.08555}{{\ttfamily
  2310.08555}}.

\bibitem{Odak:2021axr}
G.~Odak and S.~Speziale, \emph{{Brown-York charges with mixed boundary
  conditions}}, \href{https://doi.org/10.1007/JHEP11(2021)224}{\emph{JHEP}
  {\bfseries 11} (2021) 224}
  [\href{https://arxiv.org/abs/2109.02883}{{\ttfamily 2109.02883}}].

\bibitem{Bekenstein:1973ur}
J.~D. Bekenstein, \emph{{Black holes and entropy}},
  \href{https://doi.org/10.1103/PhysRevD.7.2333}{\emph{Phys. Rev. D} {\bfseries
  7} (1973) 2333}.

\bibitem{Hawking:1975vcx}
S.~W. Hawking, \emph{{Particle Creation by Black Holes}},
  \href{https://doi.org/10.1007/BF02345020}{\emph{Commun. Math. Phys.}
  {\bfseries 43} (1975) 199}.

\bibitem{Dias:2023rde}
O.~J.~C. Dias, G.~W. Gibbons, J.~E. Santos and B.~Way, \emph{{Static Black
  Binaries in de Sitter Space}},
  \href{https://doi.org/10.1103/PhysRevLett.131.131401}{\emph{Phys. Rev. Lett.}
  {\bfseries 131} (2023) 131401}
  [\href{https://arxiv.org/abs/2303.07361}{{\ttfamily 2303.07361}}].

\bibitem{1923rmp..book.....B}
G.~D. {Birkhoff} and R.~E. {Langer}, \emph{{Relativity and modern physics}}.
  Harvard University Press, 1923.

\bibitem{Gibbons:1979xm}
G.~W. Gibbons and S.~W. Hawking, \emph{{Classification of Gravitational
  Instanton Symmetries}},
  \href{https://doi.org/10.1007/BF01197189}{\emph{Commun. Math. Phys.}
  {\bfseries 66} (1979) 291}.

\bibitem{Zumino}
B.~Zumino, \emph{Supergravity},
  \href{https://doi.org/https://doi.org/10.1111/j.1749-6632.1977.tb37073.x}{\emph{Annals
  of the New York Academy of Sciences} {\bfseries 302} (1977) 545}
  [\href{https://arxiv.org/abs/https://nyaspubs.onlinelibrary.wiley.com/doi/pdf/10.1111/j.1749-6632.1977.tb37073.x}{{\ttfamily
  https://nyaspubs.onlinelibrary.wiley.com/doi/pdf/10.1111/j.1749-6632.1977.tb37073.x}}].

\bibitem{Gibbons:1976pt}
G.~W. Gibbons and M.~J. Perry, \emph{{Black Holes and Thermal Green's
  Functions}}, \href{https://doi.org/10.1098/rspa.1978.0022}{\emph{Proc. Roy.
  Soc. Lond. A} {\bfseries 358} (1978) 467}.

\bibitem{Hawking:1976jb}
S.~W. Hawking, \emph{{Gravitational Instantons}},
  \href{https://doi.org/10.1016/0375-9601(77)90386-3}{\emph{Phys. Lett. A}
  {\bfseries 60} (1977) 81}.

\bibitem{Hawking:1982dh}
S.~W. Hawking and D.~N. Page, \emph{{Thermodynamics of Black Holes in anti-De
  Sitter Space}}, \href{https://doi.org/10.1007/BF01208266}{\emph{Commun. Math.
  Phys.} {\bfseries 87} (1983) 577}.

\bibitem{Kodama:2003jz}
H.~Kodama and A.~Ishibashi, \emph{{A Master equation for gravitational
  perturbations of maximally symmetric black holes in higher dimensions}},
  \href{https://doi.org/10.1143/PTP.110.701}{\emph{Prog. Theor. Phys.}
  {\bfseries 110} (2003) 701}
  [\href{https://arxiv.org/abs/hep-th/0305147}{{\ttfamily hep-th/0305147}}].

\bibitem{Andrade:2015gja}
T.~Andrade, W.~R. Kelly, D.~Marolf and J.~E. Santos, \emph{{On the stability of
  gravity with Dirichlet walls}},
  \href{https://doi.org/10.1088/0264-9381/32/23/235006}{\emph{Class. Quant.
  Grav.} {\bfseries 32} (2015) 235006}
  [\href{https://arxiv.org/abs/1504.07580}{{\ttfamily 1504.07580}}].

\bibitem{doi:10.1098/rsta.1954.0021}
F.~W.~J. Olver and E.~C. Bullard, \emph{The asymptotic expansion of bessel
  functions of large order},
  \href{https://doi.org/10.1098/rsta.1954.0021}{\emph{Philosophical
  Transactions of the Royal Society of London. Series A, Mathematical and
  Physical Sciences} {\bfseries 247} (1954) 328}
  [\href{https://arxiv.org/abs/https://royalsocietypublishing.org/doi/pdf/10.1098/rsta.1954.0021}{{\ttfamily
  https://royalsocietypublishing.org/doi/pdf/10.1098/rsta.1954.0021}}].

\bibitem{Cardoso:2013pza}
V.~Cardoso, O.~J.~C. Dias, G.~S. Hartnett, L.~Lehner and J.~E. Santos,
  \emph{{Holographic thermalization, quasinormal modes and superradiance in
  Kerr-AdS}}, \href{https://doi.org/10.1007/JHEP04(2014)183}{\emph{JHEP}
  {\bfseries 04} (2014) 183} [\href{https://arxiv.org/abs/1312.5323}{{\ttfamily
  1312.5323}}].

\bibitem{Niehoff:2015oga}
B.~E. Niehoff, J.~E. Santos and B.~Way, \emph{{Towards a violation of cosmic
  censorship}},
  \href{https://doi.org/10.1088/0264-9381/33/18/185012}{\emph{Class. Quant.
  Grav.} {\bfseries 33} (2016) 185012}
  [\href{https://arxiv.org/abs/1510.00709}{{\ttfamily 1510.00709}}].

\bibitem{Chesler:2018txn}
P.~M. Chesler and D.~A. Lowe, \emph{{Nonlinear Evolution of the AdS$_4$
  Superradiant Instability}},
  \href{https://doi.org/10.1103/PhysRevLett.122.181101}{\emph{Phys. Rev. Lett.}
  {\bfseries 122} (2019) 181101}
  [\href{https://arxiv.org/abs/1801.09711}{{\ttfamily 1801.09711}}].

\bibitem{Chesler:2021ehz}
P.~M. Chesler, \emph{{Hairy black resonators and the AdS4 superradiant
  instability}}, \href{https://doi.org/10.1103/PhysRevD.105.024026}{\emph{Phys.
  Rev. D} {\bfseries 105} (2022) 024026}
  [\href{https://arxiv.org/abs/2109.06901}{{\ttfamily 2109.06901}}].

\bibitem{Jacobson:2013yqa}
T.~Jacobson and A.~Satz, \emph{{On the renormalization of the Gibbons-Hawking
  boundary term}},
  \href{https://doi.org/10.1103/PhysRevD.89.064034}{\emph{Phys. Rev. D}
  {\bfseries 89} (2014) 064034}
  [\href{https://arxiv.org/abs/1308.2746}{{\ttfamily 1308.2746}}].

\bibitem{Neri:2023esr}
G.~Neri and S.~Liberati, \emph{{On the resilience of the gravitational
  variational principle under renormalization}},
  \href{https://doi.org/10.1007/JHEP10(2023)054}{\emph{JHEP} {\bfseries 10}
  (2023) 054} [\href{https://arxiv.org/abs/2306.17505}{{\ttfamily
  2306.17505}}].

\bibitem{Fournodavlos:2019ckr}
G.~Fournodavlos and J.~Smulevici, \emph{{On the initial boundary value problem
  for the Einstein vacuum equations in the maximal gauge}},
  \href{https://arxiv.org/abs/1912.07338}{{\ttfamily 1912.07338}}.

\bibitem{Fournodavlos:2020wde}
G.~Fournodavlos and J.~Smulevici, \emph{{The Initial Boundary Value Problem for
  the Einstein Equations with Totally Geodesic Timelike Boundary}},
  \href{https://doi.org/10.1007/s00220-021-04141-8}{\emph{Commun. Math. Phys.}
  {\bfseries 385} (2021) 1615}
  [\href{https://arxiv.org/abs/2006.01498}{{\ttfamily 2006.01498}}].

\bibitem{Fournodavlos:2021eye}
G.~Fournodavlos and J.~Smulevici, \emph{{The Initial Boundary Value Problem in
  General Relativity: The Umbilic Case}},
  \href{https://doi.org/10.1093/imrn/rnab359}{\emph{Int. Math. Res. Not.}
  {\bfseries 2023} (2023) 3790}
  [\href{https://arxiv.org/abs/2104.08851}{{\ttfamily 2104.08851}}].

\bibitem{An:2020nfw}
Z.~An and M.~T. Anderson, \emph{{On the initial boundary value problem for the
  vacuum Einstein equations and geometric uniqueness}},
  \href{https://arxiv.org/abs/2005.01623}{{\ttfamily 2005.01623}}.

\bibitem{Anninos:2022ujl}
D.~Anninos, D.~A. Galante and B.~M\"uhlmann, \emph{{Finite features of quantum
  de Sitter space}},
  \href{https://doi.org/10.1088/1361-6382/acaba5}{\emph{Class. Quant. Grav.}
  {\bfseries 40} (2023) 025009}
  [\href{https://arxiv.org/abs/2206.14146}{{\ttfamily 2206.14146}}].

\bibitem{Regge:1957td}
T.~Regge and J.~A. Wheeler, \emph{{Stability of a Schwarzschild singularity}},
  \href{https://doi.org/10.1103/PhysRev.108.1063}{\emph{Phys. Rev.} {\bfseries
  108} (1957) 1063}.

\bibitem{Vishveshwara:1970cc}
C.~V. Vishveshwara, \emph{{Stability of the schwarzschild metric}},
  \href{https://doi.org/10.1103/PhysRevD.1.2870}{\emph{Phys. Rev. D} {\bfseries
  1} (1970) 2870}.

\bibitem{Zerilli:1970se}
F.~J. Zerilli, \emph{{Effective potential for even parity Regge-Wheeler
  gravitational perturbation equations}},
  \href{https://doi.org/10.1103/PhysRevLett.24.737}{\emph{Phys. Rev. Lett.}
  {\bfseries 24} (1970) 737}.

\bibitem{Moncrief:1974am}
V.~Moncrief, \emph{{Gravitational perturbations of spherically symmetric
  systems. I. The exterior problem.}},
  \href{https://doi.org/10.1016/0003-4916(74)90173-0}{\emph{Annals Phys.}
  {\bfseries 88} (1974) 323}.

\bibitem{Bardeen:1973xb}
J.~M. Bardeen and W.~H. Press, \emph{{Radiation fields in the schwarzschild
  background}}, \href{https://doi.org/10.1063/1.1666175}{\emph{J. Math. Phys.}
  {\bfseries 14} (1973) 7}.

\bibitem{Dafermos:2016uzj}
M.~Dafermos, G.~Holzegel and I.~Rodnianski, \emph{{The linear stability of the
  Schwarzschild solution to gravitational perturbations}},
  \href{https://doi.org/10.4310/ACTA.2019.v222.n1.a1}{\emph{Acta Math.}
  {\bfseries 222} (2019) 1} [\href{https://arxiv.org/abs/1601.06467}{{\ttfamily
  1601.06467}}].

\bibitem{Klainerman:2017nrb}
S.~Klainerman and J.~Szeftel, \emph{{Global Nonlinear Stability of
  Schwarzschild Spacetime under Polarized Perturbations}},
  \href{https://arxiv.org/abs/1711.07597}{{\ttfamily 1711.07597}}.

\bibitem{Dafermos:2021cbw}
M.~Dafermos, G.~Holzegel, I.~Rodnianski and M.~Taylor, \emph{{The non-linear
  stability of the Schwarzschild family of black holes}},
  \href{https://arxiv.org/abs/2104.08222}{{\ttfamily 2104.08222}}.

\bibitem{watson1922}
G.~N. Watson, \emph{A Treatise on the Theory of Bessel Functions}, ch.~XV,
  p.~482.
\newblock Cambridge University Press, Cambridge, 1922.

\end{thebibliography}\endgroup



\providecommand{\href}[2]{#2}\begingroup\raggedright\endgroup

\end{document}